\documentclass[preprint2]{aastex631}

\usepackage{amsmath}
\usepackage{layouts}
\usepackage{hyperref}

\newcommand{\vega}[1]{{#1}_\mathrm{Vega}}
\newcommand{\msun}{M_\odot}
\newcommand{\rimtimsim}{\texttt{RImTimSim}}

\shorttitle{Transiting Exoplanet Yields in the Roman GBTDS}
\shortauthors{Wilson et al.}

\graphicspath{{./}{figures/}}

\newcommand{\goddard}{NASA Goddard Space Flight Center, 8800 Greenbelt Rd, Greenbelt, MD, 20771}

\begin{document}

\title{
Transiting Exoplanet Yields for the Roman Galactic Bulge Time Domain Survey Predicted from Pixel-Level Simulations
}
%
%
\author[0000-0002-4235-6369]{Robert F. Wilson}
\affiliation{\goddard}

\author[0000-0001-7139-2724]{Thomas Barclay}
\affiliation{\goddard}
\affiliation{University of Maryland, Baltimore County, 1000 Hilltop Circle, Baltimore, MD 21250, USA}

\author[0000-0003-0501-2636]{Brian P. Powell}
\affiliation{\goddard}

\author[0000-0001-5347-7062]{Joshua Schlieder}
\affiliation{\goddard}

\author[0000-0002-3385-8391]{Christina Hedges}
\affiliation{\goddard}
\affiliation{University of Maryland, Baltimore County, 1000 Hilltop Circle, Baltimore, MD 21250, USA}

\author[0000-0001-7516-8308]{Benjamin T. Montet}
\affiliation{School of Physics, University of New South Wales, Sydney, NSW 2052, Australia}
\affiliation{UNSW Data Science Hub, University of New South Wales, Sydney, NSW 2052, Australia}

\author[0000-0003-1309-2904]{Elisa Quintana}
\affiliation{\goddard}

\author[0000-0003-0356-0655]{Iain Mcdonald}
\affiliation{Jodrell Bank Centre for Astrophysics, Department of Physics and Astronomy, University of Manchester, Oxford Road, Manchester M13 9PL, UK}
\affiliation{Open University, Walton Hall, Milton Keynes, MK7 6AA United Kingdom}

\author[0000-0001-7506-5640]{Matthew T. Penny}
\affiliation{Louisiana State University, Tower Dr., Baton Rouge, LA 70803-4001, USA}

\author[0000-0001-9513-1449]{N\'{e}stor Espinoza}
\affiliation{Space Telescope Science Institute, 3700 San Martin Drive, Baltimore, MD 21218, USA}
\affiliation{Department of Physics \& Astronomy, Johns Hopkins University, Baltimore, MD 21218, USA}

\author[0000-0002-1743-4468]{Eamonn Kerins}
\affiliation{Jodrell Bank Centre for Astrophysics, Department of Physics and Astronomy, University of Manchester, Oxford Road, Manchester M13 9PL, UK}


\begin{abstract}

The Nancy Grace Roman Space Telescope (Roman) is NASA's next astrophysics flagship mission, expected to launch in late 2026. As one of Roman's core community science surveys, the Galactic Bulge Time Domain Survey (GBTDS) will collect photometric and astrometric data for over 100 million stars in the Galactic bulge to search for microlensing planets. To assess the potential with which Roman can detect exoplanets via transit, we developed and conducted pixel-level simulations of transiting planets in the GBTDS. 
From these simulations, we predict that Roman will find between $\sim$60,000 and $\sim$200,000 transiting planets, over an order of magnitude more planets than are currently known. 
While the majority of these planets will be giants ($R_p>4R_\oplus$) on close-in orbits ($a<0.3$~au), the yield also includes between $\sim$7,000 and $\sim$12,000 small planets ($R_p<4 R_\oplus$). 
The yield for small planets depends sensitively on the observing cadence and season duration, with variations on the order of $\sim$10-20\% for modest changes in either parameter, but is generally insensitive to the trade between surveyed area and cadence given constant slew/settle times. 
These predictions depend sensitively on the Milky Way's metallicity distribution function, highlighting an opportunity to significantly advance our understanding of exoplanet demographics, particularly across stellar populations and Galactic environments.


\end{abstract}

\accepted{to ApJS}

%
\section{Introduction}
\label{sec:intro}
%

The Nancy Grace Roman Space Telescope
\footnote{Formerly known as the Wide-Field Infrared Survey Telescope (WFIRST)} 
(Roman) is a wide-field infrared survey observatory planned for launch in late 2026. Roman is designed to meet scientific objectives in cosmology and exoplanet demographics by executing three Core Community Surveys that will collectively account for up to 75\% of the science observing time over the 5-year Roman primary mission lifetime \citep{spergel2015,akeson2019}.
The technical capabilities that will enable new scientific advances using Roman data primarily come down to (1) a very high spatial resolution relative to the field of view and collecting area, (2) high observing efficiency, (3) exquisite calibration, and (4) large data downlink volume ($>$11 Tb/day).

The Wide Field Instrument (WFI) is the primary science payload on Roman, optimized for wide field optical and near-infrared (0.5-2.3 $\mu$m) imaging and slitless spectroscopy. It has a mosaic focal plane made up of 18 Teledyne H4RG-10 sensors \citep{mosby2020}. Each sensor will have 4096$\times$4096 pixels, of which 4088$\times$4088 will collect photons, each with a size of 10 $\mu$m and subtending $0.11\arcsec\times0.11\arcsec$ on the sky, resulting in a total effective field of view of 0.281 square degrees.

\subsection{The Galactic Bulge Time Domain Survey}

One reason for WFI's small pixel size is that it enables observations of high stellar-density regions of the sky, such as the Galactic bulge, without being overwhelmed by source confusion. This will be exploited by the Galactic Bulge Time Domain Survey (GBTDS) which has the primary aim of enabling the detection of gravitational microlensing events caused by exoplanets, but will enable a myriad of studies, including the detection of transiting planets considered in this work, by providing a versatile dataset for the science community. 
The design for the GBTDS will not be fully defined until much closer to the time of launch so that such decisions can incorporate input from the scientific community. However, most of the parameters for the GBTDS are  constrained by the requirements of the exoplanet microlensing goals \citep{penny2019}.

Due to the timescale of a microlensing event from a main sequence star towards the bulge, \citep[$t_E\approx$~10-40~days;][]{gaudi2012}, the season duration requirement has a minimum of $2t_E \approx 60$~days. The maximum cadence of approximately 15 minutes is set by the need to resolve multiple points within planetary-mass deviations, with typical timescales of $t_E\approx 2\,{\rm hr} \sqrt{M_p/M_\oplus}$ \citep{penny2019}. 
Finally, the goal of detecting 100 Earth-mass planets combined with the current best estimates for the microlensing detection rate and sensitivity set the need to continually monitor $\gtrsim 10^8$ stars for the duration of the survey \citep{penny2019}.

Combining these requirements with the observational constraints of the Roman observatory (e.g., expected slew and settle times and the orientation of the solar panels with respect to the Sun) and the goal of maximizing the number of detected microlensing events, the survey is expected to have six 60-72 day seasons centered on the Autumnal and Vernal Equinoxes, with three seasons clustered at the beginning of the 5-year primary mission lifetime and three seasons clustered at the end of the 5-year primary mission lifetime.

\begin{figure}
    \centering
    \includegraphics[width=\columnwidth]{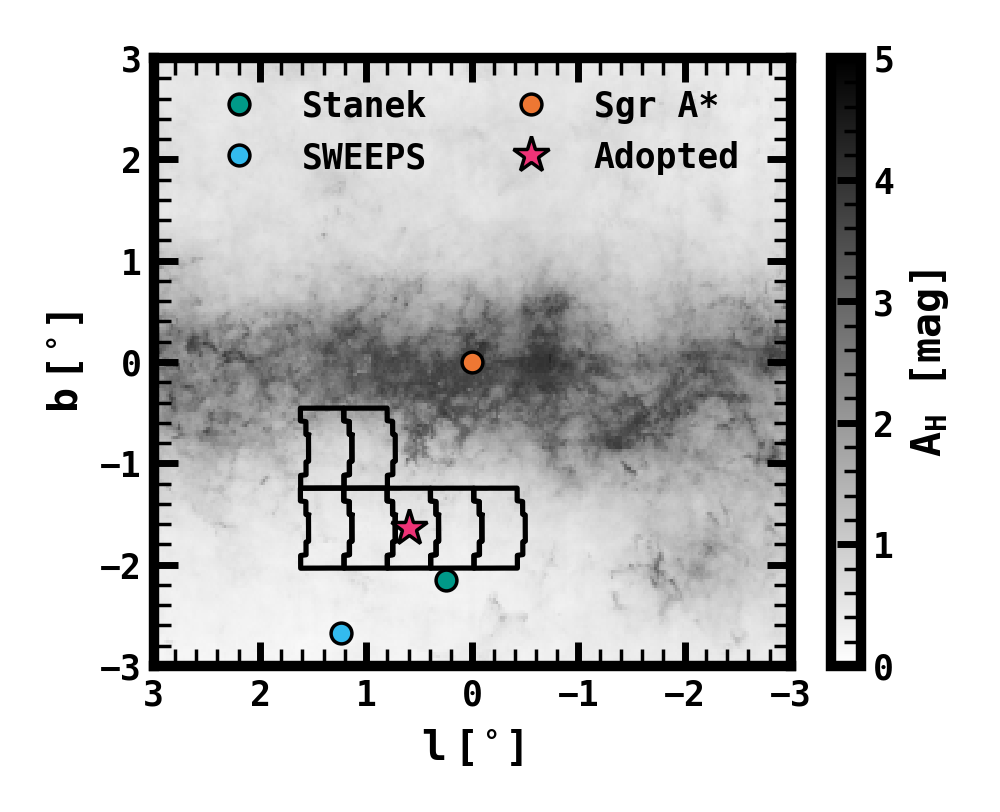}
    \caption{The field considered in this study, with the seven proposed GBTDS fields from \cite{penny2019} overlaid on an $H$-band extinction map of the Galactic Center. Also shown for the sake of comparison are a few well-studied fields, the Stanek window ($l= 0.25^\circ$, $b = -2.15^\circ$; green), the SWEEPS field ($l=1.24^\circ, b=-2.67^\circ$; blue), and Sagittarius A* (orange). The extinction map was calculated from the reddening map of \cite{surot2020} using extinction ratios from \cite{alonso-garcia2017}.  }
    \label{fig:fields}
\end{figure}

During each season Roman will observe approximately seven fields (an observing area of $\sim$2 square degrees) at a low Galactic latitude, with the location of the observed area chosen as an optimum between maximizing stellar densities and minimizing extinction (see Figure \ref{fig:fields}). 
Primary observations will be taken in the wide $F146$ filter (0.93-2.00~$\mu\mathrm{m}$) at an approximately 15-minute cadence and effective exposure time of $t_\mathrm{exp} = 54.7$~s. 
These observations will likely be supplemented with observations in at least one additional filter at a much lower cadence to measure colors for the purpose of stellar characterization, although the specifics of the observing strategy for additional filters has yet to be decided. One proposed plan is to sample two additional filters at a six-hour cadence, alternating between the $F087$ (0.76-0.99~$\mu\mathrm{m}$) and $F184$ (1.68-2.00~$\mu\mathrm{m}$) filters.

\subsection{Transiting Planets in the GBTDS}

Due to the high stellar densities, the Galactic bulge fields have long been proposed as a fruitful field for finding exoplanets via transit \citep{gaudi2000,bennett2002,mcdonald2014}.  
The OGLE-III microlensing survey supported this notion by finding dozens of candidate transiting planets despite the disadvantages of ground-based observing campaigns \citep{udalski2002a,udalski2002b}. However, due to confusion limits, the OGLE-III survey was only sensitive to transiting planets around disk stars, and from the fraction of systems that could be followed up it became clear that this catalog had a high false positive rate. As a result only a few of the candidate transiting planets have been reliably confirmed \citep[e.g.,][]{dreizler2003,bouchy2005}. 
To limit complications caused by crowding, the Sagittarius Window Eclipsing Extrasolar Planet Search \citep[SWEEPS; ][]{sahu2006} performed the first space-based transit survey of the Galactic bulge with the Hubble Space Telescope, improving both confusion limits and photometric precision. This 7 day campaign monitored $\sim$180,000 stars, resulting in 14 transiting planet candidates with an estimated false positive rate of $\lesssim$55\%, foreshadowing the scientific potential of a high spatial resolution, high cadence, photometric monitoring campaign of the Galactic bulge. The GBTDS should fully realize this potential.

The primary allure of transiting planet science in the GBTDS lies in the statistical power offered by the sheer number of expected detections. Combining early planet occurrence rates from the Kepler mission \citep{borucki2010}, estimates for Roman's photometric performance, and a Galactic population model, \cite{montet2017} predicted that the GBTDS should yield between $\sim$70,000 and $\sim$150,000 transiting planets,  an order of magnitude more planets than are currently known. 
The uncertainty in these predictions is driven by the strong correlation between planet occurrence and stellar metallicity combined with the Milky Way's  metallicity distribution function, highlighting perhaps an equally appealing facet of the GBTDS transiting planet science case: the breadth of Galactic populations surveyed.

Due to the chemical evolution of the Milky Way, stars in the local stellar neighborhood (within $\sim$1 kpc) are highly correlated between stellar age, metallicity, and more detailed chemical abundances, making it extremely difficult to test planet formation theories that predict multivariate trends between these parameters \citep{wilson2022}. 
This limitation can be overcome by surveying stellar populations outside the local stellar neighborhood, which vary significantly in age, metallicity, and detailed chemistry based on their Galactic location \citep{haywood2013,anders2014,hayden2015, bensby2017,zoccali2017,weinberg2019,griffith2021,queiroz2021,eilers2022}.
As we will show in this work, the GBTDS will be sensitive to transiting planets at distances of $\gtrsim$16-20 kpc, clear across the far side of the Galactic bulge. 
Thus, the Galactic distribution of transiting planets detected in the GBTDS will constrain correlations between planet occurrence and these multivariate parameters, providing valuable insight into the formation and evolution of giant planets.

\subsection{Goals and Organization of this Paper}

The goals of this paper are threefold. Our primary goal is to improve on the planet yield predictions made by \cite{montet2017}. We accomplish this by developing detailed pixel-level simulations of the GBTDS to estimate the photometric performance and overall efficiency with which transiting planets will be detected. These simulations are further improved with updated knowledge of planet occurrence rates and the stellar populations expected in the Galactic bulge fields.

Second, we wish to quantify the consequences of survey design trades (e.g., observing cadence vs. surveyed area) on the overall GBTDS transiting exoplanet science return and identify potential limiting systematics to inform the development of analysis software (e.g., photometric pipelines, transit search algorithms).

Our last goal is to identify unique science opportunities that will be enabled by the GBTDS transiting planet survey which cannot be achieved with existing state of the art observatories. 
Many of these science goals are enabled due to the large samples offered by the GBTDS, which will facilitate statistical studies of intrinsically rare events and planetary systems, while other science goals are enabled by the breadth of stellar populations surveyed.

The organization of this paper is as follows. In Section \ref{sec:simulations}, we describe the parameters of our simulations, including all assumptions regarding the underlying stellar and exoplanet populations, and explain our methodology for generating pixel-level simulations and synthetic data. 
In Section \ref{sec:detections} we generate and apply synthetic data to evaluate Roman's transit survey sensitivity and overall photometric performance. 
In Section \ref{sec:results} we present our estimated planet yields. Finally, we end this work with a discussion on the science enabled by the transiting planet sample and directions for future simulations in Section \ref{sec:discussion}, and reiterate our primary conclusions in Sections \ref{sec:summary} and \ref{sec:conclusions}.

%
\section{Simulating the Galactic Bulge Time Domain Survey} \label{sec:simulations}
%

In this section, we discuss our methodology for simulating the GBTDS. In section \ref{sec:defaultsurvey} we explain our adopted survey parameters. Sections \ref{sec:simstars} and \ref{sec:simplanets} describe assumptions about the observed stellar and exoplanet populations, respectively, and section \ref{sec:simdata} discusses our methodology for creating simulated data products.

\subsection{Default Survey Parameters }\label{sec:defaultsurvey}

 For this work we adopt nearly the same survey as the notional survey design presented by \citet{penny2019}, simulating only the $F146$ images at a 15-minute cadence, a maximum observing baseline of 72 days, six seasons, and seven observed fields. We ignore observations from a secondary filter, essentially making the assumption that such observations can either be interpolated over in the transit search pipeline, or that a multi-band transit search can be performed with minimal decreases in sensitivity. 
 These parameters are listed in Table \ref{tab:survey}, along with the relevant WFI technical capabilities.

\begin{table}[]
    \centering
    \caption{Assumed Survey Parameters and Detector, Telescope, and Noise Properties. }
    \begin{tabular}{lr} \hline 
    \multicolumn{2}{c}{\textbf{GBTDS Default Parameters} } \\ 
    Survey duration & 4.5 yr \\
    Seasons & 6 \\
    Fields & 7 \\
    Season Duration & 72 days \\
    Primary Bandpass  & $F146$\\
    Primary Cadence & 15 min \\
    Primary Exposure Time & 54 sec\\
    \multicolumn{2}{c}{\textbf{Number of Stars Observed in GBTDS} } \\ 
        Stars per Detector\tablenotemark{a}  & $\gtrsim 5\times10^6$\\
        Stars per Field\tablenotemark{a} & $\gtrsim 90 \times 10^6$\\
        Total Stars\tablenotemark{a}  & $\gtrsim 630 \times10^6$ \\
        Total Stars ($F146<15$) & $\sim 0.5\times 10^6$\\
        Total Stars ($F146<17$) & $\sim 1.7 \times 10^6$\\
        Total Stars ($F146<19$) & $\sim 8.9 \times 10^6$\\
        Total Stars ($F146<21$) & $\sim 59 \times 10^6$\\
        Total Stars ($F146<23$) & $\sim 220 \times 10^6$\\
    \multicolumn{2}{c}{\textbf{Wide Field Instrument Properties} } \\ 
        Field of View & $0.8^\circ \times 0.4^\circ$\\
        Detectors & 6$\times$3 \\
        Pixels per Detector & 4088$\times$4088 \\
        Plate Scale ($\arcsec$/pix) & 0.11  \\
        $F146$~Zeropoint (mag)    & 27.648 \\ 
        Readout Time (sec) & 3.04 \\
        Gain (counts/e$^-$) & 1 \\
        Read Noise (rms e$^-$/read) & 11 \\
        CDS Read Noise (rms e$^-$/read) & 16 \\
        Dark Current (e$^-$/pix/s) & $<$0.005 \\
        Bias (e$^-$/pix) & $10^3$ \\
        Saturation Limit (e$^-$/pix) & $10^5$ \\
    \multicolumn{2}{c}{\textbf{Additional Background}} \\
        $F146$~Sky (e$^-$/pix/s)\tablenotemark{b} & 4.25 \\
        $F146$~Thermal (e$^-$/pix/s) & 0.98  \\
        \hline
    \end{tabular}
    \raggedright
    \tablenotetext{a}{Lower limit based on our simulated stellar catalog with $\vega{H} < 26$. While the GBTDS has the sensitivity to detect stars as dim as $F146\sim30$, the confusion limit is  likely brighter than this. }
    \tablenotetext{b}{For this work we assume a constant of 5$\times$ minimum Zodiacal light. In reality, this quantity ranges from 2.5-7$\times$ the minimum Zodiacal light based on the time of year. However, most of the stars considered are dominated by other sources of uncertainty.}
    \label{tab:survey}
\end{table}

\subsection{Simulated Stellar Catalog}\label{sec:simstars}

To simulate a realistic population of planet-search stars, we apply the most recent version of the Besan\c{c}on Galactic population synthesis model \citep[referred to hereafter as BGM1612;][]{robin2003,robin2012,czekaj2014}, which we accessed via their webform.\footnote{\url{https://model.obs-besancon.fr}} The BGM1612 Model returns a simulated stellar catalog along some user-defined line of sight, which we choose to be $(l,b) = (0.60^\circ,-1.64^\circ)$, the location of the most central of the seven fields in the nominal survey design from \cite{penny2019}.

\begin{figure*}
    \centering
    \includegraphics[width=\textwidth]{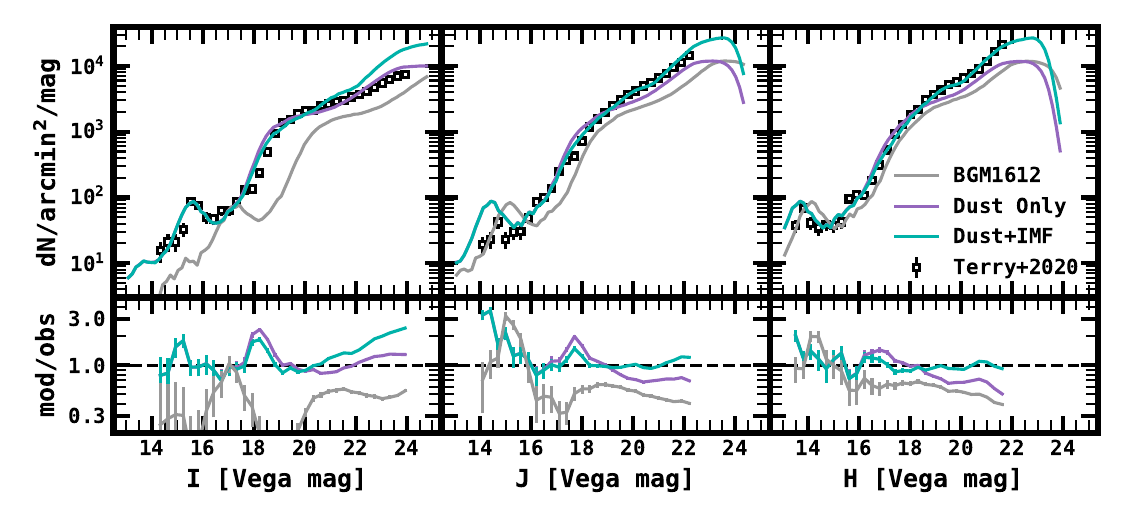}
    \caption{The Luminosity Function (LF) for the bulge population in our simulated catalog compared to observations. The top row of figures shows the empirical completeness corrected luminosity function (squares) from \cite{terry2020}, the LF from the default BGM1612 model catalog (gray), the LF after correcting for binarity and dust (purple), and our final adopted catalog LF which corrects for binarity, dust, and the bulge IMF (teal). 
    The bottom row of panels shows the star counts from the default BGM1612 catalog (gray), our catalog after correcting for dust and binarity (purple), and our adopted catalog (teal) divided by the empirical luminosity function. 
    The error bars were calculated assuming Poisson uncertainties from the empirical LF, adjusted for catalog completeness and proper motion selection efficiency. For the residuals, the empirical LF uncertainties were added in quadrature with the Poisson uncertainties from the generated catalog. The BGM1612 model under predicts NIR star counts below the main sequence turn-off (i.e., $J\gtrsim 18.5$, $H \gtrsim 18$), even after adjusting for dust (purple), which necessitates the amended bulge IMF in our final adopted catalog (teal). 
    The turnover at $H > 23$ is likely too early, and a result of the BGM1612 model excluding very low mass dwarfs ($M_\star < 0.15 M_\odot$) in the bulge population. }
    \label{fig:starcounts}
\end{figure*}

To make our simulated stellar catalog more realistic, we make several adjustments to the BGM1612 output to better agree with empirical bulge star counts from \cite{terry2020}. These authors made use of HST WFC3/UVIS and WFC3/IR observations of the Stanek window, $(l, b) = (0.25^\circ, -2.15^\circ)$, originally obtained by \cite{brown2009, brown2010} as part of the Galactic Bulge Treasury Program under GO-11664 and GO-12666 to measure stellar magnitude distributions in the Galactic bulge for several filters, and applied relative proper motions to remove disk contamination. Specifically, they measured star counts in the 
$F555W\;(V)$, $F814W\;(I)$, $F110W\;(J)$, and $F160W\;(H)$ filters\footnote{Note: The HST and Johnson-Cousins bandpasses are defined in the Vega system. The Roman filters, on the other hand, are defined in the AB system.}. For the analysis in this section, we make the simplifying assumption that the HST filters are equivalent to their corresponding filter in the Johnson-Cousins photometric system adopted by the BGM1612 model.
Because the transformation from $F555W$ to $V$ is less trivial \citep[see, e.g.,][]{harris2018} and more strongly impacted by reddening, we omit this filter from our analysis.
Our modifications are detailed below, and a comparison between our adopted stellar catalog and their completeness-corrected luminosity functions (LF) is shown in Figure \ref{fig:starcounts}. 
 Our primary adjustments include redefining the extinction law, adding binary stars in the catalog, and amending the IMF of the Galactic bulge population.

\subsubsection{Extinction}

Our first adjustment regards the treatment of dust. The extinction law adopted by the BGM1612 model assumes a universal Galactic extinction law with $R_V \approx 3.1$, where $R_V \equiv A_V / E(B-V)$ is the ratio of total to selective extinction.
Instead, we adopt the empirical law from \cite{cardelli1989} with $R_V = 2.5$, which is a more appropriate treatment for the Galactic bulge \citep{nataf2013}. Because the BGM1612 output generally over predicts the line of sight extinction, we reduce the median extinction in our catalog so that $A_I = 1.5$, which forces agreement between the location of the Red Clump in our catalog and the LF from \cite{terry2020} in the $I$, $J$, and $H$ bands.
However, it's worth noting that this is discrepant with measured values in the Stanek window, which typically have $A_I \approx 1.28$ \citep{reid2009,nataf2013}.

\subsubsection{Binarity}\label{sec:simstars_binarity}

To include the effects of stellar multiplicity in our sample, we randomly assign a fraction of stars in our catalog to be binaries. 
For simplicity, we ignore the effects of triple and higher order systems and only consider multiplicity for stars on the main sequence and subgiant branch, where transiting planets are detectable, and where the flux ratio between the primary and secondary stars in the system is most significant. 
To assign binary systems to our catalog, we follow the population statistics reported by \cite{duchene2013}, who compiled properties for stars of differing spectral types from various volume-limited imaging and spectroscopic surveys \citep[e.g.,][]{duquennoy1991,fischer1992,delfosse2004,raghavan2010,dieterich2012}.

For stars with $M_\star>0.7 \msun$, we assume a multiplicity fraction of 44\%, as measured for stars with $0.7 \msun < M_\star < 1.3 \msun$ in the Solar neighborhood. For stars with $M_\star\leq0.7 \msun$ we assume a multiplicity fraction of 26\%, typical of nearby stars with $M_\star<0.1\msun$. 
To assign an orbital period, we sample from a log-normal distribution with $\langle \log P \rangle = 5$ and $\sigma_{\log P} = 2.3$~for $M_\star>0.7 \msun$ and $\langle \log P \rangle = 3.8$ and $\sigma_{\log P} = 1.3$~for $M_\star\leq0.7 \msun$.
The distribution of binary mass ratios, $q$, with FGKM-type primaries is relatively flat ($dn/dq \sim q^{0.3}$), at least at larger orbital separations \citep{duchene2013}, so we adopt a constant mass ratio of $q=0.5$ for simplicity.

For our catalog we differentiate between three types of binary systems: resolved, unresolved, and close. A system is considered resolved if the projected orbital separation is $>$0.11$\arcsec$, the size of a WFI pixel, and a binary is considered close if the orbital separation is $<$100~au. We distinguish close binaries from unresolved binaries because evidence suggests that planet occurrence is only significantly suppressed in binary systems with orbital separations $\lesssim$100 au \citep{kraus2016,moe2021}.

For unresolved and close binaries, we add flux from the secondary star by interpolating between the mean absolute magnitudes for main sequence stars in the catalog in bins of 0.05 $\msun$, and then adjusting for extinction and distance. For resolved binaries, we make no adjustments as the presence of the secondary star should be reflected in the luminosity function.

\subsubsection{Bulge Initial Mass Function}

After the above extinction correction and inclusion of binary systems, the bulge population in the simulated catalog will still under predict the LF below the main-sequence turn off at $J \gtrsim 19$ and $H \gtrsim 18$, with an increasing discrepancy at dimmer magnitudes (purple line in Figure \ref{fig:starcounts}). This implies the need for a steeper IMF, as lower mass main sequence stars are underrepresented. 
To adjust for this we resampled the current catalog with replacement, so that the mass distribution would follow a newly defined IMF of the form $dn/dm \propto m^{-\alpha}$. We define the power law such that $\alpha=2.7$ at $M_\star>0.7 \msun$, and $\alpha=1.0$ at $0.7 \msun \leq M_\star < 0.15 \msun$.
For comparison, the default IMF used by BGM1612 adopts a power law broken at $0.7 M_\odot$~as well, but uses $\alpha = 2.3$ for the high mass sample and $\alpha=0.5$ for the low mass sample. Adopting the new IMF removes the slope in the residuals at dimmer magnitudes in the $J$ and $H$ bands. This amended IMF was only used to adjust the stars in the bulge population.

\subsubsection{Limitations and Other Discrepancies}

The LFs for our adopted catalog in the $I$, $J$, and $H$~bands are shown by the teal line in Figure \ref{fig:starcounts}. Our amendments to the original BGM1612 catalog were designed primarily to match the LF at dimmer magnitudes in the NIR. The residuals are slightly underpredicted in $H$ by 3\%, and slightly overpredicted in $J$ by 5\%, with typical scatter of 10\% in $J$ and 6\% in $H$. 
As a result of these priorities, the $I$-band LF between the two catalogs has larger discrepancies.

For example, our simulated catalog overestimates the empirical LF at 
$I \approx 18$ compared to the inferred uncertainties. 
This apparent magnitude range is dominated by subgiants and G dwarf binaries, and the discrepancy likely arises from our simplistic treatment of stellar multiplicity, where we assume a constant $q=0.5$ for all binary systems. 
Because the flux contrast between the primary and secondary star in a binary system depends more strongly on $q$ at shorter wavelengths, 
the same residual feature is reduced in the NIR (see $J\approx18$, $H\approx 16.5$ in Figure \ref{fig:starcounts}), and therefore shouldn't have a significant impact in our simulations. However, this discrepancy does lead to our adopting an IMF which reduces the inferred number of G0-2 dwarfs in the bulge, which may bias our planet yield estimates in this parameter space.
Because binary fraction is anti-correlated with metallicity, it is also possible that our model overestimates the binary fraction in the bulge as a whole \citep{badenes2018,moe2019,price-whelan2020}. This assumption is revisited in section \ref{sec:yield_binary}.

The largest discrepancy is at $\vega{I} \gtrsim 21$, where our adopted catalog overpredicts the LF by as much as a factor of $\sim$3 at $I\approx24$, with a steeply increasing slope at dimmer magnitudes. This leads to our catalog over predicting the total number of stars in the range $I=$~14--24 by 77\%. 
This is likely caused by a combination of effects, such as differences in the low mass IMF, our treatment of extinction, model isochrones for $M_\star<0.7 M_\odot$, and systematic differences in the adopted filters where at redder colors the assumption that $I \approx F814W$ diverges, with differences on the order of $\sim$0.2 mag \citep{harris2018}. 
However, a thorough examination of these effects is outside the scope of this paper as our motivations primarily require our catalog to reproduce the near infrared LF at $H \sim$~16--20 where most of the transiting planets will be detected \citep{montet2017} and from $H \lesssim 22$ where we can accurately simulate the effects of crowding. This goal is well accomplished, where the difference in total number of stars integrated over this range is $<0.1$\% between our adopted model and measurements.

One final shortcoming of our adopted catalog is the exclusion of very low mass dwarfs ($M_\star < 0.15 \, M_\odot$) from the bulge population. The exclusion of these stars results in an early turnover in the bulge LF ($H\sim23$). 
However, because we only generate light curves down to $H \approx 20$, this exclusion is unlikely to have a significant impact on our estimated transiting planet yield although it may lead us to underestimate photometric uncertainties caused by crowding and the diffuse, confusion-limited stellar background.

\subsubsection{Comparison to Other Simulated Bulge Catalogs}

Our methodology for creating a simulated catalog follows the logic of \cite{penny2019}, which focused primarily on microlensing event rates. \cite{penny2019} adopted an earlier version of the 
Besan\c{c}on Galactic population synthesis model \citep[BGM1106;][]{robin2012}, and then in a similar strategy scaled the microlensing event rate to match empirical star counts measured in the SWEEPS field \citep{calamida2015}. 
Where our catalog diverges is in the choice of empirical luminosity function and the particular bulge field with which we calibrated our stellar density estimates.

We used a Luminosity function measured from the Stanek field rather than the SWEEPS field, because it is closer to the proposed microlensing survey fields. From the choice of field alone, we should expect a 40\% increase in the total number of bulge stars \citep{terry2020}. Comparing the number of stars in our simulated stellar catalog to that used by \cite{penny2019}, we have a nearly constant $\sim$50\% increase in the number of bright ($F146<21$) stars in the full survey, 
which is approximately consistent with these expectations.

In addition, because we calibrated our catalog to $J$~and~$H$-band star counts, rather than $I$-band, we were prompted into adopting a steeper IMF for the bulge stars. 
This choice, in combination with the different choices in extinction, may inflate the number of lower mass stars ($<0.8 M_\odot$) in our catalog relative to the catalog from \cite{penny2019}.
This is apparent in a nearly $\sim$100\% increase in the number of stars with $F146<23$. This increase doesn't directly affect our results as we are only considering stars with $F146<21$, but the number of stars does directly contribute to crowding which will generally decrease the transit survey efficiency by adding additional photon noise, particularly for stars in dimmer magnitude ranges. Thus, in a somewhat contradictory manner, our increase in the number of stars, particularly at $F146\approx$~21-23, may actually lead to us predicting a transit survey efficiency that is too low. 

\subsection{Assumed Exoplanet Population}\label{sec:simplanets}

To simulate the exoplanet population, we follow the same procedure as in \cite{barclay2018}, which we outline below, but with some updates to the assumed occurrence rate distributions. The basic process is as follows: first, we draw a random number of planets from a Poisson distribution consistent with measured occurrence rates (i.e., number of planets per star). Next, each planet is assigned properties such as radius and orbital parameters.
Then each planet is randomly assigned to a main sequence or subgiant star within our stellar catalog, defined by the luminosity class assigned in the BGM1612 model. 
Even though low-luminosity red giant branch stars are known to host a similar number of large, close-in planets compared to their main sequence counterparts \citep{grunblatt2019,temmink2022}, only dwarf and subgiant stars are considered as viable planet hosts in this work because modeling RGB sources would require a more careful treatment of correlated noise from asteroseismic activity and second-order detector effects not presented here.

To be consistent with the observed demographics in planet orbital period ($P$) and radius ($R_p$), we adopt planet occurrence rates across two differing grids in the $P$-$R_p$ plane,
one for AFGK stars and one for M stars ($R_\star<0.6 R_\odot$, $T_{\rm eff}<3900\,{\rm K}$), based primarily on the occurrence rates from \cite{hsu2019} and \cite{dressing2015}, respectively. Table \ref{tab:occrates} gives an overview of the source of the assumed planet occurrence rates. While \cite{hsu2019} only report occurrence rates for FGK stars, the number of A stars in our simulated stellar catalog is small, so this is unlikely to impact our overall planet yield estimates.

Because Kepler discovered very few planets with $R_p>4\;R_\oplus$~orbiting M dwarfs, we are forced to make assumptions about the period and radius distribution of such planets. Thus, for this population, we assume the same $P$-$R_p$ distribution as was measured by \cite{hsu2019} for FGK stars out to $P=256$~days, but lower the overall occurrence by a factor of 3, consistent with the relative occurrence of giant planets between M dwarfs and G dwarfs measured by \cite{johnson2010}. 
This treatment is consistent with recent estimates of the occurrence of Hot Jupiters orbiting early M dwarfs from TESS \citep{gan2023}.

To account for long period planets around AFGK stars ($P>512$~days), we adopt the occurrence rates from \cite{herman2019}, who estimated occurrence rates for planets with orbital periods, $P=$~2--10 years and radii from $R_p = $~3.7--11.0~$R_\oplus$ from planet candidates with only one or two transits in the Kepler data, modified to have a constant occurrence rate density (i.e., number of planets per $\log P$-$\log R_p$ bin) over the period range of $P=$~512--7300 days. Because small planets are unlikely to be detected around AFGK dwarfs at long periods, we can safely ignore planets with $R_p < 3.7 R_\oplus$ at periods longer than 512 days without compromising our inferred detection yields.

Unfortunately, there is not a good estimate for the radius distribution of planets with M dwarf hosts at orbital periods of $P>200$~days. So instead, we adapt results from RV surveys for Jupiter-sized planets ($R_p>8R_\oplus$), and for small planets we extrapolate from the occurrence rates used at shorter orbital periods. 
More specifically, for smaller planets, $(R_p < 8 R_\oplus)$, we extrapolate the occurrence rate densities at the longest period bin ($P=200$~days for $R_p<4R_\oplus$ and $P=256$~days for $R_p=$~4--8~$R_\oplus$) already adopted for each radius bin, out to $P=7300$~days, similar to the methodology adopted by \cite{kunimoto2022}.

For Jupiter-sized planets, $R_p$~=~8--12 $R_\oplus$, at $P>256$~days we adapt the results from \cite{montet2014}, who used direct imaging to rule out stellar contaminants and interpret long term RV trends, and inferred an occurrence rate of 0.083$\pm$0.019 planets per star within a planet mass and semi-major axis range of $M_p$ = 1--13~$M_{\rm Jup}$, and $a$ = 0--20~au. 
Subtracting the giant planet occurrence rates adopted for $P<256$~days, and scaling the occurrence rate to preserve a constant occurrence rate density results in an inferred occurrence of 0.036 planets per star with $P$~= 256--7300 days and $R_p$~= 8-12~$R_\oplus$.

These grids result in a total inferred planet occurrence rate of  $\lambda_{\rm AFGK} = 2.69$~and $\lambda_{\rm M}=4.68$ planets per star. 
For each planet, six random values are assigned: orbital period ($P$), planet radius ($R_p$), epoch of first transit ($t_0$), eccentricity, cosine of the inclination ($\cos i$), and argument of periastron ($\Omega$). 
To determine $P$ and $R_p$, each planet is assigned a bin based on the relative occurrence of each bin in the two occurrence rate grids, as described above. 
Once the $P$-$R_p$ bin is assigned, the period and radius are drawn from a log-uniform distribution over the period and radius range of the bin size. 
For the argument of periastron and the cosine of inclination, we draw from a uniform distribution of $-\pi$ to $\pi$, and from 0 to 1, respectively. The eccentricities are drawn from a beta distribution with parameters $\alpha=1.03$ and $\beta=13.6$, consistent with findings from \cite{vaneylen2015}. The epoch of first transit is chosen from a uniform distribution between 0 and $P$.

All planets within a system are assumed to be coplanar, and we intentionally do not adjust multiple planet systems with unphysical orbits, such as those that cross each other, to remain consistent with empirical constraints on the overall number of planets per star. Because of this simplistic treatment, our methodologies will bias the inferred yield for multi-planet systems.

We remove planets with randomly assigned semi-major axes, $a$, within the stellar radius, $a \leq R_\star$. This limit may prove interesting because we consider subgiant stars while \cite{hsu2019} restricted their planet-search sample to main sequence stars, so these populations may have intrinsically different minimum semi-major axes due to planet engulfment. 
However, changing this requirement to $a\leq 2R_\star$ or $a\leq 3R_\star$ to consider the possibility of planet engulfment has a relatively minor impact ($\lesssim$3\%) on the overall planet yield due to the intrinsically low planet occurrence measured by \cite{hsu2019} at $P<1$~day, so we elect not to change this limit.

\begin{table*}
    \centering

    \begin{tabular}{llll} \hline
       {\bf Spectral Type}  & {\bf Period (day)} &{\bf Radius ($R_\oplus$)}  &  {\bf Reference}  \\ \hline
       AFGK & 0.5--512 & 0.5--16 &  \cite{hsu2019} \\ 
         & 512--7300 & 3.7--11 & \cite{herman2019} \\ \hline
       M & 0.5--200 & 0.5--4 &  \cite{dressing2015} \\
         & 200--7300 & 0.5--4 & \cite{dressing2015}\tablenotemark{a} \\
         & 0.5--256 & 4--16 &  \cite{hsu2019}, \cite{johnson2010}\tablenotemark{b} \\
         & 256--7300 & 4--8 &  \cite{hsu2019}, \cite{johnson2010}\tablenotemark{a,b} \\
         & 256--7300 & 8--12 &  \cite{montet2014} \\
       \hline
    \end{tabular}
    \tablenotetext{a}{Extrapolated for each radius bin from longest period bin.}
    \tablenotetext{b}{Radius distribution adopted from \cite{hsu2019} with the overall occurrence scaled to match the occurrence rate for M dwarf hosts relative to G dwarfs measured by \cite{johnson2010}}
    \caption{Literature sources for adopted planet occurrence rates across the $P$-$R_p$ grid.}
    \label{tab:occrates}
\end{table*}

\subsubsection{Changes in Exoplanet Population with Stellar Metallicity} \label{sec:simplanets_metals}

Because planet occurrence is intimately correlated with stellar metallicity, we also consider the yield under assumptions that the occurrence weights used above scale with host star metallicity. For this case, 
when computing the overall planet detection yields, we weight the sum of detected planets by specific planet classes based on their relative expected occurrence defined by the host star's metallicity. For this purpose we adopt the correlations measured in \cite{wilson2022}, who parameterized the occurrence rate as a power law, $f_p \propto 10^{\beta {\rm [Fe/H]}}$, and inferred $\beta$ for different planet radii and orbital period ranges (hot planets: $P\leq10$~days, warm planets: $P=$~10--100~days). The adopted values are shown in Table \ref{tab:meteffects}.
For hot, large radius planets (i.e., sub-Saturns and Jupiters) where \cite{wilson2022} did not infer a reliable dependence due to small number statistics, we instead adopt the equivalent measurements from \cite{petigura2018}. In these cases, it is worth noting that the values from \cite{wilson2022} were inferred from a magnitude limited sample in the infrared, which was deeper than that from \cite{petigura2018}, and as a result is more likely to be representative of the sample used to calculate occurrence rates by \cite{hsu2019}. Because these metallicity-planet occurrence relations were only valid for metallicities as high as ${\rm [Fe/H]} \approx 0.4$, we do not extrapolate the power law relation beyond this upper limit in metallicity.

These metallicity-based occurrence rates were calculated for FGK host stars, but we elect to use them for M dwarf hosts as well. Due to the difficulty in deriving metallicities for M dwarfs, there is not an equivalent study with a parameterized occurrence rate-metallicity dependence for a range of planet radii and orbital period, from a similarly well-characterized sample (i.e., with metallicities from high-resolution, high signal-to-noise spectra, where $\sigma_{\rm [Fe/H]} < 0.1$~dex).
However, there is some evidence that, for late-type stars, metallicity is strongly correlated with both giant planet occurrence \citep{neeves2013,montet2014} and small planet occurrence \citep{lu2020}. 
These studies suggest that the dependence between planet occurrence and stellar metallicity may actually be stronger for low-mass stars than for solar-type stars. As a result, it is possible that changes in yield due to metallicity effects may actually be underestimated for low-mass stars in this work.

\begin{table}
    \centering
    \begin{tabular}{l c c}\hline
       \bf $R_p/R_\oplus$ &  $\beta_{\rm hot}$ & $\beta_{\rm warm}$ \\ \hline
        $<$1.9 & 0.72  & -0.62 \\
        1.9-4.0 & 2.34  & 0.47 \\
        4.0-8.0 & 5.5\tablenotemark{a}  & 2.1  \\
        $>$8.0 & 3.4\tablenotemark{a} & 0.86 \\ \hline
    \end{tabular}
    \tablenotetext{a}{From \cite{petigura2018}.}
     \caption{The power law dependence of planet occurrence rate on metallicity adopted for this study.}
    \label{tab:meteffects}
\end{table}

\subsection{Simulated Data}\label{sec:simdata}

\begin{figure*}
    \centering
    \includegraphics[width=\textwidth]{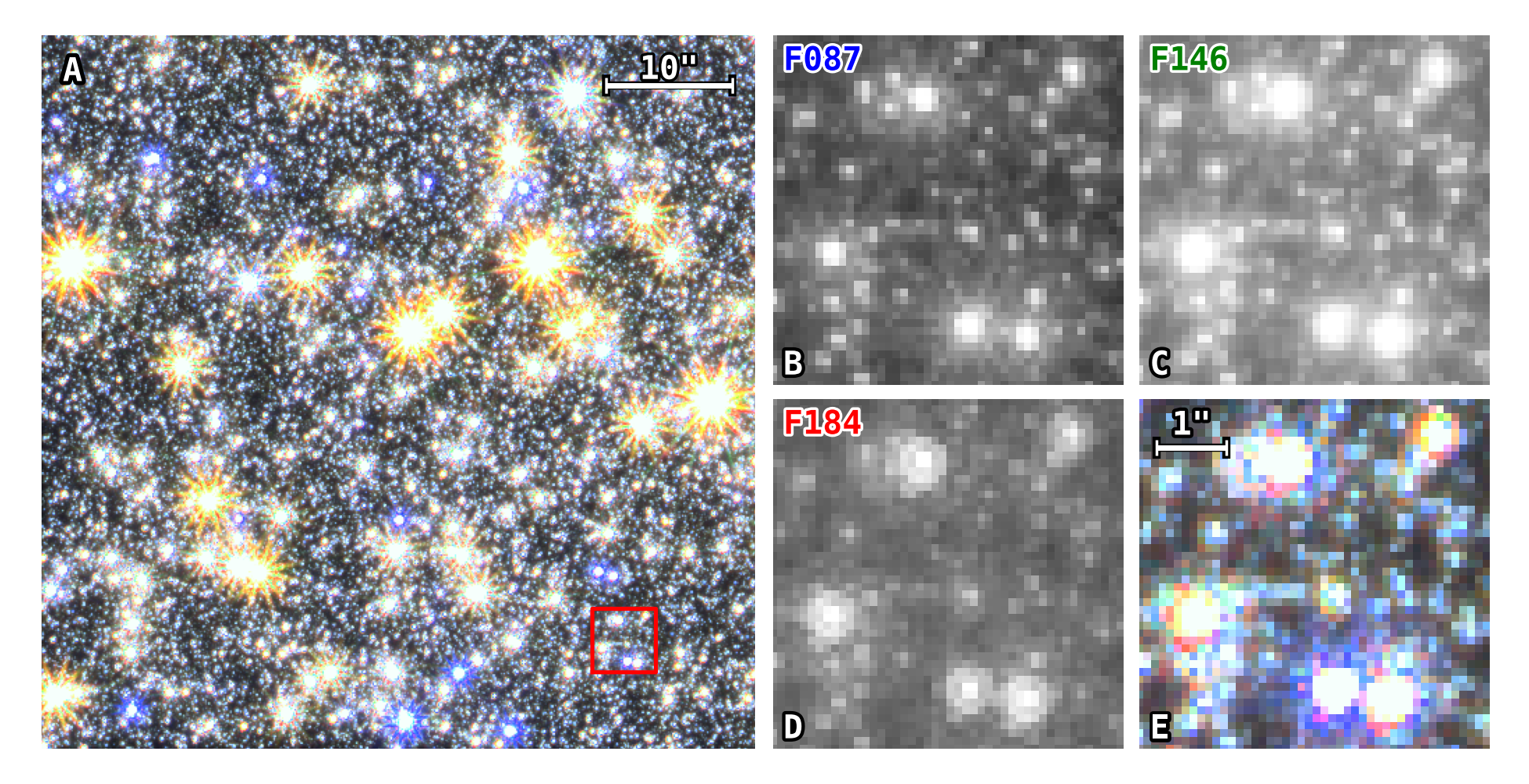}
    \caption{Simulated false color image of the adopted GBTDS field created by \rimtimsim. (A) A 512$\times$512 pixel RGB image, which corresponds to $\sim$0.09\% of the full WFI field of view. The pixel scale is demonstrated by the white bar in the top right corner. The combined image consists of a 215 second exposure in the F184 bandpass, a 46.8 second exposure in the F146 bandpass, and a 180 second exposure in the F087 bandpass for the red, green, and blue channels, respectively. 
    The red box is a 5\arcsec$\times$5\arcsec~cutout with more detail shown in subfigures B-E. (B-E) A zoomed-in version of the red box in (A), showing each individual RGB component, as well as the combined image. The individual bandpass images (B,C,D) are on a logarithmic scale, while the RGB images (A,E) are scaled to highlight the diffraction spikes using a modified version of the algorithm from \cite{lupton2004}.   } 
    \label{fig:rgb}
\end{figure*}

For the purpose of including the effects of crowding and pipeline uncertainties on the final transiting planet yield, we simulate all images for one of the detector assemblies near the edge of the Roman FOV. 
For this purpose we use a custom-created tool, the Roman IMage and TIMe-series SIMulator \citep[\rimtimsim;][]{robertfwilson_2023_8221758}\footnote{\url{https://github.com/robertfwilson/rimtimsim}}, to create simulated calibrated images.
\rimtimsim\ aims to improve upon synthetic Roman data generated in previous Galactic bulge simulations by accounting for the specific readout pattern and ``up-the-ramp'' sampling of the Roman sensors compared with prior works that modeled the instrument noise as a CCD \citep[e.g.,][]{montet2017,penny2019,johnson2020}. 
The process with which \rimtimsim\ generates simulated images is outlined in the following sections. To demonstrate the typical noise, optical, and detector properties captured by \rimtimsim, a simulated RGB color image is shown in Figure \ref{fig:rgb}.

\subsubsection{Expected Source Flux}

\rimtimsim\ calculates an expected flux per pixel based on the locations and magnitude of stars in the image. Each star is represented by a delta function that specifies its position on the detector and expected flux amplitude. The expected source flux is then calculated by convolving the sum of delta functions by an appropriately normalized point spread function (PSF) over the detector array. 
Finally other flux sources are added to the image, such as the expected sky flux and thermal background, to create the final expected source flux image. To include noise, the expected source flux is scaled according to the exposure time, and then the ``measured" flux is determined by randomly drawing from a Poisson distribution with the mean equal to the expected flux.

For this work we assume a constant PSF and define the centroid of a star on the detector to within a four by four grid per pixel. 
This oversampling is adequate to accurately model crowding from stars with sub-pixel separations and to prevent systematic photometric errors in our simulations that may otherwise correlate with the position of the source relative to the center of a pixel.
The model PSF was calculated via the python package \texttt{webbpsf}\footnote{\url{https://github.com/spacetelescope/webbpsf}} \citep{perrin2014}, where we adopted 8 mas of high-frequency pointing jitter (i.e., within a single exposure), assumed the optical center of detector assembly SCA07, an array at the edge of the WFI field of view, and an SED representative of a M0V star which is consistent with the average temperature of a star in our simulated catalog.
Because we assume a constant PSF across the detector, the expected source flux can be computed via a single Fast Fourier Transform (FFT), significantly increasing the computational efficiency of our simulated image generation pipeline. 
Thus, our methodology makes an implicit assumption that the photometric performance, and by extension the transiting planet yield, depends only weakly on PSF variations due to variables such as position on the focal plane and source SED. 
Because we do not consider the response of the pixels themselves, we are also making an implicit assumption that intra-pixel sensitivity variations are negligible.

\subsubsection{Pixel-Level Injections}\label{sec:injections}

Each star in our simulated catalog has a defined magnitude in the Johnson-Cousins $RIJHK$ photometric system. We first convert the $RIJHK$ magnitudes to $F146$, assuming the relation,
\begin{align}
    F146 = (J_{AB} + H_{AB} + \frac{K_{AB}}{2})/2.5 \;,
\end{align}
where the $JHK$ magnitudes are first converted to the AB system\footnote{For reference we adopt $J_{\rm AB}-J = 0.91$, $H_{AB}-H = 1.39$, and~$K_{AB}-K = 1.85$}. 
We use these magnitudes to define the baseline flux for the delta functions outlined in the previous section.

To inject the transiting planet signals into the simulated images at the pixel-level, we update the baseline flux for each star with a transiting planet, at each time, $t$, with a transit model light curve calculated via the python package \texttt{batman} \citep{kreidberg2015}. This model is computed with the following parameter basis: orbital period ($P$), planet-star radius ratio ($R_p/R_\star$), scaled semi-major axis ($a/R_\star$), epoch of first transit ($t_0$), inclination ($i$), eccentricity ($e$), and longitude of periastron ($\Omega$).
The light curves adopt a quadratic limb-darkening law with the $H$-band coefficients from \cite{claret2011}, with each star adopting the nearest point on a grid of $T_\mathrm{eff}$, $\log g$, and [M/H].
In this way, we modify the existing stellar catalog at each cadence to accommodate time-dependent flux variations due to the transit signal. 
Because we do not add any other source of variation, the light curves derived from this work should contain purely white noise, with the exception of transiting planet signals from the source itself or from a nearby star polluting the aperture.

\subsubsection{Detector Readout}

Because WFI sensors can be read non-destructively (e.g. without a reset), the readout strategy used by Roman will be to sample ``up-the-ramp". This method records the signal of each pixel multiple times during an exposure \citep{Fowler1990}. The whole sensor is read non-destructively every 3.04 seconds; this read is called a frame. The number of frames in an integration (which ends with a reset) can be set to different values depending on the science objectives.

\rimtimsim\ offers two choices to simulate the WFI read out scheme: ``ccd" and ``ramp". The ``ccd" mode treats the exposures as a CCD would, where pixels collect photons for a determined exposure time before being destructively read out. 
To correctly simulate the noise for bright stars that would saturate in the nominal exposure time, \rimtimsim\ increases the saturation limit by a factor of $t_\mathrm{exp}/t_\mathrm{read}$, but compensates by increasing the photon noise for pixels whose expected source flux is above the nominal saturation limit. This results in an effective noise floor set by a Poisson distribution with a mean and variance corresponding to the saturation limit.

For this study we employ the ``ramp" mode which simulates the process of several non-destructive reads \citep[i.e., ``sampling up-the-ramp" or SUTR; ][]{Fowler1990,fixsen2000} at the cost of an increased computational load. 
This allows for readout strategies which may reduce the effective read noise, as opposed to the ``ccd" mode which assumes a fixed read noise equivalent to the measured correlated double-sampling (CDS) read noise.

The ``ramp" mode in \rimtimsim\ allows the user to specify the number of read frames (i.e., a single non-destructive read) per resultant frame, where a resultant frame is an average of several consecutive read frames (often referred to as a ``group frame").  
The precise readout strategy for the GBTDS has yet to be decided, but it is expected that there will be six resultant frames, combined from 18 total non-destructive reads, corresponding to an effective exposure time of $51.7$~seconds between the first and last read frames. 
Based on these restrictions we adopt a candidate readout pattern for the GBTDS from \cite{casertano2022} which consists of a sequence of six resultant frames containing 1, 2, 3, 4, 4, and 4 read frames (i.e, the first resultant frame contains only read frame 1, the second resultant frame is the average of read frames 2 and 3, the third resultant frame is the average of read frames 4, 5, and 6, etc.).\footnote{Note: the candidate readout pattern from \cite{casertano2022} actually contains 19 read frames, where the last resultant frame contains 5 read frames instead of 4, but we elected to adopt the readout pattern specified in the Roman Design Reference Mission of 18 total read frames.} 
From the six resultant frames, we infer the slope of the accumulated flux in each pixel using a linear least squares fitting routine with the weighting scheme from \cite{casertano2022}. 
The final generated image thus consists of the fitted slope in each pixel, reported in units of electrons per second.

The primary benefit of sampling up-the-ramp is to reduce the read noise in a given exposure, so it is unlikely to provide a significant reduction in the photometric uncertainty for most of the planet search stars considered in this survey, as they will be largely limited by photon noise or crowding from nearby sources. 
However, this readout strategy still offers several advantages. 
Most important for our purposes, the sampling rate allows one to recover the signal for stars that would saturate in exposures longer than the readout time of the detector but shorter than the total integration time (hereafter referred to as ``semi-saturated"), in a way that is consistent with the actual survey resulting in a more accurate photometric treatment of bright stars.
For instance, the flux measured by a pixel which saturates in 7 seconds can be inferred from the flux measured in the first (3.04 seconds) and second (6.08 seconds) frames, even though all following frames will contain no useful information. 
Combining this with our strategy for creating resultant frames, the flux of the same pixel will be inferred from only the first resultant frame (which consists of one read frame; the equivalent of a 3.04 second exposure), because the second resultant frame (created from the average of the second and third read frames), and all following resultant frames, will be saturated.

Unlike CCDs, the WFI H4RG-10 arrays do not bleed or bloom around saturated pixels, and therefore accurate photometry can still be recovered for sources on or near such pixels \citep{gould2015,mosby2020}. 
To first order, the charge accumulated past the saturation limit on a pixel is lost (i.e., not registered in the readout electronics, or by neighboring pixels), though in practice some charge will reappear in later exposures through second-order effects such as persistence, and pixels with a lot of accumulated charge will interact more strongly with neighboring pixels (see also section \ref{sec:detector}).

\subsubsection{Photometric Uncertainty per Pixel}

\begin{figure}
    \centering
    \includegraphics[width=\columnwidth]{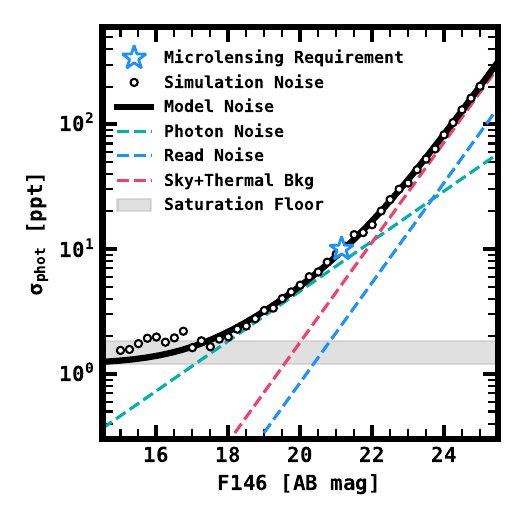}
    \caption{\rimtimsim\ generates images that yield photometry with realistic uncertainties. The circles show the measured RMS scatter from white-noise light curves generated from simulated images of isolated point sources, assuming a constant aperture. 
    The total expected noise, estimated from Equation \ref{eq:noise}, is shown by the black line, and the dashed lines show the contributions from individual noise components. 
    The grey shaded region shows the range of noise floors expected for stars with one or more nearly-saturated pixels in their aperture. 
    This floor varies based on the location of the point source relative to the center and edges of a pixel.
    The structure in the RMS scatter at $F146\lesssim18$~is due to semi-saturation effects, and is exacerbated by the relatively small aperture used to make this figure. }
    \label{fig:photnoise}
\end{figure}

The photometric uncertainty ($\sigma_\mathrm{phot}$) for each pixel can be estimated based on the model from \cite{rauscher2007}, 
\begin{align} \label{eq:noise}
    \sigma_\mathrm{phot}^2 =& \frac{12(n-1)}{mn(n+1)}\sigma_\mathrm{read}^2 + \frac{6(n^2+1)(n-1)}{5n(n+1)}t_g f \nonumber \\
    &- \frac{2(m^2-1)(n-1)}{mn(n+1)}t_f f \;\; , 
\end{align}
where $f$ is the total flux in electrons per second from the source, thermal background, sky background, and dark current on the pixel, $\sigma_\mathrm{read}$ is the read noise, not to be confused with the Correlated Double Sampling (CDS) read noise, $n$ is the number of resultant frames, $m$ is the number of data frames per resultant frame, $t_f$ is the time between frames (in this work, $t_\mathrm{read}$), and $t_g$ is the time integrated within a group (in this work, we assume $2t_\mathrm{read}$). The photometric precision for a 5$\times$5 pixel square aperture for an isolated star at varying magnitude is shown in Figure \ref{fig:photnoise} compared to this model. 

This formula is derived assuming a constant number of read frames per resultant frame. 
Although our readout pattern doesn't fit this criteria, and a more appropriate treatment would be one that considers a non-uniform sampling scheme \citep{casertano2022}, we find that this treatment accurately predicts the noise to within $\sim$5\% of our simulations when adopting $n=2$ and $m=9$ so we elect to continue with this approximation. It is only 
when semi-saturation effects cause additional scatter (as is expected) that the model begins to diverge significantly.

To estimate the effects of semi-saturation, we add $\sigma_\mathrm{phot}$ in quadrature with the photon noise of a source at the saturation limit. 
In principle WFI can perform precision photometry better than this limit with careful attention to modeling the wings of the PSF \citep{gould2015}, but given the small number of saturated sources in our simulated data, ignoring such sources will have only a minimal impact on our results.

In the context of this study, the main differences in the ``ccd" and ``ramp" modes occur for semi-saturated sources where the ``ccd" mode will underestimate the photon noise as being equal to the saturation limit. In reality, the photon noise never quite reaches the saturation limit due to the finite readout time. This is demonstrated in the sawtooth-like pattern for stars with $F146<17$ in Figure \ref{fig:photnoise}.

\subsubsection{Light Curve Generation}\label{sec:simdatalightcurve}

We apply a simple aperture photometry pipeline to generate light curves from the simulated images.
For each star we create a light curve using circular apertures with radii from 1 to 4 pixels (0.11\arcsec~to 0.44\arcsec) in steps of 0.5 pixels. Pixels that are only partially contained within an aperture are weighted by the fraction of the pixel expected within a circular radius. 
The expected contribution from the thermal and sky background (5.23 $e^{-}\,\mathrm{s}^{-1}\,\mathrm{pix}^{-1}$~in the $F146$~bandpass) is subtracted from the image before applying the aperture photometry routine. 
The uncertainty for each photometric point is estimated by adding in quadrature the variance of each pixel estimated from Equation \ref{eq:noise}, where $f$ is replaced with the total accumulated flux so as not to underestimate the uncertainty from semi-saturated pixels.

The center of the apertures are placed to match up exactly with the position of stars on the detector. This is not necessarily realistic, as positions for stars will have some uncertainty that will reflect on the aperture placement and degrade the photometric precision. However, the aperture photometry methods used in this work are an approximation to the PSF photometry methods that will likely be applied for the GBTDS data. 
As a result, even with the uncertainty neglected via the exact placement of the photometric aperture, the actual photometric noise may be overestimated compared to reality due to the limited choice of apertures, as opposed to PSF methods which will much more effectively compensate for crowding by applying lower weights to noisier pixels.

%
\section{Detecting Transiting Exoplanets} \label{sec:detections}
%

In this section we describe our methodology for simulating the transiting exoplanet search in the GBTDS. 
Our high-level strategy is to define a set of analytic formulae to model the WFI photometric performance, and then by extension the overall GBTDS transit survey efficiency. We validate the analytic formulae by generating synthetic full-frame images with injected transiting planet signals, generating light curves for each simulated transiting planet from the pixel-level data, and finally by applying a simulated transit detection pipeline for each generated transit light curve. 
The results from these pixel-level injections are then used to infer parameters in our analytic models and characterize the expected completeness of the GBTDS transiting planet survey.

This section is organized as follows: Section \ref{sec:analytic} describes our analytic model for 
estimating the efficiency of the GBTDS transit survey.
Sections \ref{sec:detectioncriteria} and \ref{sec:pipeline} describe what constitutes a detection, and our methodology for simulating a transit detection pipeline, respectively. In section \ref{sec:noise} we develop our photometric noise model, with particular attention given to crowding. 
Section \ref{sec:dataproduction} describes the synthetic data we generate to validate and fit our analytic model. 
Finally, this section ends with the results from applying our detection pipeline on the simulated full-frame images and a discussion of the overall transit detection sensitivity expected from the GBTDS in sections \ref{sec:injectiontest} and \ref{sec:sensitivity}, respectively.

\subsection{Analytic Formulae for Estimating Survey Completeness}\label{sec:analytic}

We can characterize the likelihood of an arbitrary transiting planet being detected ($p_{\rm det}$) as a function of just one parameter, the transit $S/N$, which itself depends on the properties of the planet, stellar host, and the quality of the GBTDS light curve. To first order, this value can be approximated as a step function, where $p_{\rm det}=0$ below some $S/N$ threshold, and $p_{\rm det}=1$ above some threshold. 
Due to the variance in detection efficiency caused by effects such as crowding, and measurement uncertainty in the detection pipeline, we instead define the \emph{expected} signal-to-noise ratio ($S/N_\mathrm{exp}$) of an arbitrary transiting planet. 
This value should be interpreted as the $S/N$ of a transiting planet 
in the GBTDS under several simplifying assumptions, such as circular and edge-on orbits, box-shaped transits, and the average photometric noise for a given apparent magnitude. 
We calculate this value through the following expression, 
\begin{align}
    S/N_\mathrm{exp} &= \frac{R_p^2/R_\star^2} {\sigma_{\rm dpp} \sqrt{ \Delta t/t_\mathrm{dur}}} \sqrt{n_\mathrm{tran}}\;\; , 
\end{align}
where $n_\mathrm{tran}$ is the number of transits in the data, $R_p^2/R_\star^2$ is the transit depth, $\Delta t/t_\mathrm{dur}$ is the observing cadence divided by the transit duration, and $\sigma_{\rm dpp}$ is the expected differential photometric precision for a single exposure. $\sigma_{\rm dpp}$ is a complicated function of instrumental effects and apparent magnitude and is described in more detail in  later sections.

The purpose of our detection model then is to marginalize over the distributions in these secondary parameters (e.g., crowding, measurement variance, eccentricity, impact parameter) to give a probability, $p_\mathrm{det}$, that a transit signal with a given $S/N_\mathrm{exp}$ is detected. 
We model this probability 
as a generalized logistic function of the form, 
\begin{align}
    p_\mathrm{det}(x) = C \left(\alpha + \gamma e^{-\beta(x-x_0)} \right)^{-\gamma} \;  , \label{eq:pdet}
\end{align}
where $x=S/N_\mathrm{exp}$ and $C$, $x_0$, $\alpha$, $\beta$, and $\gamma$ are free parameters. 
This function asymptotically approaches 0 at arbitrarily low $S/N_\mathrm{exp}$ and 1 (or some other arbitrary upper limit between 0 and 1) at high $S/N_\mathrm{exp}$.

\subsection{Detection Criteria}\label{sec:detectioncriteria}

Typically, transit surveys have two main criteria for considering a detection: the transiting event must meet some $S/N$ threshold, and the planet must transit at least two (or more) times to unambiguously determine the orbital period.

For the Kepler survey the threshold for detection was set to 7.1$\sigma$, and each planet candidate had to transit a minimum of three times. 
The choice of detection threshold set by the Kepler mission was made with the expectation of only one False Alarm over the course of the survey. 
Given the expectation for the Kepler mission to observe $\sim$10$^5$ stars and perform the equivalent of $1.7 \times 10^7$ independent statistical tests per star, treating each test as an independent process thus required a threshold with a false alarm probability of 1 in $1.7 \times 10^{12}$, which corresponds to a threshold of 7.1$\sigma$ \citep{jenkins2002b}. This estimate was made in the absence of red noise and instrumental systematics, so the effective threshold is actually higher than this.

Because the threshold adopted by the Kepler team is likely not appropriate for the GBTDS, we elect to determine a new threshold following a similar line of logic.
Roman should observe $\sim$10$^8$ stars 
($\sim$6$\times10^7$~stars with $F146<21$~are considered in this work) with the photometric precision necessary for detecting transiting planets. 
However, the number of effective statistical tests performed throughout the campaign is less clear given our constraints of $P<72$~days, and the large time gaps between seasons. 
For Kepler, this number was estimated by simulating transit searches for $>$10$^6$ white noise light curves, but adopting a similar Monte Carlo approach for the GBTDS would require simulating a transit search over $>$10$^9$ light curves, which would be computationally expensive. 
Instead, we adopt the same number as the Kepler mission, which should be a reasonable upper bound given that Roman will search over a shorter range of periods and will have fewer total cadences ($\sim$41,000 for Roman versus $\sim$73,000 for Kepler). 
Therefore, we adopt a false alarm probability of 1 in $1.7\times10^{15}$, which corresponds to a detection threshold of $8.0\sigma$.

Thus, for a transiting planet to be considered a detection, the planet must have $\geq$7 transits through all six seasons of the survey, and be recovered with a significance greater than $8.0 \sigma$. 
In the spirit of optimism, we also consider the case in which a planet has $\leq$1 transit per season (i.e., $\leq$6 total), as a significant fraction of planets may still be detectable with even a single transit across all seasons, although the period of such a system, even with one transit per season, may be ambiguous. Thus, throughout this work, we report planet yields and detection efficiencies based on both of these criteria. 
If not explicitly stated, planet yields over large regions of parameter space are taken to be an approximate average assuming both criteria, often rounded to one or two significant figures for simplicity.

The criteria we adopt are made under the assumption of Gaussian noise, which may not be a realistic expectation. This is particularly true for Roman, which is expected to have dithered observations that are likely to produce light curves with correlated noise due to intra-pixel and inter-pixel sensitivity variations in addition to correlated noise from astrophysical sources such as starspot modulations.

\subsection{Simulated Detection Pipeline}\label{sec:pipeline}

To simulate a detection pipeline we apply a simple matched filter. This approach has a few advantages for our application. First, the detection efficiency of a matched filter can be characterized by a single parameter, the transit $S/N$ (see section \ref{sec:analytic}), which simplifies our assessment of the GBTDS transit search completeness. Second, the formulation of the matched filter provides a simple interpretation for the detectability of a transit signal, without the need for calculating a computationally intensive periodogram, as may be required for, e.g., the box-least squares algorithm \citep{kovacs2002}, where the detection statistic is defined as a function of period.

The matched filter is formulated such that under the null hypothesis (i.e., no transit signal is present), the detection statistic returns a value with mean of zero and variance of one. In the case where a signal is present, the detection statistic returns a value also with a variance of one but with a mean proportional to the transit S/N. Thus, if the detection statistic is sufficiently larger than that expected under the null hypothesis (in this work, 8.0$\sigma$), we can reject the null hypothesis and consider the signal to be present.

The matched filter can be described mathematically by, 
\begin{align}
    Z_i &= \frac{x_i^T R s_i}{\sqrt{s_i^T R s_i}} \;\;,
\end{align}
where $Z_i$ is the detection statistic at each cadence, $i$, also referred to as the Single Event Statistic (SES), $x_i$ is the flux time series, $R$ is the autocorrelation matrix of the noise, and $s_i$ is the signal template. 
Because the photometric noise in our simulations is very nearly Gaussian and stationary, there is no need for a time-varying estimate of the noise nor is there a need to estimate the noise power spectrum. 
This is not strictly true because every bright star in our simulations contains a transit signal, so crowding from nearby bright stars will add a small amount of correlated noise into the light curve, but because the duty cycle of a transiting planet is relatively small, this should be a reasonable approximation.
Under these assumptions, the autocorrelation matrix becomes proportional to the identity matrix, $R = \sigma^2 I$, where $\sigma^2$ is the variance of the flux time series. Since the variance is a constant, we can approximate it using the median absolute deviation (MAD) of the flux,  
\begin{align}
    \sigma_{MAD} \approx 1.4826 \times MAD(x_i) \;\; ,
\end{align}
and calculating the SES simplifies to a simple cross-correlation,  
\begin{align}
    Z_i  &=  \frac{x_i \ast s_{-i}}{\sqrt{\sigma_{MAD}^2 \sum s_i^2  } } \;\; \\
    &= \frac{N_i}{\sqrt{D}},
\end{align}
where $\ast$ is the convolution operator, $s_{-i}$ is the time-reversed transit template, and the denominator is a constant over all $i$. This gives the likelihood that a transit exists at a given cadence, $i$. To test for transits at a given epoch, $t_0$, and period, $P$, we fold this statistic in the following way, 
\begin{align}
    Z_{\{t_0, P\}} =  \frac{\sum_{j} N_j}{   \sqrt{ n_{j} D} } \;\;,
\end{align}
where $j$ represents all the cadences specified by a given $t_0$ and $P$, $n_j$ is the number of cadences in $j$, and $N$ and $D$ are defined as in the previous equation. From here on, we refer to $Z_{\{t_0, P\}}$ as the Multiple Event Statistic (MES). 
It's worth noting that this formulation of a matched filter requires a uniform cadence, which may not be the case for the GBTDS as pointing errors could result in variance on the acquisition time. However, if this variance is smaller than the larger of either the exposure time or the timescale needed to properly sample the transit, it should be a reasonable approximation.

To apply this matched filter as a simulated detection pipeline the transit template is calculated using the same parameters (i.e., $P$, $R_p/R_\star$, $a/R_\star$, $i$, $e$, and $\Omega$) to match the shape of the model light curve that was injected,
and the MES is calculated for each epoch at the injected period.
One important consequence of the GBTDS survey strategy is that 
the exposure time ($\sim$54 seconds) is significantly shorter than the observing cadence of 15 minutes, and the
transit shape itself is therefore undersampled in the data. 
This effect results in a template mismatch at nearly every cadence, and a slight degradation of the transit detection sensitivity, typically on the order of $\sim$5\%, but the degradation can be more sinister for short duration transits such as those caused by planets orbiting M dwarfs at short periods or for transits with longer ingress and egress durations such as those with a high impact parameter. 
To compensate for this effect, we calculate a total of 15 templates, each of which has an epoch slightly offset from the center by $t_0 + \delta t_0$, where $\delta t_0$ ranges from 0-14 minutes in steps of one minute. We experimented with shorter steps, but found no additional increase in the recovered MES. 
Using each of these slightly offset templates to calculate a ``super-sampled" SES with a cadence of one minute,  we then calculate the MES, and consider all values within one transit duration of the injected signal. If the largest such value has $\mathrm{MES}>8.0$, we consider the transit signal detected.

The choice to adopt a transit template with the same parameters as the injected signal is made out of convenience
and is akin to assuming that the grid of transit templates used in the detection pipeline is sufficiently populated that template mismatch is negligible. 
In a real transit search with no a priori knowledge of the transit shape or period, some degree of template mismatch is inevitable, meaning we may potentially overestimate our detection sensitivity. 
As an example, for a signal with MES=8.0 and a transit duration of 3 hours, typical of a $P=10$~days planet orbiting a Sun-like star, template mismatch caused by a template with a duration of 2.75 hours (i.e., too short by one cadence) will result in an average $S/N$ decrease of 0.08$\sigma$. Although this is well below the expected variance in MES, the magnitude of this bias will depend on the methodology for constructing the template grid and may disproportionately impact certain areas of parameter space. 
Given that transits in the GBTDS will be under sampled, the strategy for constructing the transit template grid may be non-trivial, and the optimal choice of templates may vary substantially from star to star depending on the expected range of transit durations.
Simply adopting a finer template grid is not necessarily a valid solution, because more finely-sampled grids will have higher false alarm rates. 
Thus, we leave the exercise of optimizing the template grid for a later work, and 
continue with the assumption that this potential bias is negligible.

\subsection{Average Differential Photometric Precision}\label{sec:noise}

To properly predict the differential photometric uncertainty for any arbitrary star, we build upon the pixel-level uncertainties described in Equation \ref{eq:noise}. In particular, we need to differentiate between the error on the flux measured by a pixel, and the overall photometric error for a point source. 
To accomplish this, we consider three additional parameters to predict the differential photometric error in our light curves: the flux attributable to the star of interest ($f_\star$), the flux attributable to neighboring stars ($f_N$), and the number of pixels used to perform the aperture photometry, $n_\mathrm{pix}$. Using these three parameters, the differential photometric precision per cadence can be expressed as 
\begingroup
\allowdisplaybreaks
\begin{align}
    \sigma_\mathrm{dpp} &= \frac{\sqrt{  C_0 \sigma_\mathrm{read}^2 n_\mathrm{pix} + (C_1 t_g-C_2 t_f) f_\mathrm{tot}} }{f_\star} \\
     &C_0 = \frac{12(n-1)}{mn(n+1)} \approx 0.22 , \nonumber \\
     &C_1 = \frac{6(n^2+1)(n-1)}{5n(n+1)} = 1.0  ,\nonumber \\
     &C_2 = \frac{2(m^2-1)(n-1)}{mn(n+1)} \approx 2.96
\end{align}
\endgroup
where $C_0$, $C_1$, and $C_2$ are calculated assuming $n=2$ and $m=9$, and $\sigma_\mathrm{read}$, $t_g$, and $t_f$ are defined in Equation \ref{eq:noise}.  $f_\mathrm{tot}$ and $f_\star$ are expressed as, 
\begin{align}
    &f_\mathrm{tot} = f_\star + f_N + n_\mathrm{pix}f_\mathrm{bkg} \\
    &f_\star = F_\mathrm{aper} 10^{-0.4(m_{F146}-zp)} 
\end{align}
where $F_\mathrm{aper}$ is the fraction of the flux from the target star contained within the adopted aperture, $m_{F146}$ is the apparent magnitude of the target star, and $zp=27.648$ is the AB magnitude at which WFI is expected to measure a flux of one electron per second in the $F146$ bandpass.  

The terms in these last two equations depend on the adopted aperture radius, ($r_\mathrm{aper}$). This value can vary significantly by star based on random fluctuations in the density of nearby (especially bright) sources. 
To determine these properties, we run a series of simulations to understand the optimal aperture radius chosen for each star, as described below, and the resulting effects of crowding.

\subsubsection{Aperture Crowding}\label{sec:crowding}

Because of the stellar density in the Galactic bulge, there is significant photon noise from nearby stars diminishing the photometric quality of the transit search data. 
To judge the impact of crowding in the recovery of transits, we measured the crowding distribution empirically by generating a set of synthetic images with \rimtimsim.

We constructed a set of time-series images equivalent to one full season of the GBTDS (72 days, 6913 cadences) with the same stellar density, but only a fraction of the full detector size to reduce the computational overhead.
We injected a signal from a transiting planet into every star with $F146\leq21$. 
The period and radius of each injected planet was randomly chosen to be between $R_p$ = 12--16 $R_\oplus$ and $P$ = 1--20 days, which is easily detectable for most of the stars in our catalog, with the exception of giants and distant subgiants such as those on the far side of the Galactic bulge. 
The transit signals were injected via the process described in section \ref{sec:injections} and assuming circular orbits, Sun-like limb-darkening coefficients, $i=90^\circ$, and a random transit epoch, $t_0$.

From these simulated images we applied our simple aperture photometry routine to generate light curves for a randomly-selected sample of $\sim$38,000 stars. 
Because we injected the original signal, we know the transit depth a priori, so any dilution of the transit depth should be due to crowding from additional flux in the aperture. 
Thus, by fitting the light curve with an additional term to account for transit dilution we can empirically measure the crowding for each star as a function of aperture radius.

\begin{figure*}
    \centering
    \includegraphics[width=\textwidth]{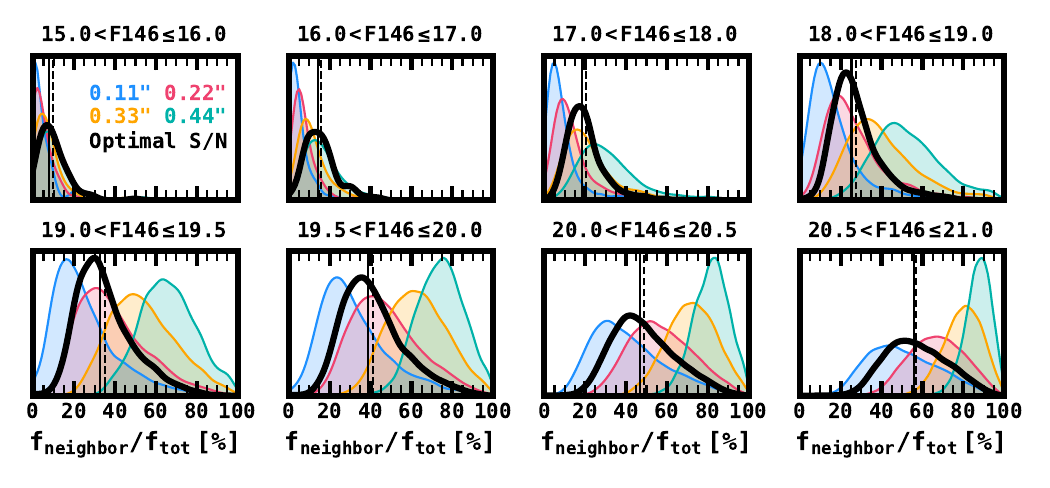}
    \caption{The distribution functions for the fraction of flux contributed by neighboring stars within circular apertures of varying radii, for different magnitude ranges. Each panel shows a specific magnitude range, and each color denotes a different aperture radius. The distribution marked by the solid black line shows the crowding in the aperture that optimized the signal-to-noise of the injected signal, measured via the multiple event statistic (MES), and is the crowding distribution that we adopt to build our noise model. The vertical solid and dashed lines show the median and mean crowding, respectively, in the optimized aperture. }
    \label{fig:crowding}
\end{figure*}

We define the flux contamination, $\Gamma_c$, as the fraction of total flux within the aperture that derives from neighboring stars, 
\begin{align}
\Gamma_c &= \frac{f_N}{f_\star + f_N }  \;\;. 
\end{align}
This term is often recast in the form of the dilution factor, $D$, which is the factor by which the apparent transit depth from a crowded source is decreased, 
\begin{align}
    D = \frac{f_\star + f_N}{f_\star} = \frac{1}{1-\Gamma_c}  \;\;. 
\end{align}
Each light curve with an injected signal can thus be modeled as, 
\begin{align}
    s'(t) = (1-\Gamma_c) \times s(t) + \Gamma_c \;\;,
\end{align}
where $s(t)$ is the injected transit signal and $s'(t)$ is the signal after adjusting for dilution. 
 We measure $\Gamma_c$ for each star for each circular aperture with radii from 1--4 pixels described in $\S$\ref{sec:simdatalightcurve}, by fitting each light curve for $s'(t)$ with a linear least-squares routine. After fitting $\Gamma_c$ for each source in each aperture, we can visualize the distribution of crowding using a Gaussian Kernel Density Estimator (KDE). The distributions of $\Gamma_c$ for each aperture and magnitude range are shown in Figure \ref{fig:crowding}.

\begin{figure*}
    \centering
    \includegraphics{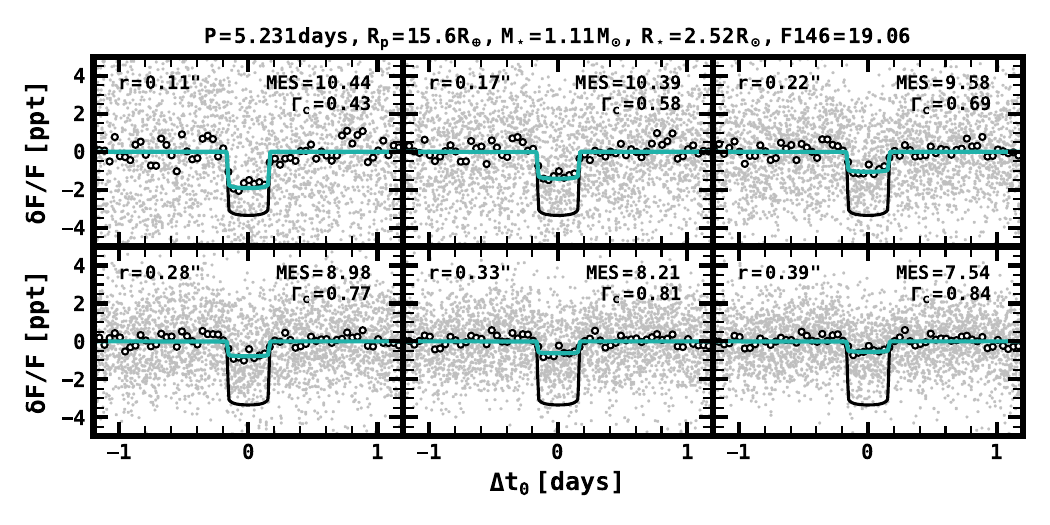}
    \caption{A light curve of a transiting planet, in a particularly crowded region, generated from multiple apertures in our synthetic images. Each panel is labeled with the aperture radius, measured multiple event statistic (MES), and crowding ($\Gamma_c$). The black line shows the injected signal, and the teal line shows the diluted transit signal after fitting for $\Gamma_c$. The white circles show the relative flux averaged in bins of 15 minutes, the observing cadence. }
    \label{fig:transit_snr}
\end{figure*}

The question still remains as to which aperture is optimized for each star. 
There is a careful trade-off in choosing the aperture size, as larger apertures will improve the signal of the source but will also add additional photon noise from neighboring stars and background flux (i.e., sky and thermal background). 
The optimal aperture is a trade-off between these two extremes, and depends strongly on the number and brightness of nearby stars. Because the crowding for each star is dependent on the selected aperture, the amount of contaminating flux should correspond to the aperture which optimizes the transit S/N. 
Thus, for each light curve, in addition to measuring the amount of contaminating flux, we also apply our simulated detection pipeline to measure the MES of the injected signal for each aperture, and adopt the flux contamination for that star as the contamination in the aperture that maximized the recovered MES  (see Figure \ref{fig:transit_snr}). 
This procedure is akin to simulating a photometry pipeline which assumes the optimized aperture, from the selected set of apertures, is always found. 
The considerations discussed here are also similar to those employed by optimal-weighting photometry, where a PSF model is used to define an aperture that optimizes the photometric $S/N$. Thus, our methodology provides a logical precursor and, at least for bright stars, a reasonable approximation to the more advanced PSF-based methods that will be used to analyze the GBTDS data.

For some of the brightest stars in our sample ($F146\lesssim$~17) the optimal aperture is larger than the adopted circular aperture with a radius of 0.44\arcsec, so the crowding is underestimated, although the overall noise will be higher. 
Due to semi-saturation effects, the photon noise limit is capped for pixels in the center of the PSF, meaning that, the relative increase in S/N from pixels outside the core of the PSF is higher in bright stars, which combined with our upper aperture radius limit, caps the light curve noise for bright stars as equal to that experienced by a $F146 \approx 17$~mag star.

Using larger apertures would lead to a somewhat paradoxical effect in which crowding would actually increase for the brightest stars in the sample, as the S/N gained by including pixels at larger aperture radii will include more contamination from other stars. 
This effect is one limitation of using aperture photometry to analyze the GBTDS data, as a more sophisticated PSF photometry pipeline would more appropriately de-emphasize pixels with contaminating sources.

Even at these larger apertures, the noise caused from crowding will grow very quickly, so this limitation in our light curve generating pipeline should be a fairly modest concession, and we do not expect the light curve noise to decrease significantly more with larger apertures for the stars considered here. In addition, sources with apparent magnitudes $F146\lesssim 17$ will be prone to other effects not considered in this work, such as classical and count-rate dependent non-linearities, the Bright-Fatter effect, and persistence \citep{mosby2020}, which we are not considering at present, so it's not clear to what degree including larger apertures would improve the S/N as expected from this discussion. 
However, because stars with $F146<17$ represent only a small fraction of the stars considered in this paper, we leave this more detailed instrumental characterization for a future work.

\subsubsection{Cumulative Aperture Flux} \label{sec:apertureflux}

\begin{figure}
    \centering
    \includegraphics[width=\columnwidth]{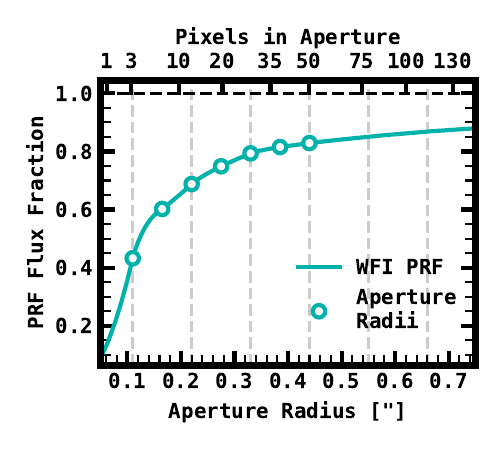}
    \caption{The fraction of total flux from the WFI Pixel-Response Function (PRF) captured as a function of aperture radius. The circles are the aperture radii considered in this work and the dashed vertical lines denote the pixel scale.}
    \label{fig:prf}
\end{figure}

As well as defining the crowding for each star in the sample, the optimal aperture also sets the cumulative flux and photon noise collected from the source. Because we are considering aperture photometry for this study, not all of the flux from a star is collected for the light curve.
As a function of apparent magnitude, we adopt the average aperture radius used to optimize the transit S/N for the light curves generated in the previous section and use that value to estimate $F_\mathrm{aper}$. $F_\mathrm{aper}$ is shown as a function of aperture radius in Figure \ref{fig:prf}.

Brighter stars on average allow for larger apertures because the background noise, primarily from neighboring sources, is much lower in comparison to the signal gained with larger apertures. This relationship is even more significant for semi-saturated sources with the WFI detectors because not all of the photons at the center of a stars PSF will be collected. 
In this case, larger apertures actually improve signal to noise more than is expected from simple crowding estimates because the center most pixels in the star's PSF are gathering only a fraction of the available photons. 
However, predicting this relationship is complicated because the exact fraction of photons collected on semi-saturated pixels depends on the position of the star in relation to the center or edges of the brightest pixel, as well as the adopted prescription for which read frames are averaged into resultant frames. 
To simplify this relationship, in our noise model we assume a constant $F_\mathrm{aper}$ for stars brighter than $F146 = 17$ of $\sim$83\%, and instead adopt a systematic limit on the differential photometric precision of $\sim$1.3 ppt, which corresponds to the photon noise limit of an isolated star in which the brightest pixel is just below the saturation limit and $\sim$14\% of the total light is lost because the brightest pixel is saturated after the first read. Accounting for semi-saturation in pixels neighboring the brightest pixel would reveal an even larger discrepancy. 
This ratio is more severe for stars on the corner of a pixel compared to that of a star on the center of a pixel, though the saturation limit itself is brighter in the case of the former. 
Thus, the relationship shown in Figure \ref{fig:prf} is true only for stars not near saturation, though it still provides a relatively good approximation for stars with $F146 \gtrsim 17$. At brighter apparent magnitudes, our noise model under predicts the photon noise in our simulations, but the model is likely closer to reality.

\subsection{Synthetic Full-Frame Data Production}\label{sec:dataproduction}

To measure the transit detection efficiency of the GBTDS, we simulated a series of full-frame images with \rimtimsim\ in which every dwarf and subgiant star with $F146 < 21$ on the detector included an injected transit, following the procedure described in section \ref{sec:simdata}. The transit signals were chosen to have a random orbital period drawn from a log-uniform distribution between 1-72 days. 
Inclinations ($i$) were chosen such that the injected transit signals will have a uniform distribution in impact parameter from 0 to 1, and the rest of the orbital parameters ( i.e., $e$, $\Omega$, $t_0$) were assigned based on the population statistics and procedure outlined in section \ref{sec:simplanets}. The radius ratio for each signal was chosen such that the transit signals are drawn from a uniform distribution in $S/N_\mathrm{exp}$ ranging from 3 to 25.
This formulation was chosen to more efficiently characterize our matched-filter detection pipeline. At arbitrarily low and high $S/N_\mathrm{exp}$, the detection efficiency should plateau at 0\% and at some maximum value just below 1, respectively. Measuring the detectability of injected transit signals outside these bounds doesn't provide significant leverage in characterizing the GBTDS transit survey sensitivity. 
In other words, our transit injection/recovery experiment is optimized by clustering our injected transits around the transition region where the likelihood of recovering a signal with MES~$\approx8.0\sigma$ is high.

Each instance of \rimtimsim\ created a 4088$\times$4088 pixel simulated image, at a file size of 256 MB.  On a 2.4 GHz Central Processing Unit (CPU), each image required an average time of $\sim$24 minutes to generate.  The full simulation required 6913 images which, if computed in serial, would have required $\sim$115 days to create.  As such, we used the {\em Discover} supercomputer at the NASA Center for Climate Simulation (NCCS)\footnote{\url{https://www.nccs.nasa.gov/systems/discover}} in order to substantially reduce the time required to generate the images.  We used nineteen 40-core {\tt skylake} nodes, running 6 simultaneous instances of \rimtimsim\ with different input parameters on each core, iterating until the 6913 images were finished.  Employing this method, we were able to complete the entire image-production effort in $\sim$1 day, after which we started light curve generation.

For the computational effort involved in light curve generation, we again returned to the NCCS {\em Discover} supercomputer. Our simulation required 445,071 light curves.  We parallelized the light curve construction similar to the image generation.  Although, due to reduced memory requirements in comparison to the image generation, we were able to scale up the effort significantly, running groups of 40 instances per node, with up to 1,600 instances running simultaneously, depending on resource availability.

\begin{figure*}
    \centering
    \includegraphics[width=\textwidth]{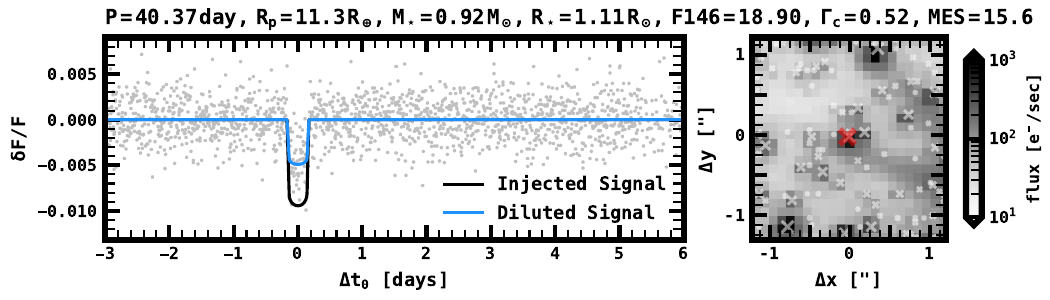}
    \includegraphics{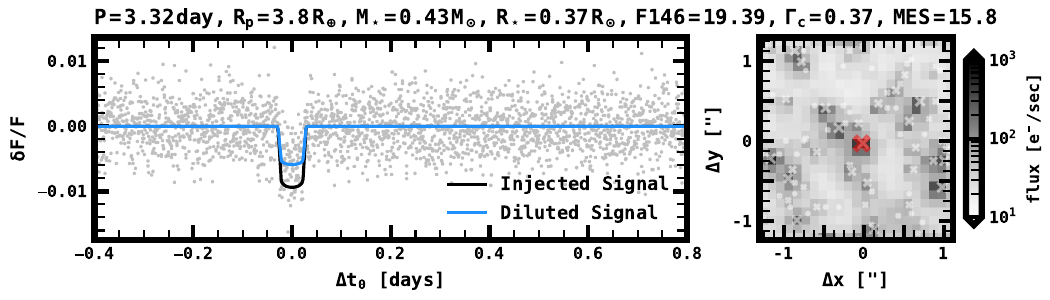}
    \caption{Example light curves for two simulated sources with transits injected at the pixel level. In each row of figures, the left panel shows the light curve (gray points), along with the injected signal before (black line) and after (blue line) adjusting for crowding. The right panels show a $\sim$2\arcsec$\times$2\arcsec\ cutout of the region surrounding the star of interest. Nearby stars brighter than $F146=24$ are denoted by a white ``x", with their sizes indicative of their brightness. The figure label gives relevant information on the simulated system, including it's recovered significance (MES) and crowding ($\Gamma_c$). }
    \label{fig:lightcurves}
\end{figure*}

In all, the simulated images contained a total of 451,185 stars with injected signals. Of these, we ignored stars within 10 pixels of the edges of the 4088$\times$4088 detector array, primarily because we don't simulate an astronomical scene beyond the 4096$\times$4096 numerical array required by the FFT, so stars near the edges of the detector assembly have unrealistically reduced noise from crowding and nearby stars. This left 445,071 sources from which we attempted to generate light curves and simulate the planet detection process using the method outlined in section \ref{sec:pipeline}. Of these, 13,745 sources (3.0\%) contained at least one fully saturated pixel on at least one image within the largest circular aperture (radius of 4~pixels or 0.44\arcsec). We consider events from these sources as non-detections. 
This left a total of 431,325 valid light curves from which to perform injection and recovery tests. Two examples of these light curves are shown in Figure \ref{fig:lightcurves}.

There appear to be a few limitations in the methodologies applied to create light curves in this section. 
For instance, we generate light curves based on the catalog of injected sources, rather than creating a source catalog from the synthetic images and then using that catalog to extract light curves. This simplification may ignore uncertainties caused by imprecise source coordinates (as discussed in section \ref{sec:simdatalightcurve}) or the likelihood that some sources will be missed (i.e., not identified for photometric extraction) or merged (i.e., two or more sources are recovered as a single source) in the catalog. 
However, we show in the following section that these biases appear as a natural consequence of our methodologies for simulating the transiting planet search, and should therefore be implicitly included in our final planet yield estimates.

\subsection{Transit Injection and Recovery Results}\label{sec:injectiontest}

\begin{figure}
    \centering
    \includegraphics[width=\columnwidth]{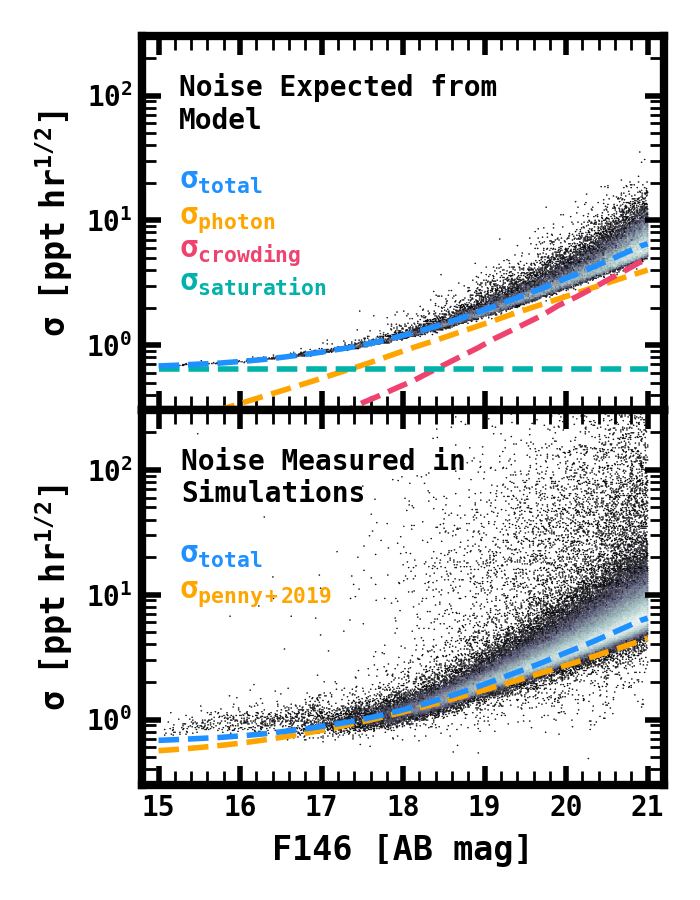}
    \caption{The combined, expected differential photometric precision over a period of one hour. {\it Top:} The expected noise distribution given the crowding at each apparent magnitude. The dashed lines show the average noise contributions from crowding (pink), photon noise (orange), and semi-saturation (teal), while the blue line gives the average total noise as a function of apparent magnitude. {\it Bottom:} The measured differential photometric precision for each injected source with $S/N_\mathrm{exp}>10$. The blue dashed line is the same as in the top panel, and the noise model for an isolated star from \cite{penny2019} is included as a comparison (orange). }
    \label{fig:instrumentmodel}
\end{figure}

For each source, we applied our detection pipeline to the light curve generated from each aperture and saved the largest recovered MES. In addition to the recovered MES, we also recorded the MAD of the flux time-series, and fit for the crowding (i.e., $\Gamma_c$) following the same methodology as in section \ref{sec:crowding}. 
The resulting photometric precision ($\sigma_\mathrm{dpp}$) for 294,149 light curves whose injected transit signal has $S/N_\mathrm{exp}>10$, where crowding could reliably be estimated, is shown in Figure \ref{fig:instrumentmodel} alongside expectations from our instrumental noise model and empirically determined crowding distributions.

As seen from Figure \ref{fig:instrumentmodel}, the noise measured from our simulations matches the expectations from our instrument model well, with a few caveats. 
The first being the average noise for bright sources ($F146 \lesssim 17$) is generally under predicted by our model, which is likely caused by our limited choice of apertures and the semi-saturation effects described in section \ref{sec:apertureflux}. 
One other apparent discrepancy is the distribution of noise which appears much more broadly than the model, and insinuates that some sources have photometric precision better than the photon limit. This is caused by increased scatter due to measurement error, since we use the measured crowding as a metric for determining $\sigma_\mathrm{dpp}$. 
The outliers are caused by events that have a low recovered MES, usually due to being near a much brighter star, and thus an unreliable crowding estimate, compared to the crowding measured from purely high $S/N$ events in section \ref{sec:crowding}, resulting in the sources appearing to have significantly worse or significantly better photometric precision than reality in this figure. 
Neglecting such sources in this figure would bias us against stars with large crowding, and force us to underestimate the photometric noise.

There are several expected deviations between the distribution of recovered MES and the injected $S/N_\mathrm{exp}$.
By far the largest source of variation is crowding. As seen from Figure \ref{fig:crowding}, the crowding for a source at a given magnitude typically varies by a factor of 2-4. Our $S/N_\mathrm{exp}$ value doesn't take this variation into account and instead only considers the average crowding as a function of magnitude. 
There are smaller effects as well. For instance, our estimate of the noise is not exact, and should be considered with measurement error, such that the typical variance of the MES may be greater than or less than 1, as opposed to the formulation in section \ref{sec:pipeline}. 
The net result of these effects is to ``smear" the distribution of recovered MES, and typically results in a lower average MES compared to $S/N_\mathrm{exp}$.

\begin{figure}
    \includegraphics[width=\columnwidth]{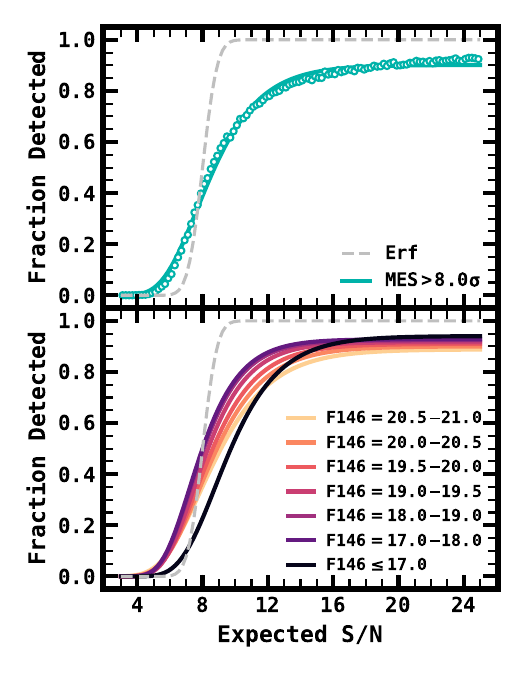}
    \caption{The fraction of injected transit signals that were detected ($p_\mathrm{det}$) as a function of expected signal-to-noise ratio ($S/N_\mathrm{exp}$). {\it Top:} The fraction of transits detected as
    $p_\mathrm{det}$ calculated from the full sample of generated light curves. The dashed grey line shows an error function with $\sigma=1$ centered on $S/N_\mathrm{exp}=8.0$, which is the expectation from the formulation of our simulated matched-filter detection pipeline. 
    The teal circles show the measured $p_\mathrm{det}$ in $S/N_\mathrm{exp}$ bins of width 0.2, and the teal line shows our fit to those points. {\it Bottom:} Our logistic fits to the injection sample split into different magnitude ranges. 
    Due primarily to crowding, $p_\mathrm{det}$ is more efficient for brighter stars, except for the brightest stars in the sample where our noise model underestimates the impact of semi-saturation effects, causing $S/N_\mathrm{exp}$ to be overestimated. }
    \label{fig:injections}
\end{figure}

For each of the 431,325 light curves, we averaged the detection rate in bins of $S/N_\mathrm{exp}$ ranging from 3-25 with width of 0.2, and measured the fraction of each bin in which $\mathrm{MES}>8.0$. We then fit these results for $C$, $x_0$, $\alpha$, $\beta$, and $\gamma$ in equation \ref{eq:pdet} to determine $p_{\rm det}$. The results of these fits are shown in Figure \ref{fig:injections}.

In agreement with our initial assessment that crowding is the dominant factor shaping $p_\mathrm{det}$, we find that the shape of our logistic function varies substantially with differing magnitude ranges as shown in the bottom panel of Figure \ref{fig:injections}. The difference in $p_\mathrm{det}$ is as large as $\sim$18\% for constant $S/N_{\rm exp}$ between the bright and dim samples. 
We also find that for brighter stars, $p_{\rm det} \approx 50\%$ for $S/N_{\rm exp}=8.0$, as is expected as crowding decreases. 
We also compared changes in $p_{\rm det}$ as a function of other parameters such as spectral type, period, and number of transits, but found that these parameters have a negligible impact in our simulations.

As mentioned in the previous section, these injection and recovery tests naturally reproduce biases caused by incompleteness that should be present in the actual transit survey, even though it is not explicitly modeled by our light curve extraction and transit search framework where we generate light curves and apply our simulated detection pipeline over every star in the injected source catalog. 
This is apparent even for the brightest stars in our simulation, where $p_{\rm det}= 95\%$ at arbitrarily high $S/N_{\rm exp}$, while for the average source, $p_{\rm det} = 90.1\%$ at arbitrarily high $S/N_{\rm exp}$. This cannot be fully explained by the exclusion of saturated sources, which account for only 3.0\% of the source catalog. 
The remaining 6.9\% of sources that are not recovered even at large $S/N_\mathrm{exp}$ represent the demographic that is rarely considered by planet-search surveys: stars that are not explicitly searched for planets. 
In other words, the asymptote for $p_\mathrm{det}$ at high $S/N_\mathrm{exp}$ reflects the relative completeness of the catalog of planet-search stars. As expected, this number is magnitude dependent, and will depend on the density of bright stars in the field. 
In support of this idea, the asymptote at high $S/N_{\rm exp}$ for stars with $F146=$~20.5-21.0 ($p_{\rm det} = 88.7\%$), is consistent with the completeness measured from deep HST imaging of the Stanek window for stars with $I\approx$~21-22 \citep[e.g., see Figure 2 of][]{terry2020}.
Although the difference in bandpass used precludes a 
direct comparison, this broad consistency over a similar apparent magnitude range with a telescope of similar aperture and spatial resolution gives some credibility that our methodology accurately accounts for biases introduced by source catalog completeness.

In light of this effect, it is important to clarify the meaning of $p_{\rm det}$ in this work, as there is a subtlety that diverges from typical planet searches. 
In the context of most transit search surveys, $p_{\rm det}$ is a measure of the completeness of the adopted transit search pipeline over the observed sample of planet-search stars. 
In our work, however, $p_{\rm det}$ is more accurately defined as a measure of the transit-search completeness over \emph{every star within the GBTDS observing footprint}, rather than every star within the GBTDS source catalog. 
While this is a subtle difference, it does have a few implications that may affect the interpretation of our results. 
For example, our yield estimates make no claim about whether a transiting planet can be accurately characterized, or even whether its host can be properly identified. This is likely to have implications for the false negative and false positive rates in the GBTDS transit survey (see also section \ref{sec:falsepositives}).

\subsection{Overall Survey Sensitivity}\label{sec:sensitivity}

\begin{figure*}
    \centering
    \includegraphics[width=\textwidth]{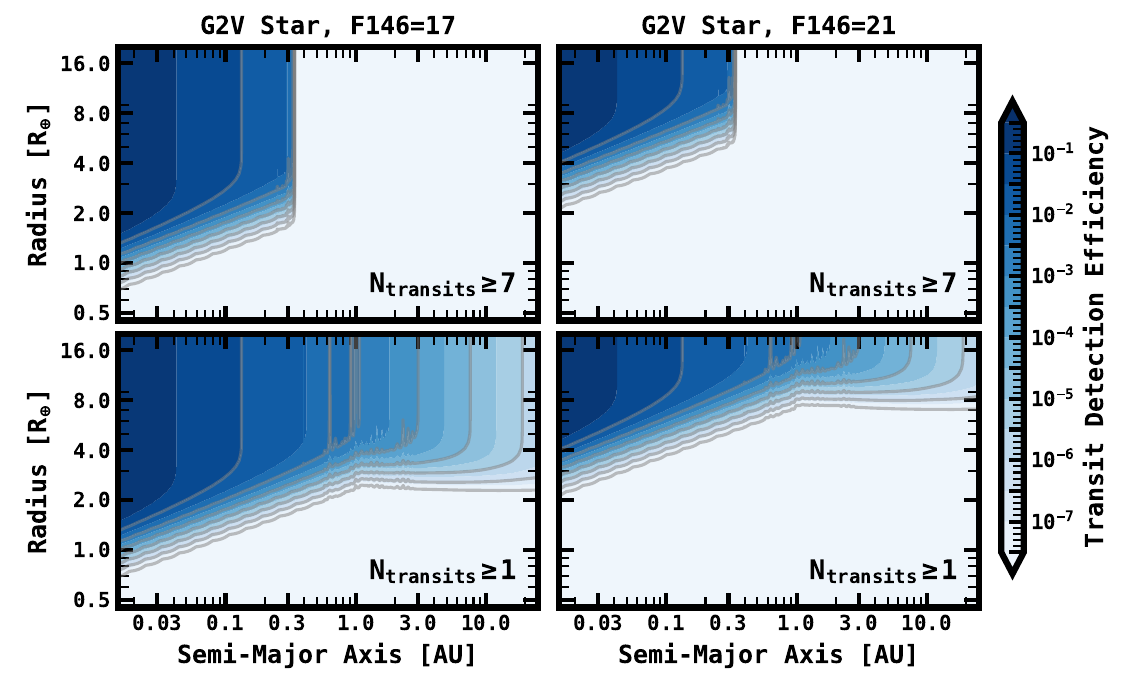}
    \caption{The overall transit survey efficiency, $\eta$, defined as the product of the probability a planet transits across our line of sight ($p_\mathrm{tra}$), the probability that the telescope observes the minimum number of required transits ($p_\mathrm{win}$), and the probability that the transit signal is detected ($p_\mathrm{det}$); $\eta = p_\mathrm{tra} \times p_\mathrm{win} \times p_\mathrm{det}$. {\it Top Left}: The survey efficiency for a bright Sun-like star requiring a minimum of 7 transits. {\it Top Right}: The survey efficiency for a dim Sun-like star requiring a minimum of 7 transits. {\it Bottom Left}: The survey efficiency for a bright Sun-like star requiring a minimum of 1 transit. {\it Bottom Right}: The survey efficiency for a dim Sun-like star requiring a minimum of 1 transit.}
    \label{fig:survey_efficiency}
\end{figure*}

The injection and recovery results give the probability that a given transit signal is detected. To understand the full survey efficiency, we must also consider the probability that a given planet has an inclination such that a transit occurs from our line of sight ($p_\mathrm{tra}$) as well as the probability that the planet transits at least the minimum number of times to be detected, also known as the window function ($p_\mathrm{win}$). In our yield simulations these probabilities are computed directly for each source, however it is worth calculating these values for the survey as a whole to understand the overall transit detection efficiency. We define the survey efficiency, $\eta$, as the total likelihood that a planet is detected in the GBTDS, expressed mathematically as, $\eta = p_\mathrm{tra} \times p_\mathrm{win} \times p_\mathrm{det}$. We show this efficiency for a Sun-like star at the bright and dim limits of our sample in Figure \ref{fig:survey_efficiency}.

For this figure we calculated the average survey efficiency for a Sun-like star with a few simplifying assumptions. Note, this is different from our yield calculations in the next section where we simulate an actual exoplanet population before determining which planets are detected. 
$p_\mathrm{det}$ is calculated for each point in the $a, R_p$ plane using the results of the injection tests in the previous section, and the model for $S/N_{\rm exp}$. 
Because we are considering an average planet, rather than $n_\mathrm{tran}$ directly, we estimate $n_\mathrm{tran}$ by dividing the product of the observing baseline and duty cycle by the orbital period.

$p_\mathrm{tra}$ is calculated assuming a circular orbit and under the simplifying assumption that $R_p\ll R_\star$, so that $p_\mathrm{tra} \approx R_\star / a$. This assumption may be reasonable even for larger planet to star radius ratios. 
While we do not consider vetting efficiency in this work, in the actual survey events that are ambiguous planet candidates are more likely to be rejected due to the large number of stars being searched, thus lowering the detection efficiency for transits with a large impact parameter. This assumption is also present in our methodology for determining whether a planet transits its host star (see section \ref{sec:simplanets}).

For orbital periods guaranteed to produce $\geq7$~transits (i.e., $P<36$~days), $p_\mathrm{win}=1$, and for orbital periods guaranteed to have fewer than two transits (i.e., $P>5$~years), $p_\mathrm{win} = T_{\rm obs}/P$, where $T_{\rm obs} = 6\times72~\mathrm{days} = 432~\mathrm{days}$ is the total observing time of the survey. For periods between 36 days and 5 years, $p_\mathrm{win}$ is calculated numerically using $10^5$ values for $t_0$ evenly spaced from $0\leq t_0<P$ to determine the probability of at least one transit occurring. For the requirement that $n_\mathrm{tran}\geq7$, we calculate $p_\mathrm{win}$ numerically for $P$ = 36--72~days and set $p_\mathrm{win}=0$ for $P>72$~days. 
Features from this window function are apparent for orbital separations of $\sim$0.3--3 au (see Figure \ref{fig:survey_efficiency}), where specific periods are more likely to have multiple transits due to aliasing effects.

These figures show that, for bright stars, Neptune-sized planets ($R_p \approx 4R_\oplus$) should be detectable at nearly any orbital period with $\geq$7 transits. On the faint end of the sample, this efficiency is significantly decreased, although Jupiter-sized planets should be easily detectable at nearly any orbital period, even for the dimmest stars in the sample. 
This detection efficiency lends credibility to the idea that a transit survey can be effective deeper than $F146=21$, and thus the yield presented in this work is a conservative limit. This is especially apparent considering that dimmer magnitudes will add more K and M dwarfs to the planet-search sample, which have more favorable detection efficiencies due to their smaller radii.

In the case that transiting planets can still be detected efficiently with only 1 or fewer transits per season ($<$7 total), the detection efficiency remains high out to several au. 
Transit surveys suffer from a reduced $p_\mathrm{tra}$ at these orbital separations. This is still the case for Roman, however the large number of planet-search stars dictates that a significant number of planets at these orbital separations should still be detected, even below survey efficiencies of $\eta \sim 10^{-5}$. 
However, this is optimistic because the detection efficiency in transit surveys drops significantly for systems with a small number of transits, an effect that we are not considering here \citep{hsu2019,christiansen2020}. This is especially true in the case that there are correlated systematics in the data, which are not yet understood for the GBTDS, and therefore not present in our simulations.

\section{Transiting Exoplanet Yields} \label{sec:results}

Combining the results of the detection efficiency model from the previous section with the Galactic model and assumed exoplanet population from section \ref{sec:simstars} and section \ref{sec:simplanets}, we are able to predict the number and types of planets that Roman will find due to the transit technique. 
For this calculation, we simulate a stellar and exoplanet population for each detector assembly (eighteen total), drawing a random number of stars from a Poisson distribution assuming the average stellar density, assign planets to those stars, and then apply our detection sensitivity model to determine which planets were detected. We do this for one field, and then multiply this yield by the number of fields (7 for our default parameters).

This Poisson process results in uncertainties on the total yield of $\sim$5\%, which is not significant compared to uncertainties in the actual planet occurrence, especially with respect to metallicity, so we generally ignore it when discussing the overall planet yield. However, when considering the planet yield for smaller subsamples of parameter space, such as in sections \ref{sec:yield_mdwarf} and \ref{sec:yield_variation}, the uncertainty is determined by randomly sampling from the simulated transiting planet population. 
We then reapply the detection criteria computed in section \ref{sec:injectiontest} to randomly determine which systems are detected as a function of $S/N_{\rm exp}$. 
We assume an overall uncertainty of 10\% on the total number of transiting planets, before considering detectability, in the GBTDS. This uncertainty should be consistent with uncertainties from the Poisson process discussed above and typical scatter in the $H$-band Luminosity Function residuals from Section \ref{sec:galactic}.
This process is repeated many times and we report the median as our yield and the inner 68\% as our uncertainty. 
Thus, these uncertainties incorporate random, Poisson fluctuations of the stellar source density and random fluctuations on the transit detection sensitivity, but do not consider uncertainties in the relative occurrence rates for different planet types themselves.

\begin{figure*}
    \centering
    \includegraphics[width=0.495\textwidth]{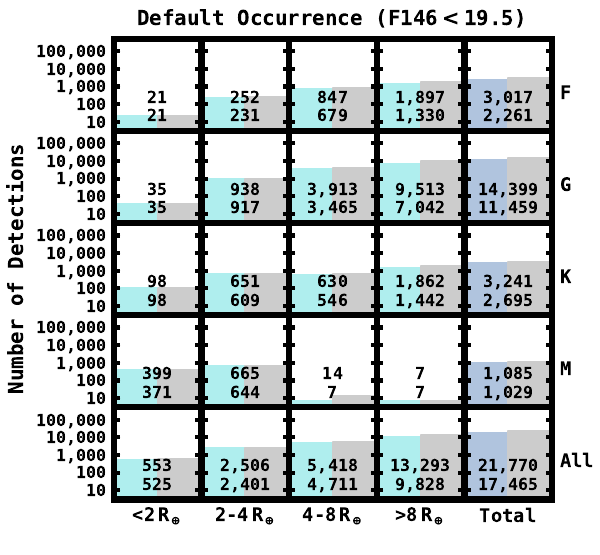}
    \includegraphics[width=0.495\textwidth]{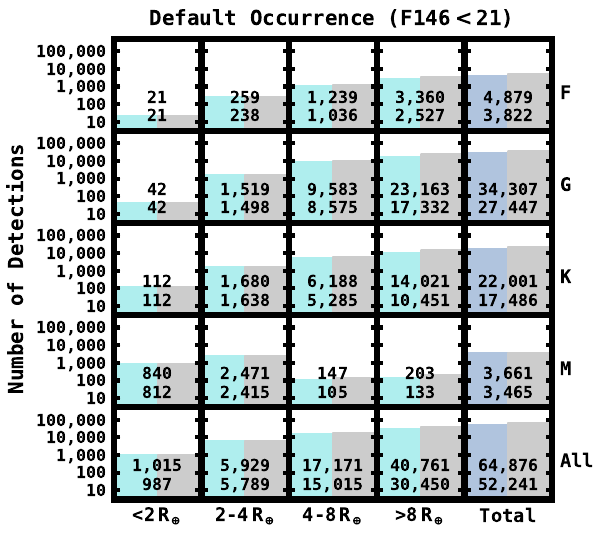}
    \includegraphics[width=0.495\textwidth]{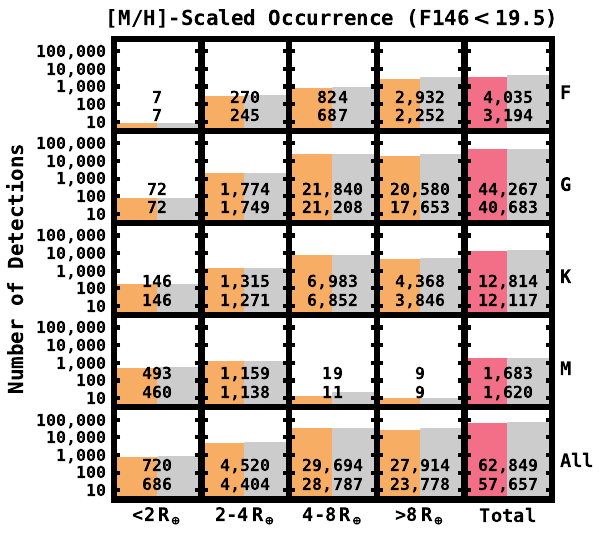}
    \includegraphics[width=0.495\textwidth]{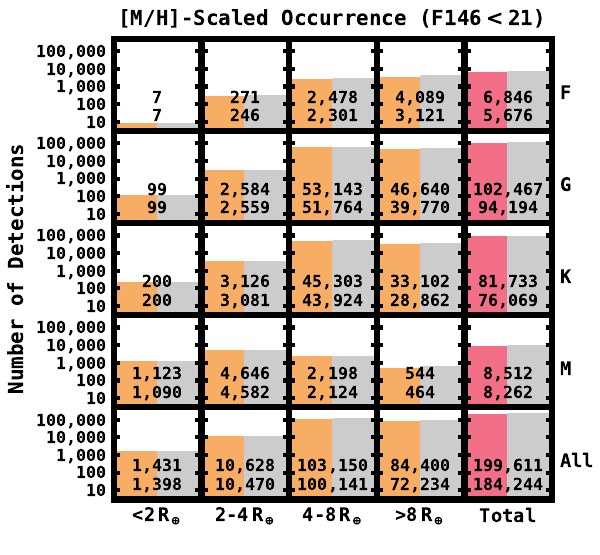}
    \caption{The total transiting planet yields for the Bulge Time Domain Survey for differing combinations of spectral type and planet size. For each combination we report two values for the yield. The bottom number (green/blue, orange/pink bar) is the yield under the conservative requirement of more than 1 transit per season ($n_{\rm tran}\geq7$), while the top number (gray bar) is the yield considering all planets with at least one transit in the full survey. 
    {\it Top Left}: The number of detected transiting planets with stellar hosts brighter than $F146=19.5$ mag. 
    {\it Top Right}: The transiting planet yields for all stars brighter than $F146=21$ mag. {\it Bottom}: The same as the top row, but assuming the relation between planet occurrence and stellar metallicity described in section \ref{sec:simplanets_metals}. }
    \label{fig:yields}
\end{figure*}

The planet yield broken down by host spectral type and planet radius is shown in Figure \ref{fig:yields}. This yield is further discussed in Sections \ref{sec:yield_default}-\ref{sec:yield_single} for the default survey parameters considered in this work. 
Because the survey design has yet to be decided, Section \ref{sec:yield_variation} considers variations on the nominal survey design and their effect on the overall planet yield.

\subsection{Exoplanet Yield for Default Simulation Parameters}\label{sec:yield_default}

The overall planet yield for our default simulation parameters and assumptions on the exoplanet and stellar populations are presented in the top row of Figure \ref{fig:yields}. Overall, we should expect that between 52,000 and 64,000 transiting planets will be discovered, depending on the requirements for $n_{\rm tran}$. 
Of these planets, about 90\% of the planets will be Jupiters ($R_p>8 R_\oplus$) or sub-Saturns ($R_p =$~4-8~$R_\oplus$).

Even under these conservative assumptions, the transiting planet yield is expected to be higher than under any other previous transit survey. Assuming the default simulation parameters, the GBTDS should find $\gtrsim$30,000 Jupiter-sized planets ($R_p > 8 R_\oplus$) and $\gtrsim$15,000 sub-Saturns, of which the vast majority will have GK-type hosts. 
These planets and stellar hosts represent the vast majority of the planet yield.

One surprising result is the relative paucity of F-type planet hosts. While the IMF assumed by the BGM1612 model results in a larger number of lower mass stars, the primary driving factor in this discrepancy derives from an under abundance of high mass stars in the Galactic bulge and older regions of the thin disk at small Galactic radii. 
Because the bulge population is modeled in the BGM1612 model as a single burst of star formation 10 Gyr ago, stars with $M_\star \gtrsim 1.1 M_\odot$ have evolved off of the main sequence and mostly occupy the Horizontal Branch. 
As a result, the number of F-type stars can serve as a tracer of the disk population. This is further discussed in section \ref{sec:galactic}.

The yield of small planets derives primarily from very short orbital periods in the case of G dwarfs, where few planets with $R_p < 4 R_\oplus$ are found at $P\gtrsim16$~days.
The total yield for small planets is higher for GK dwarfs, but only because in the brightness range considered in this work there are significantly more G dwarfs ($\sim$24$ \times 10^6$) and K dwarfs ($\sim$31$\times 10^6$) observed than M dwarfs ($\sim$1$\times 10^6$). For super-Earth sized planets ($R_p < 2 R_\oplus$), nearly 90\% of the yield should derive from M dwarf hosts, as planet-star radius ratios are not favorable for detection for any other stellar type besides perhaps, bright K dwarfs. This is discussed further in section \ref{sec:yield_mdwarf}.

These estimates for the total planet yield should be conservative for a few different reasons, the most important of which is the stellar sample considered. 
We only consider stars brighter than $F146=21$~mag, which is approximately where crowding becomes the dominant source of uncertainty in our simulated aperture photometry pipeline. However, even with these limits the average photometric precision is $\sigma_\mathrm{dpp} < 1\%$ integrated over an hour (see Figure \ref{fig:instrumentmodel}), which is adequate to discover additional planets, and
PSF photometry should improve these limits further by minimizing noise contributions from neighboring stars. 
For this reason, it is likely that a transit survey should still be successful at least as dim as $F146\approx22$, which would nearly double the number of monitored stars (see also section \ref{sec:discussion_psf}).

The predictions made here may also be conservative due to our default assumption of a constant planet occurrence.  
The yield becomes even more striking when considering correlations between metallicity and planet occurrence, as discussed in the following section.

\subsection{Change in Yield with Stellar Metallicity}\label{sec:yield_metals}

\begin{figure}
    \centering
    \includegraphics[width=\columnwidth]{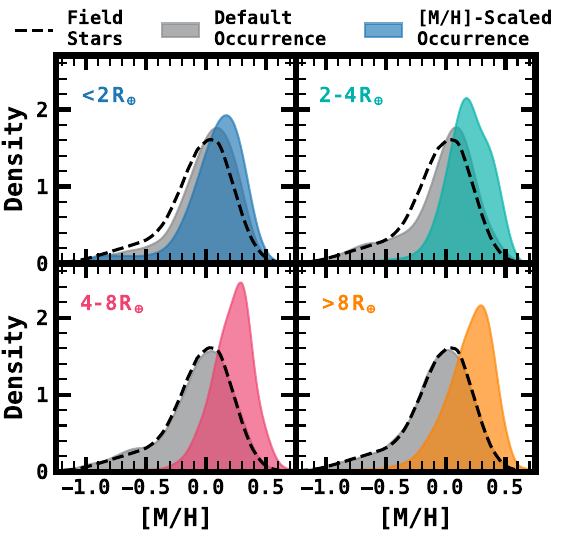}
    \caption{The stellar host metallicity distribution functions of the GBTDS transiting planet sample for differing planet radius ranges. Each panel shows
    the distribution assuming a constant occurrence rate (gray), 
    the distribution when scaling the planet occurrence as a function of metallicity (colored), and the distribution of field stars with $F146<21$~mag (dashed line). Metallicity differences in the default occurrence distributions with respect to field stars reflect the Galactic metallicity distribution function convolved with the transit detection efficiency for each particular planet size. }
    \label{fig:planetmdf}
\end{figure}

Because the occurrence of both giant and small planets, particularly at short periods, is correlated with stellar metallicity \citep{fischer2005,johnson2010,mulders2016,petigura2018,wilson2022}, one would expect planet occurrence to also vary with Galactic position. 
Therefore, the assumptions made by adopting occurrence rates from Kepler may not be strictly valid under the sample of stars considered in this work, which vary substantially in metallicity. 
To take these correlations into account, we apply weights to each of the detected planets based on the change in occurrence rate implied by its host star metallicity, as described in \ref{sec:simplanets}. This result is shown in the bottom panels of Figure \ref{fig:yields}. 
Under these scenarios, the assumed yield increases by just over a factor of $\sim$3. The stellar metallicity distribution of the planet hosts are shown in Figure \ref{fig:planetmdf} under both assumptions for the underlying planet occurrence.

Unsurprisingly, the change in yield is largest for short period ($P<10$~days) sub-Saturns, the planet archetype with the strongest correlation between planet occurrence and metallicity, which increase from $\sim$17,000 to $\sim$103,000 under these new assumptions. 
These particular planets drive the increase in total yield. 
However, because relatively few giant planets (particularly sub-Saturns) were discovered by Kepler, the precise strength of the correlation between their occurrence and stellar metallicity has large uncertainties. 
In an attempt to understand the dependence on the planet yield under different plausible scenarios for the correlation between metallicity and occurrence, we consider alternative formulations of this relation.

If it is appropriate to extrapolate the power-law dependence of planet occurrence ($f_p$) on metallicity ($f_p \propto 10^{\beta \mathrm{[M/H]}}$) indefinitely beyond [Fe/H] = 0.4, then one would expect a total planet yield of nearly $\sim$400,000, dominated by $\sim$300,000 sub-Saturns. However, extrapolating a power law is unphysical in these circumstances, especially based on the small number of detections from Kepler. At some point, it is more likely that planet occurrence will saturate. 
It is also not clear that a power law is a reliable parameterization of the dependence of planet occurrence on metallicity, even though it seems a reasonable approximation over the limited range in [Fe/H] probed by Kepler. 
Assuming instead that the hot sub-Saturn occurrence scales as the hot Jupiter occurrence, $\beta=3.5$, which is consistent within the $1\sigma$ uncertainties from \cite{petigura2018}, the overall planet yield would be $\sim$135,000, with $\sim$40,000 sub-Saturn planets. 
In all, depending on the exact relationship between the occurrence of short period sub-Saturns and stellar metallicity, the yield can reasonably be expected to range between $\sim$135,000 and $\sim$200,000 total planets for stars with $F146<21$.

Perhaps surprisingly, the expected yield for small planets increases by $\sim$70\% from nearly $\sim$7,000 to $\sim$12,000 once metallicity correlations are considered, which is almost as large as the relative increase for Jupiters ($\sim$110\%), despite the significantly smaller correlation between occurrence and host metallicity. 
This apparent discrepancy highlights the structure of the metallicity distribution in the Milky Way. 
Smaller planets are significantly more detectable around smaller and brighter stars, compared to Jupiters which are detectable around nearly every star considered in this work, which skews the effective average distance of stars amenable to small planet detection more significantly than the stars amenable to detecting large planets. 
Because the increasing (in the direction of the Galactic center) metallicity gradient in the midplane of the Milky Way applies only to the thin disk stellar population, older more metal-poor populations start to dominate the sample at distances $\gtrsim$6 kpc. Thus, the average star amenable to detecting small planets has a higher metallicity than the average star amenable to detecting large planets, according to our adopted Galactic population model (see Figure \ref{fig:planetmdf}).

\subsection{Expected Yields for M Dwarfs}\label{sec:yield_mdwarf}

According to the Galactic model adopted for this work, the GBTDS should monitor $\sim$10$^6$ M dwarfs with $F146<21$, which, while much smaller than the number of GK stars, is much larger than the number of M dwarfs monitored by previous transit surveys. 
An added benefit is that the M dwarf planet hosts have typical distances ranging from 1.1-3.0~kpc, which, depending on the impact of crowding, may be sufficient to derive parallaxes for most of the M dwarf sample in the GBTDS and infer a stellar radius given Roman's expected astrometric performance \citep[10 $\mu$as by the end of the GBTDS for $H_{AB}=21.6$;][]{sanderson2019}. 
As a result, this sample has the potential for several breakthroughs in planet demographics specific to M dwarf hosts.

For instance, under our assumptions, the expected yield for Jupiters and sub-Saturns orbiting M dwarfs is relatively small, at $329^{+51}_{-31}$, or $230^{+21}_{-27}$ if only considering systems with $n_{\rm tran}\geq7$.
However this total number is still an order of magnitude larger than the number of systems currently known, and should therefore provide a unique opportunity to measure the occurrence of such systems to high fidelity. This estimate also neglects metallicity effects, which as mentioned in section \ref{sec:simplanets_metals}, may be even more strongly correlated with planet occurrence for M dwarfs than for their FGK counterparts. Assuming the same correlation with metallicity, the yield may be as high as $\sim$2600 giant planets with M dwarf hosts. 
Also, at our dim magnitude limit of $F146=21$, the detectability of a giant planet transiting an M dwarf is high, and so this is a conservative estimate, as the number of M dwarfs increases dramatically when the magnitude limit changes from $F146=21$ to $F146=22$.

Another opportunity comes from the fact that
the season duration for the GBTDS is sufficiently long to encompass the habitable zone for M dwarfs. To estimate our yield for habitable-zone planets, we adopt the parametric relations from \cite{kopparapu2013}, who calculated the inner and outer edges of the habitable zone under varying assumptions for the planetary atmosphere. Assuming the default occurrence rate, and defining a conservative habitable zone (inner edge: max greenhouse, outer edge: moist greenhouse), the GBTDS should yield $38^{+7}_{-6}$ habitable-zone planets with  $R_p<2 R_\oplus$.  \cite{kopparapu2013} also defined an optimistic habitable zone, with the inner edge set by the instellation flux received by Venus 1 billion years ago (``recent Venus"), based on the inference that Venus has not had liquid water on its surface for at least the past 1 billion years and the outer edge set by the flux received by Mars 3.8 billion years ago (``early Mars") based on the inference that Mars had liquid water on its surface 3.8 billion years ago. Adopting this definition,
the GBTDS should yield $114^{+17}_{-14}$ optimistic habitable-zone planets with $R_p < 2 R_\oplus$. 
This yield is several times higher than the expected yield from the first seven years of the TESS mission \citep{ricker2014}, which is expected to find $9\pm3$ and $18\pm5$ small planets within the conservative and optimistic habitable zones, respectively, within the same definitions adopted here \citep{kunimoto2022}. 
Changing the radius limit to $R_p<1.5 R_\oplus$ reduces the expected yield to $12^{+4}_{-3}$ and $34^{+6}_{-7}$ small planets in the conservative and optimistic habitable zones, respectively.

\subsection{Exoplanets with One or Fewer Transits per Season}\label{sec:yield_single}

Because the number of stars searched for planets is so large, there should be a significant number of planet detections out to a few au. Under the optimistic assumption that the detection efficiency does not deteriorate for transiting planets with only one or two events, the yield for planets with one or fewer transit per season could be as high as $\sim$12,000. 
Of these planets, as many as $\sim$1,800 could be detected with only one transit across the entire survey, and have orbital separations ranging from the minimum separation of $\sim$0.3 au, to as large as $\sim$7 au. 
This area of parameter space overlaps in sensitivity with
the microlensing survey, which is most sensitive to detecting planets with orbital separations of $\sim$3 au \citep{penny2019}, offering the opportunity for an indirect comparison of the population of giant planets in the GBTDS detected via transit against those detected via microlensing. Note, that these yields assume that planets with a single, or only a few, transits can be recovered with the same significance as those with $n_{\rm tran}\geq7$. However, studies with long-period transiting planets in Kepler have found that the efficiency of detecting such planets plateaus at $\sim$50\% \citep{herman2019}, so the yields presented in this discussion should be considered optimistic, though the general principles should still apply even with a smaller sample size. 
We discuss this opportunity more in Section \ref{sec:discussion_microlensing}.

Due to the observing pattern of the GBTDS, planets with orbital periods as large as 4.5 years may have at least two transits. 
For planets with $P < 4.5$~years that have $<1$ transit per season, but $\geq$2 over the full survey, non detections in different seasons may be sufficient to rule out shorter period aliases to the degree that the most likely period has negligible uncertainty, as would be the case for a transiting planet with more than one transit in a single season. 
However, deciphering the orbital period for most planet candidates with $P>72$~days will require careful modeling of the transit shape to extract the transit duration and impact parameter, which may be easier than for optical surveys at constant transit $S/N$ because Roman is observing in the near-infrared where limb-darkening effects are less severe. In this high $S/N$ case, the likely limiting source of error will be the mean stellar density combined with unknowns in eccentricity \citep{yee2008}. 
Thus, careful modeling of the transit shape should still place reasonable constraints on the orbital period and semi-major axis that should be useful for demographics studies, even if the uncertainties are large for any one system.

\subsection{Change in Relative Yield based on Survey Variations}\label{sec:yield_variation}

\begin{figure}
    \centering
    \includegraphics[width=\columnwidth]{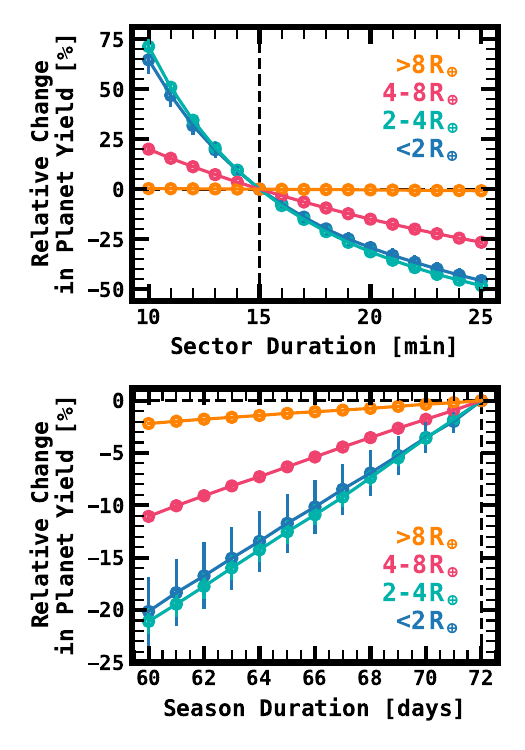}
    \caption{The relative change in yield per field as a function of planet size for changes in survey parameters. Each colors represent planets of a different size, Jupiters: orange, sub-Saturns: pink, sub-Neptunes: teal, and super-Earths: blue. {\it Top:} The relative change in yield with sector duration, which is equal to the observing cadence. {\it Bottom:} The relative change in yield as a function of season duration. } 
    \label{fig:relative_yield}
\end{figure}

The actual design of the GBTDS has yet to be decided, so it is important to test the change in transiting planet yields that come with design variations.
Changes in this design generally raise or lower the yield through one of, or the combination of the number of stars searched for planets, the photometric signal-to-noise over the duration of a transit, and the number of transits observed. 
The change in yield with respect to changes in the season duration and sector duration (i.e., cadence) of the survey are shown in Figure \ref{fig:relative_yield}.

Generally speaking, the yield of large planets is set by the number of stars searched, as the survey efficiency is not limited by $p_{\rm det}$. However, the yield for small planets at lower average $S/N$ is constrained by $p_{\rm det}$, so that improvement to the transit $S/N$ results in higher survey efficiencies. 
To estimate the change in yield for specific planets as a function of survey parameters, we simulate a single set of transiting planets for one field of the GBTDS, calculate $S/N_\mathrm{exp}$ under each survey design, and apply our detection model to estimate the yield under each set of parameters. This process is repeated 1000 times, and we report the average and standard deviation on the yield relative to the default design.

\subsubsection{Sector Duration}

The sector duration is the time it takes for the Roman space craft to observe one sector, defined as the full $\sim$2 square degrees surveyed by the GBTDS (i.e., all fields). Thus, the sector duration is equal to the observing cadence. It is limited by the sum of total exposure time per sector and the total time to slew and settle between fields. Thus, the sector duration can be reduced by decreasing the slew and settle overhead, or by observing fewer fields. Changes in sector duration lead to changes in the transit $S/N$ by increasing the sampling rate thereby obtaining more in-transit data.

The sector duration primarily impacts the yield for small planets. This impact is readily apparent, as even a one minute reduction in sector duration leads to a yield increase of 9.1$\pm$0.3\% and 9.6$\pm$0.1\% per field for sub-Neptunes and super-Earths, respectively. Conversely, an increase in cadence of one minute results in a reduction of 8.1$\pm$0.3\% and 8.2$\pm$0.1\% in the super-Earth and sub-Neptune yield. Thus, to maximize the yield for small planets, the cadence should be minimized wherever possible, even if the number of fields and surveyed area is compromised. More practically, this can be accomplished by minimizing the distance between consecutive fields.
The yield for giant planets, on the other hand, are insensitive to changes in sector duration, varying by less than 1\% from a sector duration of 10 to 25 minutes, at least within the magnitude range considered here.

\subsubsection{Season Duration}\label{sec:season_duration}

The duration of the GBTDS observing season is constrained by two factors. On the short end, it is set by the requirement to observe microlensing events with timescales up to 60 days, and on the long end it is set by the visibility window of the bulge fields, which can be observed for a maximum of 72 days. 
The net difference in yield from reducing the season duration to 60 days is a reduction in the number of observed transits, and a reduction in $p_\mathrm{win}$. The latter change is minor, but the former change has a noticeable effect on the yields of small planets. 
The change in yields with season duration is approximately linear, dropping by 1.7\% per day (-21.3$\pm$1.6\% at 60 days) and 1.8\% per day (-20.6$\pm$3.1\% at 60 days) below maximum season duration in the case of sub-Neptunes and super-Earths, respectively. The change in yield is negligible for giant planets, decreasing by a maximum of only 1.2$\pm$0.1\% from a season duration of 72 days to 60 days.

\subsubsection{Cadence and Exposure Times of Secondary Filters}

The requirements for the GBTDS specify additional observations at a regular cadence in a secondary filter, complementary to the wide $F146$ filter, for the purpose of stellar characterization. The cadence, exposure, and even the filter itself of such observations is undecided. 
\cite{montet2017} and \cite{penny2019} both assumed additional observations would be taken in the $F087$ filter ($\lambda =$~0.76--0.98~$\mu\mathrm{m}$) at a cadence of 12 hours, but \cite{montet2017} assumed a constant exposure time of 52 seconds between the two filters while \cite{penny2019} assumed an increased exposure time of 286 seconds for the $F087$ images.

If the secondary filter observations cannot be used to supplement the transit search, then the resulting loss in transit $S/N$ may have a small effect on the detectability of a putative transit signal. In this work, we make no assumptions about the cadence or observations of secondary filters, neglecting such observations altogether. 
This may not be unreasonable, as single, short exposure, observations taken at a cadence of 12 hours may be on short enough timescales that one can interpolate between the secondary observations without loss of sensitivity. Alternatively, one may be able to incorporate the secondary observations into the detection algorithm itself.

Simulating the loss of transit $S/N$ caused by omitting an observation every 6 hours, leads to a reduction of $\sim$3-5\% in the yield of small planets, having only a minor overall effect.
In fact, increasing the cadence of such observations is likely to improve the transit survey by identifying astrophysical false positive transit detections \citep{montet2017}.

\subsubsection{Variations from the Roman Design Reference Mission}\label{sec:drm}

The Roman Design Reference Mission (DRM) specifies variations of the GBTDS parameters that would accomplish the primary science goals of the survey, and could be realized by the actual mission, making reasonable (perhaps conservative) assumptions about the performance of the observatory with respect to qualities such as slew/settle times and momentum unloading. 
The DRM specifies four different strategies for the GBTDS, assuming changes primarily in the slew and settle times of the observatory, and the number of fields observed. 
The DRM thus provides a reasonable framework with which to gauge the compromise between sector duration and the number of fields observed. To understand these effects, we simulated the change in yield expected for each strategy reported in the DRM. 
The results of these simulations, as well as the different parameters of the survey strategies, are listed in Table \ref{tab:relativeyields}. 
In these simulations we make the assumption, as in our previous simulations, that one is able to either incorporate the photometry from secondary filters or interpolate over skipped cadences so that the reduction in transit $S/N$ is negligible. In reality, assuming those points are not available would result in a $\sim$5\% drop in the yields for small planets, those most affected by average $S/N$ changes. This decreases may be larger if filter changes lead to increased overhead.

\begin{table*}[]
    \centering
        \caption{GBTDS variations considered in the Roman Design Reference Mission}
    \begin{tabular}{lccccc} \hline
         & \textbf{DRM1}  & \textbf{DRM2}   & \textbf{DRM3}  &  \textbf{DRM4} & \textbf{This Work} \\ 
        Number of Reaction Wheels & 6 & 6  & 5  & 5 & -- \\
        Fields per Sector & 8 & 7  & 8  & 7 & 7 \\
        Surveyed Area (deg$^2$) & 2.247 & 1.966  & 2.247  & 1.966 &  1.966 \\
        Stars with $F146<21$ ($\times10^6$) & $\sim$67 & $\sim$59 & $\sim$67 & $\sim$59  & $\sim$59 \\
        Sector Duration (min)  & 15.4 & 13.7 & 16.6  & 14.8 & 15.0 \\
        Cadences per Season\tablenotemark{a} ($F146$) & 6458 & 7293  & 5954 & 6710  & 6913  \\
        Cadences per Season\tablenotemark{a} ($F087/F184$) & 288 & 288 & 288 & 288 & 0 \\
        \multicolumn{6}{l}{\textbf{Change in Exoplanet Yield Relative to Default (\%)} } \\
        super-Earths ($R_p\leq2 R_\oplus$) &  +10.7 &  +11.9 &  +1.4 &  +1.6  & --   \\ 
        sub-Neptunes ($2 R_\oplus < R_p \leq 4 R_\oplus$) &  +10.4 &  +12.5 &  +0.1 &  +1.7  & --   \\ 
        sub-Saturns ($4 R_\oplus < R_p \leq 8 R_\oplus$) &  +12.8 &  +4.5 &  +8.5 &  +0.7   & --  \\ 
        Jupiters ($R_p > 8 R_\oplus$) &  +14.3 &  +0.1 &  +14.2 &  +0.0  & --   \\ 
        Total &  +13.3 &  +3.2 &  +10.5 &  +0.5  & --   \\ 
         \hline 
    \end{tabular}
    \raggedright
    \tablenotetext{a}{These estimates ignore data gaps from momentum unloading and station keeping common across all survey strategies, and assume a 72-day season duration. }
    \label{tab:relativeyields}
\end{table*}

As may be expected from previous simulations, the relative yield is not significantly affected overall, and the increase in yield is primarily driven by the increase in yield of Jupiters and sub-Saturns specifically. 
As expected from our simulations in sector duration, the yield of small planets is most affected by the slew/settle times. 
Comparing the relative yields for reference designs with a constant number of reaction wheels (i.e., slew and settle speeds), the tradeoff between sector duration and number of fields observed is minimial for small planets. 
For large planets on the other hand, a reduced cadence has a minimal effect and the primary driver of the exoplanet yield is the number of fields searched.

This picture only holds up for the magnitude limits imposed by this work, where we do not consider stars dimmer than $F146=21$~mag. In this limit, the Jupiter yields are not limited by transit $S/N$, so it stands to reason that the yields would improve were we to consider a dimmer magnitude limit. 
With this in mind, it becomes clear that the changes in yield between higher cadence observations and additional observed stars have two distinct sources. 
In the former, we are able to probe stars that are dimmer, more distant, and thus reach a larger range of Galactic populations. With the latter, the survey design would improve the statistics for the closer, nearby sample of stars. 
Simulating planet transits across dimmer stars wold be necessary to understand to what extent higher cadence observations would allow one to probe planet demographics within differing Galactic populations.

\subsection{Impact in Yield Estimates due to Simplifying Assumptions }

In this work we make a number of simplifying assumptions in estimating the number of detectable transiting planets in the GBTDS. In this section we attempt to quantify the impact of some of these assumptions to understand the systematic uncertainty in our yield predictions.

\subsubsection{Correlated Instrumental Noise}\label{sec:correlated_noise}

In developing our transit detection and photometric noise model, we only explicitly model two types of noise in our simulations: photon noise from the source, neighboring stars, and background flux, and Gaussian noise from each non-destructive read (see section \ref{sec:simdata}). However, the actual GBTDS light curves will have some amount of correlated noise, which we define here as any noise source which prevents the transit $S/N$ from increasing as quickly as $ S/N_{\rm exp} \propto n_p^{1/2}$ in the light curve, where $n_p \approx n_{\rm tran}\times (t_{\rm dur}/\Delta t)$ is the number of photometric data points collected in transit. 
Sources of such noise may be astrophysical or instrumental. Causes for the latter may be imperfections in the hardware (e.g., detector non-linearities, intra- and inter-pixel sensitivity variations, pointing uncertainties) or the photometry and transit search pipeline (e.g., poor detrending or removal of ensemble noise). Such limitations are difficult to predict at this stage due to ongoing development of the Roman analysis software and detector characterization.

We can place a conservative limit on the impact of correlated detector noise by imposing a strict transit depth limit for all planet detections. 
The choice of limit is somewhat arbitrary, but a reasonable upper-bound may be the impact of detector non-linearities. \cite{mosby2020} characterized these limits for a group of candidate WFI arrays, finding that for the median pixel, the maximum deviation of a single read frame from an arbitrary non-linear correction was 1.5 ppt. 
It is unclear the degree to which this effect should be correlated with nearby pixels, so we take a conservative approach and assume that this limit applies per light curve, and represents the smallest amplitude at which a noisy signal can be recovered.
Imposing this limit as our criterion for a minimum detectable transit depth across all sources results in a yield decrease of $\gtrsim$30\% for small planets, but the impact is much smaller for sub-Saturns, and nearly imperceptible for Jupiters  (see Table \ref{tab:correlated}). For completeness we also consider minimum detectable transit depths ranging from 1.0 ppt to 0.1 ppt.

Considering that non-deterministic non-linear detector responses are typically most important for sources near the saturation or semi-saturation limit ($F146 \lesssim 18$), these changes in yield are extremely pessimistic.  Applying these limits to only bright sources, the inferred decrease in yield becomes $<$10\% for $(R_p/R_\star)^2_{\rm min} = 1.5$~ppt for all planet sizes except super-Earths, where the decrease in yield is instead 16\%. 
Even this estimate is pessimistic, as it makes an implicit assumption that the photometry of such sources is derived from a small number of brightly-illuminated pixels, which is not the case for bright sources as discussed in section \ref{sec:apertureflux}, 
where most of the signal will actually derive from the sum of many less illuminated pixels away from the PSF core which should be more robust to detector non-linearities because the ramp-fitted flux in these pixels will derive from more reads. Even in the worst (unrealistic) case scenario, the total transiting planet yield should decrease by only 7\%. This is due, in part, to the nature of transiting planets susceptible to discovery in the GBTDS which have large transit depths.

\begin{table}[]
    \centering
    \caption{The relative decrease in the planet yield under the assumption that correlated instrumental noise places a constraint on the minimum detectable transit depth. }
    \begin{tabular}{cccccc} \hline
         & \multicolumn{5}{c}{Minimum Transit Depth (ppt)} \\
         $R_p\;(R_\oplus)$ &  1.5 & 1.0 & 0.5 & 0.3 & 0.1\\ \hline
        $\leq$2&  $36\%$ & 23\% & 8.2\% & 4.3\% &$<$1\%\\
        2-4  &  31\% & 16\% & 2.8\% & $<$1\% &$<$1\% \\
        4-8  & 12\% & 5.7\% & 1.1\% & $<$1\% &$<$1\% \\ 
        $>$8 & 1.3\% & $<$1\% & $<$1\% & $<$1\% &$<$1\% \\ 
        All & 7.0\% & 3.6\% & $<$1\% & $<$1\% &$<$1\% \\ \hline         
    \end{tabular}
    \label{tab:correlated}
\end{table}

Overall, correlated instrumental noise is unlikely to have a major impact on these results, except in the case of small planets ($R_p < 4 R_\oplus$), in which the detection yields may decrease by $\sim$10-20\% if instrumental systematics are not controlled to better than 0.5-1.0 ppt.
Thus, for the purpose of detecting small planets, we advocate for a goal of controlling systematics from correlated instrumental noise on bright sources to a limit of 0.5 ppt over transit duration timescales ($\lesssim$12 hours). This robust response in the overall planet yields to correlated systematics can likely be attributed to our conservative saturation floor of $\sim$1.3 ppt per exposure. If the actual single-exposure noise floor for the GBTDS is significantly lower than our model, instrumental systematics will be more limiting for bright sources and will disproportionately impact other science cases such as the detection of secondary eclipses and phase curves (see section \ref{sec:secondary_eclipse}).

All of these calculations assume the default planet occurrence rate. Performing the same calculations using the [M/H]-scaled planet occurrence rate instead results in negligible differences at constant planet radius, but overall is even less sensitive to these limits due to the increased yield of giant planets.

\subsubsection{Correlated Astrophysical Noise}\label{sec:astrophysical_noise}

Astrophysical sources may also introduce correlated noise to the GBTDS light curves with timescales similar to the transit duration ($\lesssim$1-2~days). For the purpose of this discussion, we define correlated noise in the same way as in the previous section, but restrict our discussion in which the source of such noise is astrophysical rather than instrumental in nature.  
Such sources of noise could include background variable stars with large amplitudes (e.g., close binaries with large ellipsoidal variations, $\delta$ Scuti stars, accreting systems) as well as lower-amplitude noise from the target star itself (e.g., rotational modulations, flares, asteroseismic activity). Correlated astrophysical noise can lead to a decreased catalog reliability and higher rate of false alarms (i.e., where statistical fluctuations or coherent non-transiting signals such as stellar variability are mistakenly identified as a transit-like signal, as opposed to false positives, where a transit-like signal derives from a source other than a transiting planet), either because the true light curve noise may be underestimated or because the shape of the non-transiting signal may confuse the transit detection algorithm. 
However, in this section we neglect the impact of correlated noise on reliability, and focus solely on completeness, i.e., what is the fraction of transiting planets in our simulated planet catalog that correlated noise would otherwise prevent us from detecting?

For two primary reasons, correlated astrophysical noise is unlikely to have a large impact on these overall results. First, for most GBTDS stars, Roman's differential photometric precision is not sensitive to the low-amplitude ($\sim$0.1-1 ppt) modulations that typically plague transit surveys, so photon noise will dominate over the most common sources of stellar variability such as rotational modulations and asteroseismic activity. 
Second, the fraction of intrinsically large-amplitude variable stars is small, particularly on timescales relevant to transit detection, so the fraction of planet search stars that either employ such variability, or are polluted by such variability from nearby stars, is likely below the Poisson uncertainty from our methodology for generating the simulated planet catalog.

Predicting the fraction of sources which employ large-amplitude astrophysical correlated noise would require a more sophisticated Galactic population model than what we've adopted, but we can justify the first point by comparing the impact of correlated noise in Kepler. In Kepler, the majority of light curves with large correlated noise can be attributed to either asteroseismic oscillations or rotational modulations.
As a result, the impact of correlated noise in Kepler should serve as a reasonable upper bound for the expected impact of correlated noise in the GBTDS, because Kepler's lower photon noise limit and optical bandpass exacerbate the impact of stellar variability on transit detectability. The Kepler population is also younger than the expected GBTDS population, so stars in the GBTDS planet-search sample should have less stellar activity and therefore most brightness modulations will have lower amplitudes and lower frequencies.

Comparing the completeness of the Kepler detection pipeline on light curves where the noise increases at longer timescales from 7.5-15 hours against light curves where the noise decreases on those same timescales, it was shown that Kepler's transit detection efficiency decreased by 15-20\% for light curves with strong correlated noise, which represent $<$20\% of all observed Kepler sources \citep{burke2017,christiansen2020}. 
This implies a degradation in transit sensitivity of $\sim$3-5\% over the full survey. 
Although the GBTDS light curves may have a larger fraction of correlated noise caused by the variability of background sources, this should still be a conservative limit because 
a large fraction of the Kepler light curves with correlated noise (not necessarily from astrophysical sources) can be attributed to asteroseismic oscillations in red giants, which are not considered in this work and for which the amplitude of such variations is, in most cases and in the magnitude range considered here, well below Roman's photon noise limit anyway. 
Thus, assuming a detection pipeline with a Kepler-like response to correlated noise, the impact of intrinsic stellar variability, which makes up the majority of correlated astrophysical noise, on the overall planet yield should be $<$3\%, which is below the Poisson uncertainties from our process for generating a simulated planet catalog.

\subsubsection{Change in Binarity with Galactic Environment} \label{sec:yield_binary}

In this work we assume that there are no planets in binary systems, and that the binary fraction is constant. 
However, because the binary fraction is anti-correlated with both stellar metallicity \citep{badenes2018,moe2019,price-whelan2020} and $\alpha$ abundance \citep{mazzola2020}, binarity will also vary based on Galactic environment. Thus, one might expect a higher fraction of single stars in metal-rich environments such as the inner disk, which would inflate the transiting planet yields.

To estimate the impact that variable binarity and multiplicity may have on our predicted transiting planet yield, we adopt the results from \cite{moe2019}, who found that the close ($P\lesssim 10^3$~days) binary fraction ($f_{\rm bin}$) of Solar-like stars can be represented as a piece-wise linear function of the form, 
\begin{align}
f_{\rm bin}(Z) &=
   \begin{cases}
        -0.065(Z+1) + 0.4,& \text{if } Z<-1\\
        \;-0.20(Z+1) + 0.4 ,& \text{if } Z\geq -1 
    \end{cases}
\end{align} 
where $Z=$~[Fe/H] over the range $-3<{\rm [Fe/H]}<+0.5$. Using this relation we can estimate the relative increase in single stars as a function of [M/H] in our simulated catalog, $\Delta f_{\rm bin}(Z) = f_{\rm bin}(Z) - \langle f_{\rm bin} \rangle$, where $\langle f_{\rm bin} \rangle = 21.5\%$ is the average close binary fraction implied by integrating this relation over the metallicity distribution of dwarfs and subgiants with $F146<21$ in our simulated stellar catalog. 
Note, based on our definition of close binary ($a<100$~au) and adopted period distribution from section \ref{sec:simstars_binarity}, our catalog has a close binary fraction of 27.4\%. 
Using the above estimate, we calculate a weighting function for each planet host in our simulated planet catalog of the form, 
\begin{align}
     & w_{\rm bin}(Z) = \frac{f_{\rm single}-\Delta f_{\rm bin}(Z)}{f_{\rm single}} 
\end{align}
where $f_{\rm single}$ is the fraction of single stars (56\% for $M_\star \geq 0.7 M_\odot$, 74\% for $M_\star<0.7 M_\odot$; see  section \ref{sec:simstars_binarity}).

Applying this framework assuming the default planet occurrence, we estimate a minor relative increase of 2\% and 1\% in the yield of super-Earths and sub-Neptunes, respectively. There is no discernable change ($<$0.1\%) in the relative yield for giant planets caused by a metallicity dependent binary fraction. 
Under the [M/H]-scaled planet occurrence rates, we find that the relative yield increases by 5\% for super-Earths, and between 9-10\% for sub-Neptunes, sub-Saturns, and Jupiters. 
Thus, a variable close binary fraction will be a significant confounding variable for the detailed demographics studies the GBTDS transiting planet sample should enable, particularly when controlling for Galactic environment. However, the effect should be significantly smaller than the change in the underlying planet occurrence rate itself, although both effects will be correlated.

\subsection{Comparison to Other GBTDS Transiting Planet Yield Estimates}

\cite{montet2017} estimated the transiting planet yield of the GBTDS to be between $\sim$70,000 and $\sim$150,000, depending on the strength of the correlation between planet occurrence and stellar metallicity. 
While these are similar to our overall results, the breakdown of planet size and host spectral type varies significantly between our catalogs. 
For example, we predict fewer planets with F-type hosts ($\sim$4,000-6,000 compared to 13,000-25,000), more planets with K-type hosts ($\sim$17,000-76,000 compared to 3,000-52,000), and an order of magnitude more small planets overall ($\sim$7,000 in this work compared to $\sim$800).

The source for each of these differences is unclear, but is likely a combination of several factors, including differences in the assumed planet population, simulated stellar sample, and noise model. \cite{montet2017} generally predict a higher occurrence of giant planets, while we predict more small planets. 
In the case of GK dwarfs, the difference in small planet yield may partly derive from differences in the assumed underlying planet occurrence rates. \cite{montet2017} adopted occurrence rates for FGK dwarfs from \cite{howard2012} who underestimated the occurrence of sub-Neptunes compared to \cite{hsu2019}, possibly due to a combination of systematic biases in the inferred stellar radii \citep{berger2018} and/or early uncertainties in the completeness of Kepler's transit detection pipeline. 
These differences, combined with the lower number of GK dwarfs assumed by their Galactic population model ($\sim$17$\times$10$^6$ compared to $\sim$55$\times$10$^6$ in this work) may explain the drastically different small planet yields between these two comparisons.

Another particularly interesting discrepancy between the two catalogs is the lack of F dwarf planet hosts in our yield estimates compared to \cite{montet2017}. This can be mostly explained by the relative number of F stars in our simulated stellar population, which is 2$\times$ smaller ($\sim$1.6$\times$10$^6$ in this work compared to $\sim$3.3$\times$10$^6$) and consists of a large fraction of approximately solar mass, metal-poor subdwarfs.

Recently, \cite{tamburo2023} estimated the GBTDS yield specifically for small planets ($R_p \leq 4 R_\oplus$) orbiting cool and ultra cool dwarfs ranging from spectral types M3-T9.
These authors created a simulated stellar catalog by extrapolating from volume-limited surveys of cool dwarfs in the local stellar neighborhood, and estimated
that Roman should observe $75,500^{+11,800}_{-7,000}$ sources within this spectral range. Then applying this new catalog with the methodology from \cite{montet2017}, the authors predicted that the GBTDS should yield $1347^{+208}_{-124}$ such transiting planets. In our results, 
considering spectral types from M3V to M9V, the latest spectral type considered in the BGM1612 model, 
we predict that the GBTDS should monitor $\sim$74,000 M3-M9 sources with $F146<21$ which will yield $610^{+58}_{-66}$ small planets ($595^{+63}_{-64}$ if only considering $n_{\rm tran}\geq7$). 
\cite{tamburo2023} also predict that Roman should find $37^{+8}_{-7}$ potentially rocky ($R_p < 1.48 \, R_\oplus$) planets in the conservative habitable zone, which is 3$\times$ our estimate of $12^{+4}_{-3}$ planets in the conservative habitable zone with $R_p<1.5 R_\oplus$. 
Although the BGM1612 model doesn't consider sources with spectral types later than M9, the planet yield for these spectral types is $\ll$1\% of the total and does not impact this comparison.

This discrepancy in cool dwarf planet yields primarily derives from the different treatment of noise and simulated detection pipelines by our work and that of \cite{tamburo2023}. Our estimates are more conservative because we include crowding in our noise model and require a significance of $8.0\sigma$ for a detection, while these authors assume isolated sources and adopt a significance threshold of $7.1\sigma$. 
Thus, the small planet yield for cool dwarfs from these authors may be optimistic, but it is not necessarily discrepant with the predictions presented in this work.

%
\section{Discussion} \label{sec:discussion}
%

The overwhelming transit yield from the GBTDS presents a number of opportunities to further exoplanet demographics, some of which we have already discussed in previous sections. In this section, we explore two of the most unique cases: the breadth of Galactic populations probed by the expected planet hosts (\S\ref{sec:galactic}), and the science to be extracted from jointly analyzing the transiting and microlensing GBTDS planet catalogs (\S\ref{sec:discussion_microlensing}). 
Finally, we end with discussions on controlling for false positive detections (\S\ref{sec:falsepositives}), the prospects for follow-up observations (\S\ref{sec:followup}), and opportunities to improve future simulations (\S\ref{sec:future}).

\subsection{Exoplanet Demographics across the Milky Way}
\label{sec:galactic}

As mentioned in the introduction, there are several reasons why the exoplanet population should vary across the Galaxy. Perhaps the most compelling is the correlation planet occurrence and stellar metallicity  \citep{santos2004,fischer2005,sousa2008,johnson2010,ghezzi2010,wang2015,mulders2016,wilson2018,petigura2018,ghezzi2021,wilson2022}. 
For giant planets, this signature is typically interpreted as support for the core accretion model of planet formation, in which enhanced stellar metallicity is assumed to correlate with higher solid surface densities in the protoplanetary disk which facilitates the rapid growth of planetary cores up to a threshold mass of $\sim$10$\,M_\oplus$, allowing the accretion of a gaseous envelope \citep[][]{pollack1997,ida2004,mordasini2012,chabrier2014}. 
There is a theoretical basis that core accretion can be more effectively facilitated by the enhancement of specific $\alpha$ elements (elements with an even atomic number primarily generated in core-collapse supernovae; e.g., C, O, Mg, Si), rather than bulk metallicity. For example, silicates may be needed to seed planetary cores \citep{natta2007}, and CNO ices constitute a significant fraction of the mass in planetesimals beyond the water ice line \citep[][]{pontoppidan2014}, where core accretion is most efficient \citep[see, e.g.,][and references therein]{dawson2018}.

Taking this idea to its extreme conclusion would imply that the correlation between giant planet occurrence and bulk stellar metallicity is actually a correlation with $\alpha$ abundance, and bulk metallicity is simply a confounding variable that happens to correlate with $\alpha$ abundances in the local stellar neighborhood, making it difficult to disentangle this multivariate trend. 
However, it should be possible to isolate correlations between planet occurrence and these two parameters by searching for planets in multiple Galactic populations, with differing star formation histories and therefore differing metallicity, $\alpha$, and age distributions. 
While ground-based transit surveys are currently being performed to test this hypothesis \citep{penny2020}, the GBTDS will accomplish this on a larger scale by effectively searching for exoplanets across all major Galactic populations.

The particular methodology for isolating these correlations is not obvious given that reliable metallicities, and especially $\alpha$-abundances, will be difficult to obtain for the sources in the GBTDS. 
Because of this, the transiting planet yield from the GBTDS may best be interpreted in the context of Galactic populations, perhaps as a hierarchical model built upon a Galactic population synthesis model. 
Forward modeling planet catalogs have proven successful in analyzing the Kepler planet catalog \citep[e.g.,][]{mulders2018,mulders2019,he2019}, and we propose that
adopting a similar approach, with key parameters in the planet demographics that vary as a function of stellar metallicity, alpha abundance, and stellar age, layered over a reliable Galactic model offers the best opportunity to disentangle correlations with these parameters.

\begin{figure}
    \centering
    \includegraphics[width=\columnwidth]{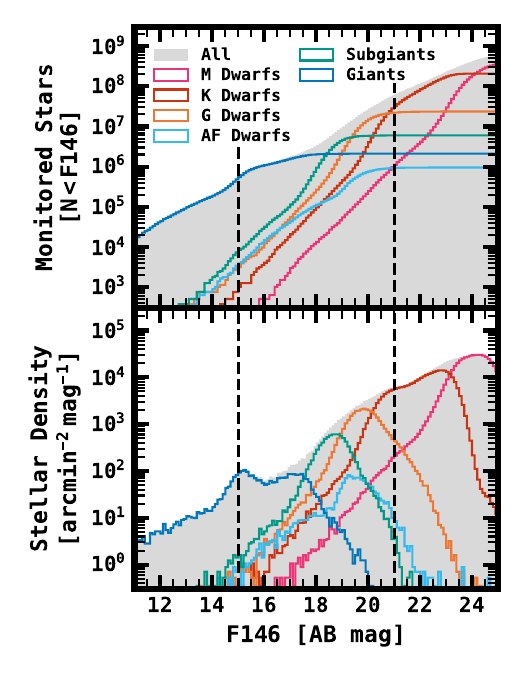}
    \caption{The $F146$ distribution of stars in the GBTDS, by stellar type. In both figures the vertical dashed lines show the magnitude range considered in this study. {\it Top:} The number of monitored stars below a given magnitude. {\it Bottom:} The stellar density, in units of stars per unit solid angle per magnitude.  }
    \label{fig:magdistribution}
\end{figure}

This approach has several advantages. One advantage is in stellar characterization. Due to the deep near-infrared imaging that the GBTDS will yield, Malmquist bias is severely reduced allowing one to place reasonable priors on the stellar properties of a source with only its apparent magnitude (see Figure \ref{fig:magdistribution}). 
For example, there are virtually no giants with $F146 > 20$, i.e., the GBTDS will be nearly complete for giant stars in the Milky Way along this line of sight, so any such sources are almost guaranteed to be dwarfs, with the exception of very small numbers of halo stars beyond the far side of the disk, which in this instance are negligible. 
One particularly difficult source of degeneracies will be distinguishing between subgiants and G dwarfs. This was apparent in the Kepler survey, originally underestimating the fraction of subgiants in their source catalog due to Malmquist bias \citep{gaidos2013, berger2018}. 
However, as shown in Figure \ref{fig:magdistribution}, subgiants and G dwarfs occupy different, albeit overlapping, ranges of apparent magnitude, and therefore such degeneracies can be minimized with a prior from a reliable Galactic population synthesis model.

This same principle can be used to estimate likely distributions in [Fe/H], [$\alpha$/Fe] and age from the sample of planet-search stars, and identify key metrics to distinguish exoplanet demographics models. 
The BGM1612 Galactic model is mostly described by three Galactic populations with differing chemical signatures: the thin disk ($\alpha$-poor, Fe-rich), the thick disk ($\alpha$-rich, Fe-poor), and the bar ($\alpha$-rich, both Fe-rich and Fe-poor). 
The BGM1612 model doesn't adopt a classical bulge model, instead modeling the bulge regions as a combination of the thick disk and a bar component. For the purposes of this discussion, when referring to the bulge, we are referring to the bar population, and excluding the thick disk component at very small Galactocentric radii.

\begin{figure*}
    \centering
    \includegraphics{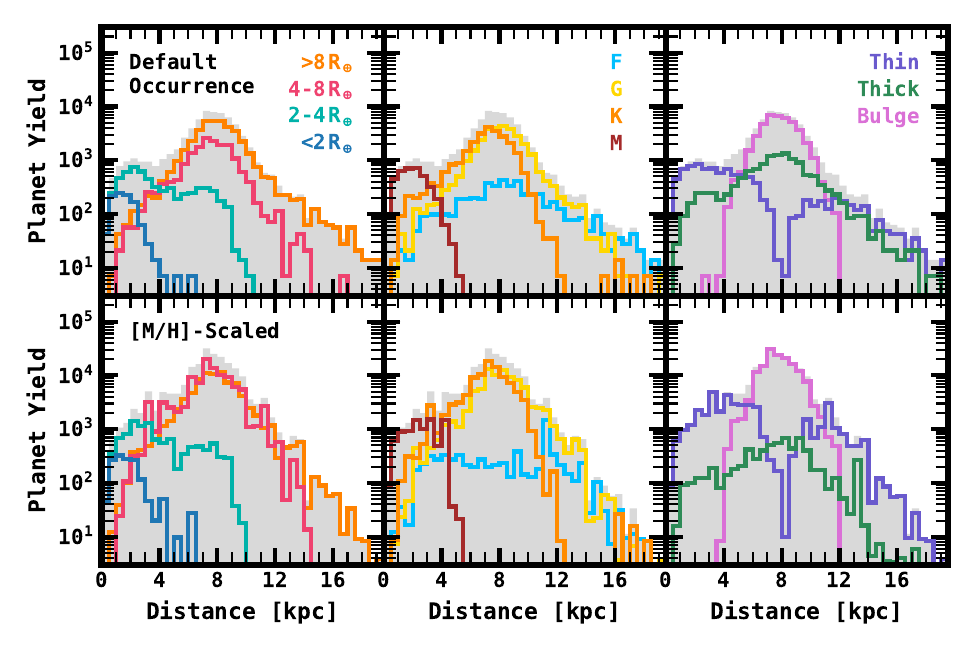}
    \caption{The transiting planet yield as a function of distance. The top row is the transit yield assuming the default planet occurrence, and the bottom row is the yield after taking metallicity correlations into account. {\it Left Panels:} The planet yield labeled by planet size. {\it Middle Panels:} The planet yield labeled by spectral type. {\it Right Panels:} The planet yield labeled by Galactic stellar population. }
    \label{fig:distyield}
\end{figure*}

Thick disk stars begin to dominate the stellar population at distances $\gtrsim$500 pc above and below the Galactic midplane, and become more prominent closer to the Galactic center. 
Within the context of the GBTDS, the thin disk generally dominates the stellar population at $d\lesssim 5$~kpc, the bulge dominates at distances of $\sim$5-11 kpc, 
and at $d\gtrsim11$~kpc, the thin and thick disks contribute similarly to the stellar populations. 
Due to these variations, the planet yield as a function of distance should be a powerful tool for distinguishing between exoplanet demographics models.  
This idea is highlighted in Figure \ref{fig:distyield}. The top row of these figures shows the planet yield from this work as a function of distance under the assumption that the planet occurrence is constant with respect to metallicity (i.e., default simulation parameters), and therefore not impacted by Galactic population, while the bottom row shows the yield after accounting for metallicity effects as described in section \ref{sec:simplanets_metals}.

While our simplistic Galactic population model (i.e., single pointing, constant source density, constant extinction) prevents us from making quantitative predictions, there are a few trends worth discussing. 
First, the yield of sub-Saturns and Jupiters have a strong dependence on the average stellar metallicity. This is most apparent in the slope of the planet yield with distance out to $\sim$6 kpc. Because the average metallicity increases more than the average alpha abundances, this slope should be more prominent in the case that planet occurrence more strongly depends on metallicity than alpha abundance. However, because the bulge population is alpha-rich, the yield should be higher from $d\sim$~6-10 kpc in the case that $\alpha$ abundances are more strongly correlated with planet occurrence.
Thus, comparing the distribution of planet detections at these distance ranges may highlight the relative role of $\alpha$ abundances and bulk metallicity.

Another strong result is the relative occurrence of planets in the thin and thick disks as a function of Galactic population, as shown in the rightmost panels of Figure \ref{fig:distyield}. At $d \gtrsim$~12 kpc, the planet yield is dominated by the thin disk when planet occurrence is considered as a function of metallicity, but nearly equal when ignoring such effects. 
Because stars at these distances are outside the Galactic midplane ($|Z_{\rm Gal}|\approx$~340-450 pc), the thick disk becomes a significant fraction of the stellar population, and thus measuring the relative planet occurrence for stars with $d \approx $~0-4~kpc and $d \approx $~12-16~kpc, which are at the same Galactic radii, should reflect the relative planet occurrence in the thin and thick disk populations. 
These inferences may be further evaluated by comparing proper motion dispersion between the field stars and planet hosts as a function of distance.  
In this vein, performing a transit search over stars in fields at varying Galactic latitude should lead to fields with differing combinations of thin and thick disk stars, and predictable, model-dependent changes in the planet yield as a function of distance.

One last parameter of note is the age distribution for each sub-population in our adopted Galactic population model. 
The mean age of thin disk stars increases toward the Galactic center, while the thick disk and bulge have a more narrow, well-mixed distribution of ages, because they are modeled as a single burst of star formation $\sim$10-12 Gyr ago. This has a few interesting implications. 
Because the main sequence turn off in the bulge occurs at $M_\star \approx 1.1 M_\odot$, the dominant source of F-type stars at small Galactic radii ($d\sim$~6-10 kpc) will be low metallicity subdwarfs ([M/H]~$\lesssim -0.5$). Thus, the planet yield for F-type stars should vary extremely sensitively on the inferred planet occurrence as a function of metallicity. This is reflected in the middle-panel of Figure \ref{fig:distyield}, in which the planet yield for F stars is roughly constant at distances of $d \approx$~5-11 kpc, despite the increasing stellar densities in the Galactic bulge.

\subsection{The Transit Survey as a Complement to the Microlensing Survey}\label{sec:discussion_microlensing}

\begin{figure*}
    \centering
    \includegraphics[width=\textwidth]{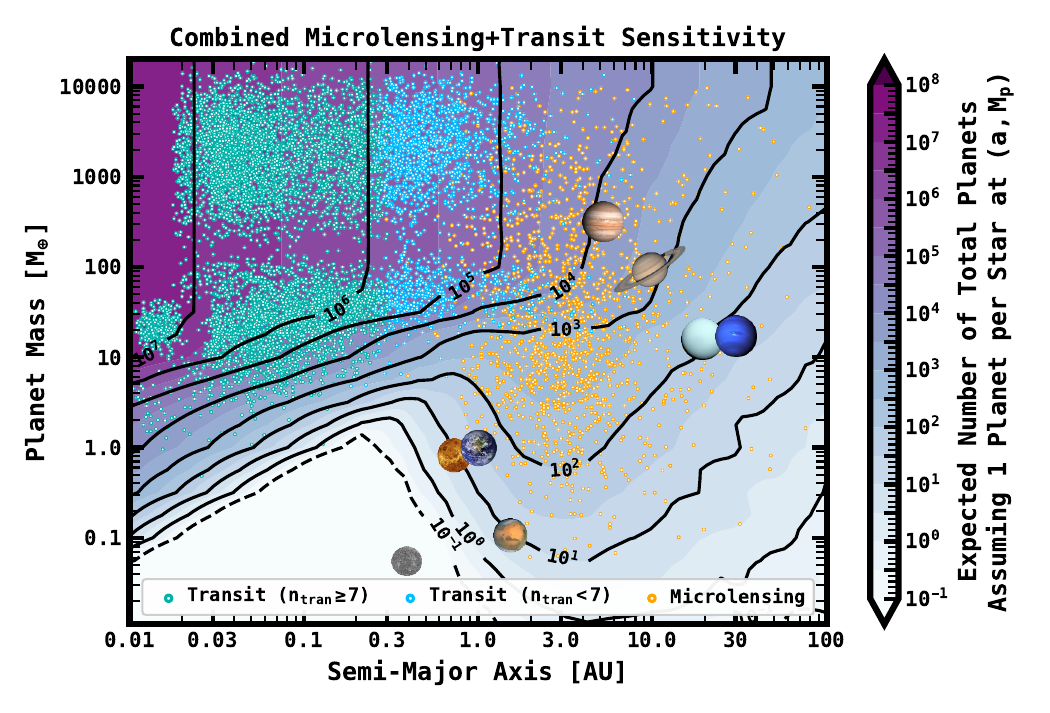}
    \caption{The combined planet detection sensitivity via transit and microlensing in the Roman Galactic Bulge Time Domain Survey. The contours show the sum of the transit survey efficiency multiplied by the number of stars searched and the microlensing survey efficiency from \cite{penny2019}. 
    The green/blue points show our sample of detected planets for one field of the GBTDS. The green points denote conservative requirements of $>$1 transit per season ($\geq$7 total) and the blue points denote liberal detection requirements of $\leq$1 transit per season ($<$7 total), including single-transiting systems. Both transiting planet yields assume our default simulation parameters and adopt the mass-radius relationship from \cite{chen2017} to convert from planet radius to planet mass. The orange points show the simulated catalog of detected microlensing planets from \cite{penny2019}. 
    The Solar system planets are also included at their respective mass and semi-major axis for reference. The planets with fewer than 7 transits should overlap significantly with the microlensing sample in this parameter space. }
    \label{fig:combined_yield}
\end{figure*}

For many reasons, the transiting planet survey discussed in this work will complement and enhance the microlensing planet survey.
Figure \ref{fig:combined_yield} shows the combined transit and microlensing sensitivity to detecting planets as a function of semi-major axis and planet mass. The transit sensitivity was calculated by averaging over a representative sample of the stars considered in this work using the methodology in section \ref{sec:sensitivity}, and planet masses were inferred from planet radii using the mass-radius relation from \cite{chen2017}.

Particularly for giant planets, the transiting and microlensing planet yields should overlap significantly in parameter space from $a \sim $~0.3-3 au, and $M_p \gtrsim 30 M_\oplus$, which offers a few interesting opportunities. 
Though, as mentioned in section \ref{sec:yield_single}, our yield estimates for planets within this parameter space are optimistic, and are likely to be reduced anywhere from $\sim$20-70\% due to difficulties in detecting planets with few transits. Even at these reduced detection rates, the sample of planets with only one or two transits should still be large enough for demographics studies. 
Because transit surveys will infer precisely the planet radius, while microlensing surveys will infer the planet mass, the joint distribution of long-period planets from the microlensing and transiting samples may be used to infer mass-radius distributions. 
Because, in the case of single transits, the semi-major axis can be inferred by the transit duration and inference on stellar density, while the projected separation is measured for microlensing planets, a combined analysis of the population of single-transiting planets with the microlensing population can constrain the eccentricity distribution of cool, Jupiter analogs, potentially revealing information on their migration histories.

Another interesting consideration is the relative frequency of detected multi-planet systems between the microlensing and transiting planet searches. Because the detectability of multi-planet systems via transit is subject to the dispersion in mutual inclination, unlike in microlensing, the joint analysis of both samples should constrain the multi-planet frequency and distribution of mutual inclinations within such systems, at least for planets within the joint survey sensitivity. 
With these population-level distributions on mutual inclination and eccentricity, the importance of planet-planet Kozai scattering can be statistically probed as a formation and migration pathway in such systems \citep[][and references therein]{dawson2018}.

Comparisons of the long period transiting planet sample with long-term studies of RV-detected planets \citep[e.g.,][]{rosenthal2021} could yield valuable results as well.
For instance, one interesting result from RV-planet search surveys is a sudden increase and drop in the occurrence of giant planets at $\sim$1-3 au \citep[][]{fernandes2019,fulton2021}. It is currently debated whether this sudden increase is due to a more efficient growth of planetary cores \citep{schoonenberg2017}, or the increased efficiency in which cores grow their gaseous envelopes.
The transiting planet sample could include as many as $\sim$2,000 planets within this semi-major axis range allowing for a systematic comparison in planet occurrence between both samples, and when analyzed jointly with the microlensing sample, population-level demographics on the mean planet density and eccentricity distributions. 
Such constraints may prove insightful in inferring bulk compositions, and by extension, formation pathways. 
Note, it is extremely unlikely for a single planet (and possibly multiple planets within the same system) to be detected via both transit and microlensing, so the comparisons discussed above will be limited to a statistical population analysis, the effectiveness of which will depend sensitively on the accuracy and precision with which the completeness and reliability of each respective exoplanet catalog can be assessed.

In addition to providing a complementary set of population parameters for microlensing giant planets on $\sim$0.3-3 au orbits, the transiting planet sample may act as a key variable in interpreting the small planet sample of microlensing planets against transit surveys in the local neighborhood, such as Kepler. 
Because the Kepler sample was sensitive to small planets just interior to $\sim$1 au, and the microlensing survey will be sensitive to small planets just exterior to $\sim$1 au, they provide a natural complement to each other. 
However, there is a significant confounding variable in the differing Galactic environments and stellar masses probed by these two surveys. Thus, the $\sim$1000s of small planets detected in the GBTDS via transit should provide an essential keystone in placing the homogeneous Kepler sample into a greater Galactic context, anchoring the interpolation needed to combine the Kepler transit and Roman microlensing planets into a single, joint analysis framework for testing planet formation theories.

\subsection{Prospects for False Positive Rejection and Planet Validation}
\label{sec:falsepositives}

For any transit survey, a primary concern is removing astrophysical false positives caused by, e.g., eclipsing binaries (EBs). In this work, we did not consider false positives contaminating the transiting planet catalog,  but this should be a concern for the GBTDS transit survey, particularly because giant planets are the most difficult transiting planet candidates to validate without additional data (e.g., radial velocities) because the light curve shape of transiting Jupiter-sized planets is often consistent with the light curve shape of eclipsing brown dwarfs and low-mass M dwarfs.
For this reason, a planet candidate vetting framework for the GBTDS transit search needs to be more robust than previous attempts, and producing a transiting planet catalog with high reliability from the GBTDS will require a monumental effort. This need becomes even more imperative at the dimmer magnitudes needed to probe the transiting planet population beyond the Galactic bulge, as proposed in section \ref{sec:galactic}.

The majority of false positives will be due to EBs in some form. The most common tests to identify EBs include comparing the transit depth in odd and even transits to identify events that are actually EBs at half of the reported period, measuring the significance of secondary events at the folded period to identify secondary eclipses, and measuring the PSF centroid, either directly or via difference imaging, in and out of transit to identify signals that are off-target. 
Some variation of each of the above tests are a staple in any transiting planet search \citep[e.g.,][]{coughlin2016,thompson2018,twicken2018,kostov2019,zink2020,guerrero2021}, and should be again for the GBTDS. However, due to the differences in observational capabilities between Roman and the Kepler and TESS observatories, there are unique variations of these tests that will provide additional benefit in discriminating false positive scenarios from transiting planets.

Many of these scenarios were simulated by \cite{montet2017} to provide a proof of concept, though some tests may only be applicable for bright sources with high transit S/N. 
However, providing a full assessment of the ability for each test to identify EBs or validate planets in the GBTDS data would require more detailed simulations of the false positive population, which is outside the scope of this paper. 
Thus, rather than assessing the effectiveness of each test, we limit this section to a brief discussion of some unique considerations needed to improve the reliability of the GBTDS transiting planet candidate catalog and leave detailed simulations for a later work.

\subsubsection{Direct Validation}

For a few specific scenarios, it should be possible to directly validate a transiting planet candidate. \cite{montet2017} give a few examples, such as the detection of a secondary eclipse (discussed in more detail below), the detection of transit timing variations, or systems with multiple transiting planets. 
Because multi-planet systems are difficult to mimic by multiple EBs, the vast majority of such systems can be reliably confirmed as planetary in nature, provided the transit source is consistent with the apparent source \citep{morton2016}. For additional reliability, some of these systems should have detectable transit timing variations. The GBTDS should be sensitive to timing variations for planets as small as $R_p\approx 3 R_\oplus$ and $M_p \approx 10 M_\oplus$ for the brightest Sun-like stars in the GBTDS \citep{montet2017}. Although timing variations measured in these cases will likely not be sensitive enough to infer masses, the combination of timing amplitude and period should be adequate to confirm the planetary nature of an object.  
Unfortunately, this is likely only possible for a small subsample of planets in the GBTDS, as Jupiter-sized planets are known to have less detectable transit timing variations.

\subsubsection{Secondary Eclipses and Emergent Planetary Flux}\label{sec:secondary_eclipse}

The detection of a secondary eclipse typically is a near definitive identification of a false positive in optical transit surveys, but because Roman will primarily observe from 0.93-2.00 $\mu$m it will be sensitive to planetary emission for $\sim$1000s of Hot Jupiters in the GBTDS sample \citep{montet2017}.

In the case that a secondary eclipse is detected, jointly modeling the depth of the primary and secondary eclipse should place strict constraints on the combination of temperature and radius of the occulting body, which should identify self-luminous objects with surface temperatures that are inconsistent with substellar mass objects.
In other words, because secondary eclipse depth is so strongly correlated with surface temperature, it should be extremely difficult for a stellar-mass object to imitate the secondary eclipse of a planetary mass object. This is not necessarily true for some ultra-hot Jupiters which can have surface temperatures approaching and exceeding that of M dwarfs (e.g., Kelt-9b: \citealt{gaudi2017}; TOI-2109b: \citealt{wong2021}), but such systems should be identifiable by their high instellation flux.
For most cases, the only other plausible scenario would be in a hierarchical multiple system, in which the original transit signal is extremely diluted. Such systems may be identified by multi-color observations, as discussed in the following section.

Another effect that could help validate hot planetary candidates is the phase variations of the out-of-transit light curve. Eclipsing binaries show variations on or at half the orbital period due to effects such as ellipsoidal variations and doppler beaming. 
These effects can be described to first-order by the BEaming, Ellipsoidal, and Reflection/heating (BEER) model \citep{faigler2011}. At large companion masses the dominant terms in this model, i.e., ellipsoidal variations and doppler beaming, are in phase with the primary and secondary eclipses of the transiting/eclipsing companion. However, this may not be the case for hot Jupiters, in which the dominant variation in the GBTDS will likely derive from thermal emission from a hot spot on the dayside of the planet, except in the case that a hot Jupiter has a high albedo. 
Due to atmospheric circulation, the brightest point of the phase curve is often out of phase with the timing of the secondary eclipse \citep{showman2009}, an effect that cannot be imitated by ellipsoidal variations allowing for additional discrimination in identifying false positives from eclipsing stellar mass companions, provided that Roman has the long-term photometric stability to obtain such a measurement.

\subsubsection{Transit Depth Chromaticity}

The GBTDS will take observations in a second, and possibly third, bandpass for the purpose of characterizing stars in the GBTDS. Depending on the frequency and specific bandpasses chosen, these observations can provide supplementary information useful for validating planet candidates by measuring wavelength-dependent eclipse depth variations caused by the differing portions of the stellar SEDs sampled in each bandpass.

\cite{montet2017} simulated GBTDS observations of a hot Jupiter with $P=3$~days and $R_p/R_\star = 0.1$ around a 15th magnitude Sun-like star, assuming secondary observations in the $F087$~(0.76-0.99 $\mu$m) bandpass at a cadence of 12 hours.
In this scenario, variations as large as $\sim$11\% can be identified at a significance of 3$\sigma$. While this is adequate for identifying many stellar sources, it is likely not sufficient for the majority of the catalog, which will be significantly dimmer. 
In the case of discriminating between low mass stars and planets with similar radii, secondary eclipse detections will likely be more significant. 
However, increasing this cadence would improve the limits on identifying chromatic variations, and the choice of a third, redder wavelength, such as the $F184$ (1.68-2.00 $\mu$m) or $F213$ (1.95-2.30 $\mu$m) bandpass may provide additional discriminating power.

Interpreting this measurement will also depend on the degree to which the chromatic, confusion-limited background is characterized, as discussed in section \ref{sec:discussion_psf}, and the degree to which blended sources can be identified. In the latter, real planets in hierarchical systems may be misidentified as stellar mass companions in an unblended system. Thus, extra care will be needed in interpreting chromatic variations of this nature.

\subsubsection{Astrometry In/Out of Transit}

Background EBs (BEBs) may masquerade as planet candidates due to their diluted eclipse depth. Identifying BEBs typically requires measuring the centroid, either directly or via difference imaging, in and out of transit to identify transit signals that are inconsistent with the position of the apparent source. 
Applying these methods for identifying BEBs in Kepler and TESS are typically sensitive to sub-pixel offsets.
However, driven by science requirements from the other core community surveys, Roman will have much better spatial resolution and PSF stability than any previous transit surveys, with a PSF characterized to many orders of magnitude. 
As a result, the GBTDS should yield single-exposure astrometry to a systematics limit of $\sim$1/100 pixel \citep{gaudi2019}. For high-S/N transits, performing astrometry on the difference images should identify the source of transit with a precision up to this systematics limit.

Even in the worst case scenario, i.e., in which a background EB and apparent source of transit are in perfect alignment, relative proper motions between the background EB and apparent source should allow for many systems to be separable by the end of the survey.
If it can be assumed that the distribution of relative proper motions between the background EB and apparent transit source are the same as those for source-lens pairs in the simulated microlensing sample from \cite{penny2019}, then the majority of pairs should have relative proper motions ranging from 5-15 mas/yr, sufficient to separate the two sources by $\sim$20-60\% of a pixel over the course of the 4.5-year baseline, which should be detectable for high S/N transit candidates.

Thus, despite the crowded bulge fields, the imaging quality of the WFI combined with the relative proper motions of stars in the GBTDS fields should enable effective identification of false positives from unassociated background EBs, at least for high $S/N$ transiting planet candidates.

\subsubsection{Probabilistic Validation}

Finally, the most likely assessment of reliability for the majority of the candidates in the sample will come from some form of probabilistic validation, where the light curve is modeled assuming a number of astrophysical false positive scenarios, the relative probability of each scenario is computed, and if the probability that the data derive from a false positive is below some threshold, the candidate is considered validated \citep[e.g.,][]{morton2016,giacalone2021}.

Typically, this strategy is most effective at high S/N, and for small planets. In the case where there is some ambiguity, the interpretation of the results depend strongly on the priors assumed for properties such as the likelihood of a transiting planet, the density of background sources, and the occurrence of EBs. 
Thus, using systems validated in this way for demographics studies can lead to a degree of circular logic that can bias any inferred trends. 
In the case that the false positive rate is relatively high for the GBTDS, any inferred demographics trends will either be limited to bright stars that can be adequately confirmed, or subject to the priors assumed for the likelihood of false positive scenarios.

It is likely that a large fraction of the planet candidates in the GBTDS will be ambiguous. To compensate for this, a probabilistic validation framework for the GBTDS planet search should incorporate simulated false positives at the catalog level, as advocated in section \ref{sec:galactic}, applying constraints for the close binary fraction combined with estimates for the vetting sensitivity, the false positive yield can be jointly modeled with the planet yield. This strategy has proven effective in inferring occurrence rates in Kepler \citep[e.g.,][]{fressin2013} and conceptually similar strategies applied to modeling false alarms have proven critical to inferring the occurrence of Earth-like planets \citep{bryson2020}. 
It is in this hierarchical approach that the GBTDS sample can be most efficiently exploited for demographics studies. Thus, even if any one planet candidate is not reliable, a population level analysis should still prove powerful.

\subsection{Prospects for Spectroscopic Follow-up}\label{sec:followup}
 

Spectroscopic follow up of the GBTDS transiting planet sample could prove invaluable to understanding the [Fe/H] and [$\alpha$/Fe] distributions of the Galactic bulge sample. 
To survey all of the G dwarf planet hosts in the bulge and nearside of the Galactic disk, a program would need the capability to observe $\sim$24,000-49,000 sources, depending on correlations between  metallicity and planet occurrence,  
with magnitude limits of $I \lesssim 20$~mag ($R \lesssim 21$~mag). Near-infrared wavlengths would help compensate for extinction and increase S/N, but will be more susceptible to confusion limits. However, in this case, one would only need to be sensitive to $J\lesssim 19$ ($H\lesssim 18.5$).
One strategy could be to follow up sources at moderate spectral resolution ($R \sim 5,000$) and signal-to-noise ($S/N \sim 30$), where one can derive stellar atmospheric parameters -- including [Fe/H] -- at a reasonable significance, and measure radial velocities to a precision of $\sim$1 km/s, sufficient to rule out stellar-mass companions.

\begin{figure}
    \centering
    \includegraphics[width=\columnwidth]{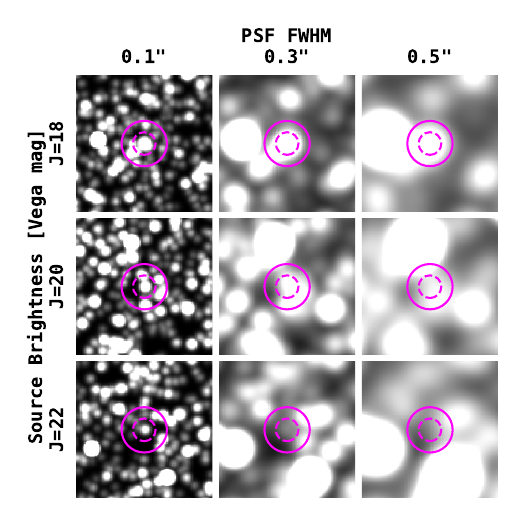}
    \caption{The seeing-limited flux distribution of sources in the GBTDS. Each row is centered on a source of different magnitude, and each column represents a different seeing limit. The pink circles represent fiber entrance apertures of diameters 0.5\arcsec (dashed) and 1.0\arcsec\ (solid). In good seeing conditions ($\sim$0.5\arcsec), ground-based observations are limited to sources with $J\lesssim 18$. Sources with $J <20$ can be followed up with a 0.5\arcsec\ diameter fiber if the PSF can be controlled to ${\rm FWHM}\approx 0.3\arcsec$.}
    \label{fig:fibersize}
\end{figure}

While such a survey would be resource-intensive, the next generation of 10+ meter multiplexed spectroscopic observatories \citep[e.g.,][]{mcconnachie2016,pasquini2018} will not be constrained by this faint magnitude limit and sample size. If the sources were isolated, then these observatories would be able to conduct this survey in $\sim$10-100 hours of observation. However, the biggest limitation will be source confusion in the low-latitude Galactic bulge fields. 
As a result, fiber-fed multi-object spectroscopic surveys to follow up the GBTDS transiting planet hosts will likely need a way to compensate for image degradation from atmospheric turbulence (i.e., ``seeing") to reliably observe sources dimmer than $J \approx 18$ (see Figure \ref{fig:fibersize}).

Observing dimmer sources should be possible with a fiber entrance aperture of 0.5\arcsec, if the PSF FWHM can be controlled to $\sim$0.3\arcsec. 
This could be accomplished with MOSAIC \citep{kelz2015}, a multi-object spectrograph planned for the European Extremely Large Telescope \citep[E-ELT;][]{gilmozzi2008}, which can sufficiently achieve the desired spatial resolution with a Ground-Layer Adaptive Optics system over a 40 arcmin$^2$ field of view. 
Given the average density of transiting planet candidates considered ($\sim$150-300 per 40 arcmin$^2$) and the number of fibers available (200), each object can be observed once in $\sim$200 pointings. This may be accomplished at high spectral resolution ($R\sim18,000$) due to the large telescope aperture. 
Alternatively, a carefully selected subsample of planet candidates with multiple visits to measure their RV variability may be sufficient to characterize the false positive rate in the GBTDS planet catalog, improving the demographics analysis.

\subsection{Caveats and Implementations to Improve Future Simulations} \label{sec:future}

Here we discuss a few areas where these simulations can be further improved and additional effects that can be added to make these simulations more realistic.

\subsubsection{Crowding and PSF/PRF Photometry}\label{sec:discussion_psf}

The noise model developed in section \ref{sec:noise} is specific to the aperture photometry pipeline that we implemented, but with PSF photometry the effects of contaminating flux are likely to be reduced. For this reason, we omitted stars dimmer than $F146=21$~mag, approximately when crowding becomes the dominant source of photometric noise. 
However, such stars should still be bright enough to effectively detect planets.
Thus, implementing a PSF photometry routine into these simulations is likely to improve the inferred transit survey efficiency for dim stars, which is particularly important for detecting distant ($d\gtrsim 10$~kpc) planets on the far side of the Galactic bulge, as discussed in section \ref{sec:galactic}.

Because we adopted an aperture photometry pipeline and made no attempt at characterizing the planet candidates themselves, we neglected difficulties such as accurately inferring a transit depth. 
The approach to modeling the transit depth will be an important concern for the actual survey, and improper treatment of crowding may lead to underestimating the transit depth by a factor of $\sim$2-4 for the dimmest stars considered here (see Figure \ref{fig:crowding}). 
These uncertainties may be mitigated by observations in bluer filters (e.g., $F087$) which will have a smaller diffraction limit and lower sky background, and should therefore aid in deblending the $F146$ images. 
However, such methods will likely have to account for some kind of diffuse, chromatic stellar background from dim and/or unresolved stars below the confusion limit that are unlikely to be fully modelled out.
This diffuse background could be estimated in conjunction with a Galactic synthesis model, reducing some of these uncertainties. Further constraints on the density of dim sources below the confusion limit may also be inferred from the empirical microlensing event rate.

By the end of the survey there will be $\gtrsim$40,000 cadences from which to measure relative positions, parallaxes and proper motions to build a reasonable background star model. 
Therefore, the photometry should be significantly improved as the survey continues, and background stars will become more precisely modeled as uncertainties in proper motions and positions are reduced in later seasons of the survey, allowing for tighter priors on the source position and reducing degeneracies between position and flux in the PSF modeling.

\subsubsection{Galactic Population Synthesis Model}

In this work we adopted a simple Galactic model, with many of the limitations described in section \ref{sec:simstars}. We assumed in this work that each field has the same stellar populations, neglecting the spatial dependence on source density, as well as reddening and extinction that are variable on sub-arcminute scales \citep[e.g.,][]{gonzalez2012,surot2020}. 
As a result there should be some variation in the planet yield, and photometric uncertainty caused by crowding for each field. The metallicity and $\alpha$-abundance distribution functions in the Galactic bulge also change on $\sim$degree spatial scales \citep{zoccali2017}, so different fields are likely to have different underlying planet occurrences.  
By incorporating such a model, we could test more detailed theories of the dependence of Galactic environment on the transiting planet yield since different fields would presumably have differing fractions of thin disk vs. thick disk vs. bulge stars.

Another improvement to a Galactic population synthesis model would be a detailed treatment of binarity and multistar systems. In this work, we injected binary systems into our Galactic model, but made no attempt to determine the yield of planets in such systems, though planet candidates in hierarchical systems should be fairly common. 
The statistics we use to inject binary systems in our sample are also limited to studies of the local stellar neighborhood, which is an imperfect treatment of multistar systems in the GBTDS, and the prevalence of such systems likely leads to uncertainties in our yield estimates of a few percent, particularly when adopting the metallicity-scaled planet occurrence (see section \ref{sec:yield_binary}).

\subsubsection{Intrinsic Stellar Variability}

In this work we neglected stellar variability in our yield estimates and noise model.  
Although this is unlikely to have a significant impact on our overall results, as discussed in section \ref{sec:astrophysical_noise},
this is likely an important consideration for bright sources, such as low-luminosity giants and nearby G dwarfs.  
For evolved stars, Solar-like oscillations may introduce higher frequency noise on timescales comparable to the transit duration, but the amplitude of these signals is likely below the white noise threshold for the majority of giants in the sample with radii small enough to detect transiting planets \citep{gould2015}. 
On the other hand, because charge doesn't bleed in the WFI H4RG-10 arrays, like in a CCD \citep{mosby2020}, extremely precise photometry can be derived from saturated sources. In this case there may be an opportunity to detect transiting planets among some of the brightest giant stars in the sample where asteroseismic oscillations may introduce noise with amplitudes large enough to impact the detectability of transits even without a confident detection of the Solar-like oscillations themselves. 
Modeling evolved stars in this way requires a more detailed treatment than is presented in this work, as we ignore saturated sources and most second-order detector effects that would become important for accurately modeling the brightest sources in the GBTDS, as discussed in section \ref{sec:apertureflux}.

\subsubsection{Detector Systematics}\label{sec:detector}

There are additional detector effects and observing details that are not included in our analysis. 
Because high-level data products from the Roman mission will be calibrated based on continuous measurements to remove the worst of the effects below, we believe these systematics will have only a minor effect on our results (see section \ref{sec:correlated_noise}). 
However, these effects may be important to include in specific situations, or for other scientific motivations, such as modeling pixel-level data for sources brighter than $F146 \approx 18$. 
Below we discuss effects likely to affect this photometric performance, particularly for bright stars.

\textit{Persistence:} Near-IR arrays are susceptible to the incompletely understood effect of persistence. Persistence acts as a memory effect, where trapped charge from previous exposures is slowly released into subsequent exposures. 
Despite the attention to reducing these effects in the WFI arrays, persistence can remain a significant problem for longer exposures and for fainter sources \citep{mosby2020}. 
Given the shorter exposures of the GBTDS, persistence should only be significant for a small fraction of stars whose position is correlated with bright stars in a previous field (i.e., a similar number to the $\sim$3\% of stars near saturated pixels should be affected by persistence). Even in this case, the median magnitude of the persistence signal has been measured to be $\sim$0.02\% of the full well depth 150 seconds after a bright exposure, which is unlikely to impact these detected planet yields, as shown in section \ref{sec:correlated_noise}. For timescales of $\sim$50-70 seconds, the typical slew/settle time between exposures of adjacent fields in the GBTDS, this signal is $\sim$1-3~$\, \mathrm{e}^- /\mathrm{s}$ \citep{mosby2020}, which for the dimmest stars in this study translates to $\sim$0.2-0.6\% of the source flux, which is similar in magnitude to shot noise from the thermal and sky background, but below that from crowding. 
Thus, if uncorrected, persistence may affect $\sim$3\% of the dimmest stars in this sample, but is unlikely to significantly impact our overall results.
However, due to the goals of the other core community surveys, particularly for flux measurements of supernovae as dim as $\sim$26 mag in the High-Latitude survey, modeling persistence at the data calibration level is a focal point for the Roman science team \citep{spergel2015}.

{\it Non-Linearities:} Non-linearities arise from a number of sources, including count-rate dependent non-linearities (i.e., reciprocity failure) and classical non-linearities. While the latter is typically understood to be deterministic, the former is likely correlated with less stable effects such as persistence and burn-in. 
\cite{mosby2020} measured the linearity performance of candidate WFI H4RG-10 detectors, and found that a typical pixel has a maximum deviation from an arbitrary linear correction of 0.15\%, implying that at worst the error from non-linearities should be on par with the semi-saturation limit considered in this work. 
However, these estimates are likely to be improved due to the calibration requirements of the other core community surveys, would primarily impact the brightest stars in this study, and even in the worst case scenario should only have a major impact on the detectability of small planets, as discussed in section \ref{sec:correlated_noise}.

{\it Other detector systematic effects}: The work presented here considered simulations on which the per-pixel noise comes from photon-counting plus an uncorrelated read-noise component derived from up-the-ramp sampling (see equation \ref{eq:noise}). In reality, spatially correlated noise (e.g., $1/f$ noise, alternating column noise, intra- and inter-pixel sensitivity variations) in H4RG detectors will also impact the achievable photometric precision by the Roman mission \citep[see, e.g., ][and references therein]{rauscher2015}.
While studying the effects of such components is outside the scope of this work, techniques to mitigate those at the data calibration-level will be valuable to study in detail in the near future. 
As advocated in section \ref{sec:correlated_noise}, limiting these effects to below $<$0.5 ppt on timescales of $\lesssim$12 hours should be adequate to not impact the GBTDS transiting planet yield, although the relative magnitude of this uncertainty will depend on the noise floor for saturated and semi-saturated sources which we conservatively predict to be $\sim$1.3 ppt.

%
\section{Summary and Implications} \label{sec:summary}
%

NASA's next flagship mission, the Nancy Grace Roman Space Telescope, will conduct a high-cadence survey of the Galactic bulge with the primary objective of discovering $\sim$1000s of exoplanets on wide orbits ($\sim$1-10 au) via microlensing. However, given the GBTDS observing mode, these data will also facilitate the detection of transiting exoplanets.

To fully assess this potential and the scientific opportunities afforded by transiting planets in the GBTDS, we developed the Roman IMage and TIMe-series SIMulator (\rimtimsim) to perform detailed pixel-level simulations of the GBTDS.
Using synthetic full-frame images from this simulation framework, we generated $>$430,000 light curves with transit signals injected at the pixel level, applied a simulated transit detection pipeline, characterized the expected photometric performance of the GBTDS light curves, and directly inferred the transit detection efficiency over all stars in the GBTDS observing footprint with $F146<21$.

Combining the results from these detailed transit search simulations with planet occurrence rates from Kepler and a simulated stellar population adjusted to match the empirical NIR luminosity function of the Galactic bulge, we predict that the GBTDS should observe 59$\times$10$^6$ bright stars, which should yield the discovery of between $\sim$60,000 and $\sim$200,000 transiting planets.

In this section, we present the major results and implications of this work, including the scientific opportunities afforded by the GBTDS transiting planet survey (section \ref{sec:science}), implications for mission design (section \ref{sec:mission_design}), and finally the expected quality of the GBTDS light curves and implications of the GBTDS observing strategy (section \ref{sec:performance}).

\subsection{Predicted Yields and Implications for Exoplanet Demographics}\label{sec:science}

The allure of the GBTDS transiting planet science case primarily derives from two sources. First, the stellar densities of the Galactic bulge which, combined with Roman's field of view, offer the largest homogeneously analyzed collection of planet-search stars to date. 
The result will be an extensive planet sample which will yield statistically significant results, even for intrinsically rare objects and events.  
The following predictions and opportunities are direct results of the large number of stars searched in the GBTDS. 
\begin{enumerate}
    \item Of the $\sim$60,000-200,000 total transiting exoplanet detections, 90-95\% ($\sim$54,000-180,000) will be giants ($R_p > 4 R_\oplus$), potentially increasing the number of known giant planets by as much as two orders of magnitude.  
    \item The large number of surveyed M dwarfs should yield the detection of $329^{+51}_{-31}$ giant planets with such host stars, which is over an order of magnitude more such planets than are currently known. Assuming that the occurrence of giant planets around M dwarfs scales with metallicity similarly to giant planets around FGK hosts, this yield would instead be $\sim$2600, over two orders of magnitude more than are currently known. Given the deep transit depths in such systems and the self-imposed limit of $F146<21$ in this study, this yield estimate is likely conservative. 
    \item The GBTDS transiting planet yield will primarily consist of planets on close-in orbits ($a \lesssim 0.3$~au). However, the large number of surveyed stars should combat the low transit probability for stars at large orbital separations, facilitating the discovery of an additional $\sim$12,000 planets at larger orbital separations ($a \gtrsim 0.3$~au), including $\sim$1,800 single-transiting planets with orbital separations as large as $a\sim7$~au. 
    \item The GBTDS transit sensitivity for planets with $M_p \gtrsim 30 M_\oplus$ and $a\sim$~0.3-3~au should overlap significantly with the microlensing survey (see Figure \ref{fig:combined_yield}). 
    If the completeness and reliability of both surveys can be adequately characterized, then a joint analysis of both surveys could constrain the eccentricity distributions, planet density distributions, mutual inclinations, and multi-planet frequency at large orbital separations for giant planets at the population level. Such inferences will provide fundamental tests of core accretion and constrain the most prominent migration pathways for close-in giant planets (see section \ref{sec:discussion_microlensing}).  
\end{enumerate}

The second allure of the GBTDS transit survey is the depth of observations, which yield high detection efficiencies for transiting planets orbiting stars with $F146 = 21$~mag, of which many are at large distance. Although we only simulated the planet population down to this limit, transit searches over dimmer stars should still be effective, with the primary limitation being crowding and source catalog completeness. 
This depth translates to a planet population over an extensive range of Galactic environments and stellar host parameters, allowing for demographics analyses not possible in the stellar neighborhood. The following results are direct consequences of either the breadth of Galactic environments surveyed or the high detection efficiencies for dim stars. 
\begin{enumerate}
    \item The overall transiting planet yield has a strong dependence on the Milky Way's metallicity distribution function.
    Adjusting for the correlation between stellar metallicity and planet occurrence, the total transiting planet yield could realistically range from 135,000 to 200,000. Even for small planets, adjusting for the Milky Way's metallicity distribution function results in a significant increased yield of $\sim$7,000 to $\sim$12,000, highlighting an opportunity to constrain the role of bulk metallicity and stellar composition as a whole on the demographics of both small and giant planets (see section \ref{sec:yield_metals}). 
    \item The GBTDS data will be sensitive to transiting planets at distances of $\gtrsim$16-20 kpc for giants ($R_p>4 R_\oplus$), distances of $\sim$8-10 kpc for sub-Neptunes ($R_p=$~2-4~$R_\oplus$), and distances of $\sim$3-4 kpc for super-Earths ($R_p<2 R_\oplus$).   These distances will enable the detection of $>$1000s of planets across each of the Galactic thin disk, thick disk, and bulge (see Figure \ref{fig:distyield}).    
    \item A typical M dwarf planet will have a distance of $\sim$1-3~kpc, nearly an order of magnitude further than most M dwarf planet candidates in current state of the art transit surveys. 
    \item The GBTDS will be sensitive to small habitable zone planets, particularly with low mass host stars ($M_\star \lesssim 0.3 M_\odot$). More specifically, the GBTDS should yield $39^{+7}_{-6}$ super-Earths in the conservative habitable zone and $114^{+17}_{-14}$ super-Earths in the optimistic habitable zone (see section \ref{sec:yield_mdwarf}). 
    \item Finally, the  breadth of Galactic populations surveyed should be sufficient to disentangle the relationships between stellar age, bulk metallicity, and $\alpha$ abundances on giant planet occurrence, which are degenerate in the local stellar neighborhood due to Galactic chemical evolution. To accomplish this task, we advocate for a forward modeling approach, with an exoplanet demographics model layered over a Galactic population model. Such a model would marginalize over [Fe/H] and [$\alpha$/Fe] distributions in the Milky Way and allow the inference of key observables such as the planet yield as a function of distance, Galactic radius, height above the mid plane, and proper motion distributions (see section \ref{sec:galactic}). 
\end{enumerate}

\subsection{Implications for Mission Design}\label{sec:mission_design}

To quantify survey design trades on the expected transiting exoplanet yield, we applied an analytic transit detection model, developed through detailed simulations, to our simulated exoplanet population while varying survey parameters such as the observing cadence, number of fields surveyed, and the season duration. 
We present these findings below. 
\begin{enumerate}
    \item The overall yield for Jupiters ($R_p > 8 R_\oplus$) with bright ($F146<21$) host stars is insensitive to all survey parameters except surveyed area (i.e., number of fields), implying that the yield of giant planets is not limited by $S/N$ and transit searches for giant planets will be effective even for significantly dimmer stars (see section \ref{sec:yield_variation}). 
    \item  For small plantes ($R_p < 4 R_\oplus$), a larger surveyed area yields more nearby small planets, while higher cadence observations yield more distant small planets. These effects nearly cancel for constant slew/settle times, making the overall yield for small planets ($R_p < 4 R_\oplus$) insensitive to the trade between surveyed area and observing cadence (see section \ref{sec:drm}).
    \item The small planet ($R_p < 4 R_\oplus$) yield depends sensitively on season duration, changing approximately linearly by nearly 2\% per day, for a total decrease of 21\% from a maximum season duration of 72 days to the minimum of 60 days (see section \ref{sec:season_duration}).
    \item Overall, these planet yield estimates are robust to the expected sources of correlated noise, with the exception of super-Earths ($R_p < 2 R_\oplus$). To prevent significant losses in the yield of super-Earths, instrumental noise (e.g., intra-pixel variations, detector non-linearities) should be controlled to $<$0.5 ppt, with a stretch goal of $<$0.3 ppt (see section \ref{sec:correlated_noise}). 
\end{enumerate}

\subsection{Expected GBTDS Performance and Implications for Analysis Strategies}\label{sec:performance}

Our transit detection sensitivity model was developed through detailed, pixel-level simulations of transiting planets in the GBTDS. 
These simulations incorporated a realistic stellar population to accurately model crowding as well as an authentic treatment of the Roman Wide Field Instrument detectors, including a readout scheme and ramp-fitting algorithm to properly model photometric uncertainties from bright and semi-saturated sources. 
Below are some implications from these simulations. 
\begin{enumerate}
    \item Crowding will dominate the photometric error budget for stars with $F146\gtrsim21$ (see Figure \ref{fig:photnoise}), and, due to semi-saturation effects, will be counter intuitively prominent for bright stars as well (see section \ref{sec:crowding}). This motivates the use of PSF photometry in the GBTDS, and the need to properly model the stellar background to avoid biases in the inferred planet radii caused by dilution (see section \ref{sec:discussion_psf}). 
    \item For stars with $F146<18$, semi-saturation effects will dominate the error budget. As a result, studies focused on such stars will need to characterize second-order detector effects, such as classical and count rate dependent non-linearities, which will need to be controlled to reach single-exposure photon-limit uncertainties of $\sigma_{\rm dpp} \lesssim 1$~ppt. 
    \item The GBTDS observing cadence is significantly longer than the exposure time.  As a result, high-frequency signals will be undersampled but not lost. In the context of transiting exoplanet searches, this could result in template mismatches that reduce detectability, particularly for short duration transits (see section \ref{sec:pipeline}).     
\end{enumerate}

%
\section{Conclusions} \label{sec:conclusions}
%

The goals of this work were to (1) estimate the yield of transiting exoplanets in the GBTDS, (2) characterize the expected performance of the GBTDS and quantify survey design trades on the overall transiting exoplanet science return, and (3) identify unique science opportunities in the GBTDS that cannot be conducted with current state of the art observatories.

To accomplish these goals, we generated a synthetic stellar catalog of the GBTDS observing footprint, adjusted to match the empirical NIR luminosity function of the Galactic bulge population. We combined this synthetic stellar catalog with planet occurrence rates from Kepler to simulate a transiting exoplanet population in the GBTDS observing fields. Finally, we combined these simulated exoplanet and stellar populations with an analytic model of the expected GBTDS transit detection sensitivity, which was validated against detailed simulations of a GBTDS transit-search simulated at the pixel level, to estimate the overall yield of transiting exoplanets.

Our results indicate that the GBTDS should yield at least $\sim$60,000 and as many as $\sim$200,000 transiting exoplanets, consisting of 90-95\% giant planets ($R_p > 4 R_\oplus$) on primarily close-in ($a \lesssim 0.3$~au) orbits. In addition to discovering nearly an order of magnitude more transiting exoplanets than the combined output of every other transit survey to date, the GBTDS should find planets around stars with distances as far as $\sim$16-20 kpc, probing significant fractions of nearly every major Galactic population in the Milky Way.

We quantified survey design trades on the transiting exoplanet yield by simulating variations on the default survey design. We found that the yield for small planets is insensitive to the number of fields searched but improves noticeably with longer seasons and higher cadence observations which improve the overall transit detection sensitivity. The yield for giant planets on the other hand is insensitive to cadence and season length but increases directly with the number of fields searched, implying that giant planets should still be efficiently detected for stars dimmer than $F146=21$~mag.

Finally, we identified several scientific opportunities for the GBTDS transiting exoplanet sample, which are driven by either the large number of stars searched, the breadth of Galactic environments surveyed, or a combination of both. 
Among these goals is quantifying the change in planet yield as a function of distance, which will provide unprecedented information regarding the effects of Galactic environment with varying age, metallicity, and $\alpha$ abundance distributions on planet demographics, a feat not possible with present-day transit surveys that are primarily limited to bright stars in the local stellar neighborhood.

\begin{acknowledgments}
We wish to thank the anonymous referee, whose suggestions significantly improved the organization and overall quality of this manuscript.
RFW acknowledges support by the NASA Postdoctoral Program at the NASA Goddard Space Flight Center, administered by Oak Ridge Associated Universities under contract with NASA. 
This material is based upon work supported by NASA under award number 80GSFC21M0002.
Resources supporting this work were provided by the NASA High-End Computing (HEC) Program through the NASA Center for Climate Simulation (NCCS) at Goddard Space Flight Center.
\end{acknowledgments}

\software{{\tt astropy} \citep{astropy}, {\tt batman} \citep{kreidberg2015}, {\tt matplotlib} \citep{matplotlib}, {\tt numpy} \citep{numpy}, {\tt pandas} \citep{pandas}, {\tt photutils} \citep{photutils}, {\tt RImTimSim} \citep{robertfwilson_2023_8221758}, {\tt scipy} \citep{scipy}, {\tt webbpsf} \citep{perrin2014}}

\bibliography{references}{}
\bibliographystyle{aasjournal}

\end{document}